\documentclass[reqno,11pt,letterpaper]{amsart}

\usepackage{mathtools} \mathtoolsset{showonlyrefs=true}
\usepackage{accents}
\usepackage{amsfonts,amsmath,amssymb,amsthm}
\usepackage{array}
\usepackage[english]{babel}
\usepackage{bbm}
\usepackage[font=sl,labelfont=bf]{caption}
\usepackage[usenames,dvipsnames]{color} 
\usepackage{colortbl}
\usepackage{enumerate}
\usepackage{enumitem}
\usepackage{float} \restylefloat{table}
\usepackage[left=1in,top=1in,bottom=1in,right=1in]{geometry} 
\usepackage{graphicx}
\usepackage[colorlinks=true,pdfstartview=FitV,linkcolor=blue,citecolor=blue,urlcolor=blue]{hyperref}
\usepackage{mathrsfs}
\usepackage[framemethod=default]{mdframed}
\usepackage{subcaption}
\usepackage{url}
\usepackage{verbatim}

\newtheorem{theorem}{Theorem}
\newtheorem{proposition}[theorem]{Proposition}
\newtheorem{lemma}[theorem]{Lemma}

\newtheorem{assumption}[theorem]{Assumption}
\theoremstyle{definition}

\newtheorem{remark}[theorem]{Remark}

\numberwithin{equation}{section}
\numberwithin{theorem}{section}

\def\cB{\mathcal{B}}
\def\cC{\mathcal{C}}

\def\cI{\mathcal{I}}
\def\cJ{\mathcal{J}}

\def\cU{\mathcal{U}}

\def\bE{\mathbb{E}}
\def\bF{\mathbb{F}}

\def\bN{\mathbb{N}}

\def\bP{\mathbb{P}}

\def\bR{\mathbb{R}}

\def\sA{\mathscr{A}}

\def\sF{\mathscr{F}}

\def\sT{\mathscr{T}}

\newcommand{\wh}{\widehat}
\newcommand{\wt}{\widetilde}

\def\cd{\;{\stackrel{\mathfrak {D}}{\longrightarrow}}\;}
\def\cp{\;{\stackrel{\mathbb{P}}{\longrightarrow}}\;}
\def\ed{{\,\stackrel{\mathfrak {D}}{=}\,}}

\definecolor{Red}{rgb}{1,0,0} \newcommand{\Red}{\color{Red}}
\definecolor{DRed}{rgb}{0.7,0.3,0} 
\definecolor{Green}{rgb}{0.2,0.5,0.2} 
\definecolor{Blue}{rgb}{0,0,1} 
\definecolor{PaleGrey}{rgb}{0.6,0.6,0.6} 
\definecolor{Purple}{rgb}{0.5,0.00,1} 

\title{Near-Maturity Asymptotics of Critical Prices of American Put Options under Exponential L\'{e}vy Models}

\author{Jos\'{e} E. Figueroa-L\'{o}pez}
\address{Department of Statistics and Data Science, Washington University in St. Louis, St. Louis, MO, 63130, USA}
\email{\tt figueroa-lopez@wustl.edu}

\author{Ruoting Gong}
\address{Mathematical Reviews, American Mathematical Society, Ann Arbor, MI, 48103, USA}
\email{\tt rxg@ams.org}

\date{\today}

\begin{document}
\begin{abstract}
In the present paper, we study the near-maturity ($t\rightarrow T^{-}$) convergence rate of the optimal early-exercise price $b(t)$ of an American put under an exponential L\'{e}vy model with a {\it nonzero} Brownian component. Two important settings, not previous covered in the literature, are considered. In the case that the optimal exercise price converges to the strike price ($b(T^{-})=K$), we contemplate models with negative jumps of unbounded variation (i.e., processes that exhibit high activity of negative jumps or sudden falls in asset prices). In the second case, when the optimal exercise price tend to a value lower than $K$, we consider infinite activity jumps (though still of bounded variations), extending existing results for models with finite jump activity (finitely many jumps in any finite interval). In both cases, we show that $b(T^{-})-b(t)$ is of order $\sqrt{T-t}$ with explicit constants proportionality. Furthermore, we also derive the second-order near-maturity expansion of the American put price around the critical price along a certain parabolic branch.



\bigskip
\noindent
\textbf{JEL Classification}: C6
\end{abstract}

\maketitle

\section{Introduction}\label{sec:Intro}

\vspace{0.3cm}
\noindent
It is generally recognized that the standard Black-Scholes option pricing model is inconsistent with options data, while remaining a widely used model in practice because of its simplicity. Exponential L\'{e}vy models provides a tractable extension of the classical Black-Scholes setup by allowing jumps in stock prices and heavy-tailed return distributions, while preserving the independence and stationarity of returns. We refer the readers to the monograph \cite{ContTankov:2004} for further motivations and literature on the use of jump processes in financial modeling. A large number of publications have been devoted to the pricing of European options under various L\'evy-based models. In this paper, we revisit the problem of American put option pricing with finite maturity under general exponential L\'{e}vy models and, especially, on the study of the near-maturity asymptotics of the critical price (the exercise boundary) in this setting. No closed-form solution are known for American options and, as a result, numerical and approximation methods are employed in practice.

The near-maturity behavior of the critical price of the American put is well understood in the Black-Scholes model. In the pioneer work of Moerbeke \cite{Moerbeke:1976}, the near-maturity limit of the critical price for American call options is investigated, which can be easily transferred to American puts. Moreover, the conclusion in \cite{Moerbeke:1976} suggests a parabolic behavior for the convergence rate without any restrictions on the model parameters. However, Barles et. al. \cite{BarlesBurdeauRomanoSansoen:1995} show that a parabolic behavior cannot occur in some situations. Indeed, they show that, in the absence of dividends,
\begin{align}\label{eq:ConvRateBSIntoverDiv}
\lim_{t\rightarrow T^{-}}\frac{K-b^{\text{BS}}(t)}{\sigma K\sqrt{-(T-t)\ln(T-t)}}=1,
\end{align}
where $T$ is the maturity, $K$ is the strike price, $\sigma$ is the volatility, and $b^{\text{BS}}(t)$ is the critical price at time $t$. This asymptotic behavior remains valid when $0\leq\delta<r$, where $\delta$ and $r$ denote the respective constant dividend and interest rates, and can be proved by the method of \cite{BarlesBurdeauRomanoSansoen:1995}. Lamberton and Villeneuve \cite{LambertonVilleneuve:2003} provide the near-maturity behavior of the critical price for the cases $r<\delta$ and $r=\delta$. More precisely, they show that the parabolic behavior stated by Moerbeke \cite{Moerbeke:1976} holds in the case $r<\delta$. They also prove that, when $r=\delta$, the critical price satisfies the following estimate:
\begin{align}\label{eq:ConvRateBSIntEquDiv}
\lim_{t\rightarrow T^{-}}\frac{K-b^{\text{BS}}(t)}{\sigma K\sqrt{-(T-t)\ln(T-t)}}=\sqrt{2}.
\end{align}

The above results have been generalized to exponential L\'{e}vy models under various conditions. Let $\nu$ be the L\'{e}vy measure of the underlying L\'{e}vy process $X$, and let
\begin{align}\label{eq:Defd}
d:=r-\delta-\int_{0+}^{\infty}\big(e^{z}-1\big)\nu(dz).
\end{align}
Note that in the Black-Scholes model, the quantity $d$ reduces to $d=r-\delta$. Lamberton and Mikou \cite{LambertonMikou:2008} provide the near-maturity limits of the critical price of the American put for both the cases $d\geq 0$ and $d<0$ (see Theorem \ref{thm:LimitAmerCritPrice} below). When $X$ has finite jump activity (i.e., $\nu(\mathbb{R}\backslash\{0\})<\infty$ or, equivalently, the process exhibits finitely many jumps on each finite time interval), and a nonzero Brownian component, the convergence rates of the critical price for all three cases $d>0$, $d=0$, and $d<0$ have been fully studied by Bouselmi and Lamberton \cite{BouselmiLamberton:2016} (see also \cite{Pham:1997}). In the former two cases, they recover the same convergence rates as in \eqref{eq:ConvRateBSIntoverDiv} and \eqref{eq:ConvRateBSIntEquDiv}, respectively, while in the last case, they show that the critical price exhibits an analogous parabolic behavior to the Black-Scholes framework (see Theorem \ref{thm:FinJumpActAmerCritPriceConvRate} below). When $X$ has infinite jump activity, Lamberton and Mikou \cite{LambertonMikou:2013} obtain the convergence rates of the critical price in the following scenarios with $d>0$ (see Theorem \ref{thm:InfVarTSNegJumpAmerCritPriceConvRate} below):
\begin{itemize}[leftmargin=2.0em]
\item $X$ is a pure-jump L\'{e}vy process with a jump component of finite variation;
\item $X$ has a nonzero Brownian component and a jump component of finite variation;
\item $X$ is a pure-jump L\'{e}vy process with tempered stable negative small jumps of infinite variation.
\end{itemize}
In particular, they recover the convergence rate \eqref{eq:ConvRateBSIntoverDiv} in the second case above.

In the present paper, we study the near-maturity convergence rate of the critical price of the American put under an exponential L\'{e}vy model with a {\it nonzero} Brownian component, which is arguably more relevant for financial modelling. Firstly, we consider the scenario of $d>0$ without imposing any restriction on the jump activity of $X$\footnote{Even though, by definition, $d>0$ implies that the positive jump component of the process is of bounded variation.}, for which we recover the convergence rate \eqref{eq:ConvRateBSIntoverDiv}, thus extending the corresponding results in \cite{BarlesBurdeauRomanoSansoen:1995,BouselmiLamberton:2016,LambertonMikou:2013} to exponential L\'{e}vy models with a possibly negative jumps of infinite variation. Our analysis combines a careful decomposition of the jump component of $X$ with a comparison argument between the European and American critical prices analogous to \cite{LambertonMikou:2013}. Secondly, we consider the case of $d<0$ and assume that the jump component of $X$ is of finite variation, for which we obtain a parabolic behavior similar to those derived in \cite{BouselmiLamberton:2016,LambertonVilleneuve:2003}. As a byproduct, we also derive the second-order near-maturity expansion of the American put price around the critical price along a certain parabolic branch.

The rest of the article is organized as follows. Section \ref{sec:Prelim} provides preliminary definitions and properties of the American put option price and the corresponding critical price under an exponential L\'{e}vy model. Section \ref{sec:KnownAmerCritPriceConvRate} reviews some existing results on near-maturity asymptotics of the critical price which are related to our study. Sections \ref{sec:NonZeroBMPosdAmerCritPriceConvRate} establishes the convergence rate of the critical price under an exponential L\'{e}vy model with a nonzero Brownian component and $d>0$. Section \ref{sec:NonZeroBMNegdAmerCritPriceConvRate} provides the convergence rate of the critical price under a similar model but with $d<0$ and a jump component of finite variations. Some technical proofs are deferred to the appendices.

\vspace{0.2cm}

\section{Setup and Preliminary Results}\label{sec:Prelim}

\vspace{0.3cm}
\noindent
Throughout this paper, we consider a risky asset with price process $S:=(S_{t})_{t\in\bR_{+}}$, where $\bR_{+}:=[0,\infty)$, defined on a complete filtered probability space $(\Omega,\sF,\bF,\bP)$, where $\bF:=(\sF_{t})_{t\in\bR_{+}}$ and
\begin{align}\label{eq:ExpLevyModel}
S_{t}=S_{0}\,e^{(r-\delta)t+X_{t}},\quad t\in\bR_{+}.
\end{align}
Above, $r\in\bR_{+}$ is the interest rate, $\delta\in\bR_{+}$ is the dividend yield, and $X:=(X_{t})_{t\in\bR_{+}}$ is a L\'{e}vy process with L\'{e}vy triplet $(b,\sigma^{2},\nu)$.

\begin{assumption}\label{assump:NoArbCondsLevy}
Throughout we will always assume that $X$ satisfies at least one of the conditions in each of the following two categories:
\begin{itemize}[leftmargin=2.0em]

\vspace{0.1cm}
\item [(i)] $\sigma\neq 0$, $\nu((-\infty,0))>0$, or $\displaystyle{\int_{(0,1]}z\,\nu(dz)=\infty}$;

\vspace{0.2cm}
\item [(ii)] $\sigma\neq 0$, $\nu((0,\infty))>0$, or $\displaystyle{\int_{[-1,0)}z\,\nu(dz)=\infty}$.
\end{itemize}
\end{assumption}

\begin{remark}
By \cite[Theorem 21.5]{Sato:1999}, Assumption \ref{assump:NoArbCondsLevy} implies that the trajectories of $X$ are neither almost surely increasing nor almost surely decreasing. This, together with \cite[Proposition 9.9]{ContTankov:2004}, ensures that the exponential L\'{e}vy model \eqref{eq:ExpLevyModel} is arbitrage-free (or equivalently, the existence of martingale measures).
\end{remark}

\begin{assumption}\label{assump:MartMeas}
To guarantee that $\bP$ is a martingale measure for the discounted price process $(e^{-(r-\delta)t}S_{t})_{t\in\bR_{+}}$ (or equivalently, $(e^{X_{t}})_{t\in\bR_{+}}$), we impose the following two assumptions on $X$:
\begin{align*}
\text{(i)}\,\,\,\int_{1}^{\infty}e^{z}\,\nu(dz)<\infty;\qquad\text{(ii)}\,\,\,\,b=-\frac{\sigma^{2}}{2}-\int_{\bR_{0}}\big(e^{z}-1-z{\bf 1}_{\{|z|\leq 1\}}\big)\nu(dz),
\end{align*}
where $\bR_{0}:=\bR\setminus\{0\}$.
\end{assumption}

In view of Assumption \ref{assump:MartMeas}, we have the L\'{e}vy-It\^{o} decomposition of $X$ as
\begin{align*}
X_{t}=\sigma W_{t}+L_{t}:=\sigma W_{t}-\frac{\sigma^{2}t}{2}-t\int_{\bR_{0}}\big(e^{z}-1-z\big)\nu(dz)+\int_{0}^{t}\int_{\bR_{0}}z\,\wt{N}(ds,dz),\quad t\in\bR_{+},
\end{align*}
where $W:=(W_{t})_{t\in\bR_{+}}$ is a standard Brownian motion, $N(ds,dz)$ is a Poisson random measure on {$\bR_{+}\times\bR_{0}$} with intensity measure $ds\nu(dz)$, independent of $W$, and $\wt{N}(ds,dz):=N(ds,dz)-ds\nu(dz)$ is the compensated measure of $N$.

Let $P_{e}(t,s)$ and $P(t,s)$ be the respective time-$t$ risk-neutral prices of the European and American put options on $S$, with strike $K\in(0,\infty)$ and maturity $T\in(0,\infty)$, namely,
\begin{align}\label{eq:EuroPutPrice}
P_{e}(t,s)&:=\bE\Big(e^{-r(T-t)}\big(K-s\,e^{(r-\delta)(T-t)+X_{T-t}}\big)^{+}\Big),\quad (t,s)\in[0,T]\times\bR_{+},\\
\label{eq:AmerPutPrice} P(t,s)&:=\sup_{\tau\in\sT_{0,T-t}}\bE\Big(e^{-r\tau}\big(K-s\,e^{(r-\delta)\tau+X_{\tau}}\big)^{+}\Big),\quad (t,s)\in[0,T]\times\bR_{+},
\end{align}
where $\sT_{u,v}$ denotes the collection of $\bF$-stopping times taking values in $[u,v]$, for any $0\leq u\leq v$. Assumption \ref{assump:NoArbCondsLevy} ensures that $P(t,s)\geq P_{e}(t,s)>0$ for all $(t,s)\in[0,T]\times\bR_{+}$. Moreover, we define the European and American critical prices, respectively, as
\begin{equation}\label{CrtclPrcs}
\begin{aligned}
b_{e}(t)&:=\inf\big\{s\in\bR_{+}\!:\,P_{e}(t,s)>(K-s)^{+}\big\},\quad t\in[0,T],\\
b(t)&:=\inf\big\{s\in\bR_{+}\!:\,P(t,s)>(K-s)^{+}\big\},\quad t\in[0,T].
\end{aligned}
\end{equation}
Clearly, $b(T)=b_{e}(T)=K$. The following proposition summaries some regularity properties of $P$ (cf. \cite[Section 12.1.3]{ContTankov:2004}, \cite[Proposition 3.2]{LambertonMikou:2008}, \cite[Proposition 2.2]{LambertonMikou:2013}, and \cite[Proposition 2.1]{Pham:1997}).

\begin{proposition}\label{prop:RegPropAmerPutPrice}
\begin{itemize}[leftmargin=2.0em]
\item [(a)] For each $t\in[0,T]$, $P(t,\cdot)$ is nonincreasing and convex on $\bR_{+}$, and satisfies
    \begin{align*}
    \big|P(t,s_{1})-P(t,s_{2})\big|\leq|s_{1}-s_{2}|,\quad s_{1},s_{2}\in\bR_{+}.
    \end{align*}

\item [(b)] For each $s\in\bR_{+}$, $P(\cdot,s)$ is continuous and nonincreasing on $[0,T]$.
\end{itemize}
\end{proposition}

It follows that (cf. \cite[Proposition 4.1]{LambertonMikou:2008} and \cite[Proposition 2.2]{Pham:1997}) the American critical price $b$ is nondecreasing on $[0,T]$ and that
\begin{align*}
b(t)\in(0,K),\quad P\big(t,b(t)\big)=K-b(t),\quad\text{for any }\,t\in[0,T).
\end{align*}
Similar results holds for $b_{e}(t)$ and $P_{e}(t,s)$, namely, $b_{e}$ is nondecreasing on $[0,T]$ and
\begin{align}\label{eq:PropEuroCritPrice}
b_{e}(t)\in(0,K),\quad P_{e}\big(t,b_{e}(t)\big)=K-b_{e}(t),\quad\text{for any }\,t\in[0,T).
\end{align}
Moreover, since $P(t,s)\geq P_{e}(t,s)$ for all $(t,s)\in[0,T]\times\bR_{+}$, we have
\begin{align}\label{eq:RelEuroAmerCritPrice}
0<b(t)\leq b_{e}(t)<K,\quad\text{for all }\,t\in[0,T).
\end{align}

We are interested in the near-maturity asymptotic behavior of $b(t)$ and $P(t,s)$, i.e., when $t\rightarrow T^{-}$. The following result provides the limit of the critical price near maturity (cf. \cite[Theorem 4.4]{LambertonMikou:2008}).

\begin{theorem}\label{thm:LimitAmerCritPrice}
Let $d$ be defined as in \eqref{eq:Defd}. Then, the following assertions hold:
\begin{itemize}[leftmargin=2.0em]

\vspace{0.1cm}
\item [(a)] If $d\geq 0$, then we have
    \begin{align*}
    b(T):=\lim_{t\rightarrow T^{-}}b(t)=K.
    \end{align*}
\item [(b)] If $d<0$, then we have
    \begin{align*}
    b(T):=\lim_{t\rightarrow T^{-}}b(t)=\xi,
    \end{align*}
    where $\xi$ is the unique solution in $(0,K)$ to the following equation:
    \begin{align}\label{eq:LimitAmerCritPriceNegd}
    rK-\delta\xi-\int_{\bR_{0}}\big(\xi e^{z}-K\big)^{+}\nu(dz)=0.
    \end{align}
\end{itemize}
\end{theorem}

\begin{remark}\label{rem:PosdPosJumpFinVar}
When $d\geq 0$, it is intrinsically assumed that
\begin{align*}
\int_{0+}^{\infty}\big(e^{z}-1\big)\nu(dz)<\infty{\Red ,}
\end{align*}
and, thus, this case intrinsically entails that the positive jump part of $X$ has finite variation.
\end{remark}

\vspace{0.2cm}

\section{The Known Cases of the Convergence Rate of the Critical Price}\label{sec:KnownAmerCritPriceConvRate}

\vspace{0.4cm}
\noindent
In this section, we review some known results on the asymptotic behavior of the critical exercise price $b$ as defined in \eqref{CrtclPrcs} near maturity.

\vspace{0.2cm}

\subsection{Finite Jump Activity Case}\label{subsec:FinJumpActAmerCritPriceConvRate}$\,$

\vspace{0.3cm}
\noindent
The following result is a combination of \cite[Theorem 4.2]{Pham:1997} and \cite[Theorems 3.2 \& 4.1]{BouselmiLamberton:2016}.

\begin{theorem}\label{thm:FinJumpActAmerCritPriceConvRate}
Assume that $\sigma>0$ and $\nu(\bR_{0})<\infty$.
\begin{itemize}[leftmargin=2.0em]

\vspace{0.1cm}
\item [(a)] When $d>0$, we have
    \begin{align*}
    K-b(t)\sim\sigma K\sqrt{(T-t)\big|\ln(T-t)\big|},\quad t\rightarrow T^{-}.
    \end{align*}

\item [(b)] When $d=0$, we have
    \begin{align*}
    K-b(t)\sim\sqrt{2}\,\sigma K\sqrt{(T-t)\big|\ln(T-t)\big|},\quad t\rightarrow T^{-}.
    \end{align*}

\item [(c)] Suppose that $d<0$. Let $\xi$ be given as in Theorem \ref{thm:LimitAmerCritPrice}-(b), and denote the local time of $W$ at $x\in\bR$ by $L^{W}(x)$. For any $\lambda,\beta\in\bR_{+}$ and $y\in\bR$, we define
    \begin{align}\label{eq:DefFuntvlambdabeta}
    v_{\lambda,\beta}(y)\!:=\!\!\sup_{\tau\in\sT_{0,1}^{W,\overline{N}}}\!\bE\bigg(e^{-\lambda\tau}{\bf 1}_{\{\overline{N}_{\tau}=0\}}\!\int_{0}^{\tau}\!f_{\lambda\beta}(y\!+\!W_{s})\,ds+\frac{\beta e^{\lambda\tau}}{2}{\bf 1}_{\{\overline{N}_{\tau}=1\}}\!\Big(L_{\tau}^{W}(-y)\!-\!L_{\overline{T}_{1}}^{W}(-y)\Big)\!\bigg),\,\,
    \end{align}
    where $\overline{N}:=(\overline{N}_{t})_{t\in\bR_{+}}$ is a Poisson process with intensity $\lambda$ which is independent of $W$, $\overline{T}_{1}$ is the first jump time of $\overline{N}$, $\sT_{0,1}^{W,\overline{N}}$ is the collection of $\bF^{W,\overline{N}}$-stopping times taking values in $[0,1]$, and $f_{a}(x):=x+ax^{+}$. Let $y_{\lambda,\beta}:=-\inf\{x\in\bR:\,v_{\lambda,\beta}(x)>0\}$.
    \begin{itemize}

    \vspace{0.1cm}
    \item [(c.1)] If $\nu(\{\ln(K/\xi)\})=0$, then we have
        \begin{align*}
        \xi-b(t)\sim y_{0,0}\sigma\xi\sqrt{T-t},\quad t\rightarrow T^{-}.
        \end{align*}

    \item [(c.2)] If $\nu(\{\ln(K/\xi)\})>0$, then we have
        \begin{align*}
        \xi-b(t)\sim y_{\lambda,\beta}\sigma\xi\sqrt{T-t},\quad t\rightarrow T^{-},
        \end{align*}
        with $\lambda:=\nu(\{\ln(K/\xi)\})$ and $\beta:=K/\xi(\delta+\int_{(\ln(K/\xi),\infty)}e^{z}\nu(dz))$.
    \end{itemize}
\end{itemize}
\end{theorem}

\vspace{0.1cm}

\subsection{Finite Variation Case}\label{subsec:FinVarAmerCritPriceConvRate}$\,$

\vspace{0.3cm}
\noindent
In this section, we consider in \eqref{eq:ExpLevyModel} a L\'{e}vy process $X$ with a finite variation jump component, i.e.,
\begin{align}\label{eq:FinVarLevy}
\int_{(-1,1)\setminus\{0\}}|z|\,\nu(dz)<\infty.
\end{align}
In this case, the convergence rate of the American critical price is only known when $d>0$.

For a pure-jump ($\sigma=0$) L\'{e}vy process $X$, the following result is due to \cite[Theorems 5.1 \& 5.2]{LambertonMikou:2013}.

\begin{theorem}\label{thm:PureJumpFinVarAmerCritPriceConvRate}
Assume $\sigma=0$ and that \eqref{eq:FinVarLevy} holds true. Then, we have
\begin{align*}
\lim_{t\rightarrow T^{-}}\frac{1}{T-t}\bigg(\frac{K}{b_{e}(t)}-1\bigg)=\int_{\bR_{0}}\big(e^{x}-1\big)^{-}\nu(dx)\quad\text{and}\quad\lim_{t\rightarrow T^{-}}\frac{b_{e}(t)-b(t)}{T-t}=0.
\end{align*}
Consequently, we have
\begin{align*}
\lim_{t\rightarrow  T^{-}}\frac{1}{T-t}\bigg(\frac{K}{b(t)}-1\bigg)=\int_{\bR_{0}}\big(e^{x}-1\big)^{-}\nu(dx).
\end{align*}
\end{theorem}

Next, we consider the case when $X$ has a non-zero Brownian component (i.e., $\sigma>0$). Let $b^{\text{BS}}$ be the American critical price under the Black-Scholes model. It was shown in \cite{BarlesBurdeauRomanoSansoen:1995} that, when $d>0$ (which simply reduces to $r>\delta$ in the Black-Scholes model),
\begin{align}\label{eq:BSAmerCritPriceConvRate}
K-b^{\text{BS}}(t)\sim\sigma K\sqrt{-(T-t)\ln(T-t)},\quad t\rightarrow T^{-}.
\end{align}
The following result follows from \cite[Theorem 4.1 \& Corollary 4.1]{LambertonMikou:2013}.

\begin{theorem}\label{thm:NonZeroBMFinVarAmerCritPriceConvRate}
Assume $\sigma>0$, $d>0$, and that \eqref{eq:FinVarLevy} holds true. Then, there exists $C\in(0,\infty)$ such that
\begin{align*}
0\leq b^{\emph{BS}}(t)-b(t)\leq C\sqrt{T-t},\quad t\rightarrow T^{-}.
\end{align*}
Together with \eqref{eq:BSAmerCritPriceConvRate}, we obtain that
\begin{align*}
K-b(t)\sim\sigma K\sqrt{-(T-t)\ln(T-t)},\quad t\rightarrow T^{-}.
\end{align*}
\end{theorem}

\vspace{0.1cm}

\subsection{Infinite Variation Case}\label{subsec:InfVarAmerCritPriceConvRate}$\,$

\vspace{0.3cm}
\noindent
In this section, we consider in \eqref{eq:ExpLevyModel} a L\'{e}vy process $X$ with an infinite variation jump part, i.e.
\begin{align}\label{eq:InfVarLevy}
\int_{(-1,1)\setminus\{0\}}|z|\,\nu(dz)=\infty.
\end{align}
In this case, the convergence rate of the American critical price is only known when $d>0$ with some additional assumption on the negative jumps of $X$. Note that when $d>0$, in view of Remark \ref{rem:PosdPosJumpFinVar}, only the negative jump component of $X$ has infinite variation.

\begin{remark}\label{rem:LinearAmerCritPriceConvRate}
When $d\geq 0$, it was shown in \cite[Theorem 6.1]{LambertonMikou:2013} that the convergence rate of $b(t)$ to $K$ cannot be linear for an infinite variation L\'{e}vy process $X$ (i.e., either $\sigma>0$ or \eqref{eq:InfVarLevy} holds). Indeed, in this case we have
\begin{align*}
\lim_{t\rightarrow T^{-}}\frac{1}{T-t}\bigg(\frac{K}{b(t)}-1\bigg)=\infty.
\end{align*}
\end{remark}

The following result, due to \cite[Theorem 7.1]{LambertonMikou:2013}, provides the convergence rate of $b$ when $d>0$ and $X$ is a pure-jump L\'{e}vy process with infinite variation tempered stable negative small jumps.

\begin{theorem}\label{thm:InfVarTSNegJumpAmerCritPriceConvRate}
Assume that $\sigma=0$, $d>0$, and that there exist $z_{0}\in(-\infty,0)$, a positive bounded Borel measurable function $\eta$ on $[z_{0},0)$ satisfying $\lim_{z\rightarrow 0}\eta(z)=\eta_{0}>0$, and $\alpha\in(1,2)$, such that
\begin{align*}
{\bf 1}_{(z_{0},0)}(z)\,\nu(dz)=\frac{\eta(z)}{|z|^{1+\alpha}}{\bf 1}_{(z_{0},0)}(z)\,dz.
\end{align*}
Then we have
\begin{align*}
\lim_{t\rightarrow T^{-}}\frac{K-b(t)}{(T-t)^{1/\alpha}\big|\ln(T-t)\big|^{1-1/\alpha}}=K\bigg(\frac{\eta_{0}\,\Gamma(2-\alpha)}{\alpha-1}\bigg)^{1/\alpha}.
\end{align*}
\end{theorem}

\vspace{0.2cm}

\section{New Results on the {Convergence Rate} of the Critical Price when \texorpdfstring{$d>0$}{}}\label{sec:NonZeroBMPosdAmerCritPriceConvRate}

\vspace{0.4cm}
\noindent
In this section, we consider the rate of convergence of $b$ near maturity when $\sigma>0$ and $d>0$.

\begin{theorem}\label{thm:NonZeroBMPosdAmerCritPriceConvRate}
Assume that $\sigma>0$ and $d>0$. Then we have
\begin{align*}
K-b(t)=\sigma K\sqrt{-(T-t)\ln(T-t)}+O\big(\sqrt{T-t}\big),\quad t\rightarrow T^{-}.
\end{align*}
\end{theorem}

The proof of Theorem \ref{thm:NonZeroBMPosdAmerCritPriceConvRate} follows a similar plan as in \cite[Sections 5 \& 7]{LambertonMikou:2013}. Namely, we first characterize the rate of convergence of the European critical price (Proposition \ref{prop:NonZeroBMPosdEuroCritPriceConvRate} below), and then, we {proceed to} estimate the difference between the European and the American critical prices (Proposition \ref{prop:NonZeroBMPosdDiffEuroAmerCritPrices} below).

To begin with, recalling that $X$ has a finite variation positive jump component when $d>0$ (Remark \ref{rem:PosdPosJumpFinVar}), we introduce the following decomposition of $X$:
\begin{align}\label{eq:NonZeroBMPosdDecompLevy}
X_{t}=X_{t}^{W}-t\int_{\bR_{0}}\big(e^{z}-1\big)^{+}\nu(dz)+J_{t}^{+}+J_{t}^{-},\quad t\in\bR_{+},
\end{align}
where
\begin{align*}
X_{t}^{W}\!:=\sigma W_{t}\!-\!\frac{\sigma^{2}t}{2},\,\,\,\,J_{t}^{+}\!:=\!\!\int_{0}^{t}\!\!\int_{(0,\infty)}\!\!zN(ds,dz),\,\,\,\,J_{t}^{-}\!:=\!\!\int_{0}^{t}\!\!\int_{(-\infty,0)}\!\!z\wt{N}(ds,dz)-t\!\int_{-\infty}^{0}\!\!\big(e^{z}\!-\!1\!-\!z\big)\nu(dz).
\end{align*}
Clearly, $X^{W}:=(X_{t}^{W})_{t\in\bR_{+}}$, $J^{+}:=(J^{+}_{t})_{t\in\bR_{+}}$, and $J^{-}:=(J^{-}_{t})_{t\in\bR_{+}}$ are independent of each other. Note that for any $p\in\bR_{+}$ (cf. \cite[Lemma 7.3]{LambertonMikou:2013}),
\begin{align}\label{eq:MGFJMinus}
\bE\big(e^{pJ_{t}^{-}}\big)=\exp\bigg(t\int_{(-\infty,0)}\big(e^{pz}-1-p(e^{z}-1)\big)\nu(dz)\bigg),\quad t\in\bR_{+}.
\end{align}
The following lemma characterizes the small-time behavior for a L\'{e}vy process with $\sigma>0$ and a finite variation positive jump component,
the proof of which is deferred to Appendix \ref{sec:AppendixA}.

\begin{lemma}\label{lem:NonZeroBMPosdLevySmallTimeAsymNormal}
Assume that $\sigma>0$ and $d>0$, then $X_{t}/\sqrt{t}\cd\sigma W_{1}$, as $t\rightarrow 0^{+}$.
\end{lemma}

\begin{remark}\label{rem:SmallTimeLimitX}
Lemma \ref{lem:NonZeroBMPosdLevySmallTimeAsymNormal} is valid regardless of the value of $d$. Indeed, in the proof of Theorem 8.1-(ii) in Sato \cite{Sato:1999}, it is shown that  for any infinitely divisible distribution $\mu$ with Gaussian component $\sigma^{2}\in[0,\infty)$, it holds that (see p. 40 in \cite{Sato:1999})
\begin{align*}
\lim_{s\rightarrow\infty}s^{-2}\ln\wh{\mu}(sz)= -\frac{1}{2}\sigma^{2}z^{2},\quad z\in\bR,
\end{align*}
where $\wh{\mu}$ denotes the characteristic function of $\mu$. Taking $t=1/s^2$, it follows that
\begin{align*}
\lim_{t\rightarrow 0^{+}}t\ln\wh{\mu}(z/\sqrt{t})= -\frac{1}{2}\sigma^{2}z^{2},\quad z\in\bR.
\end{align*}
Now, let $\mu$ be the distribution of $X_{1}$, then the distribution of $X_{t}$ is $\mu^{t}$ due to the infinite divisibility, meaning that its characteristic function is given by $\wh{\mu}^{\,t}$. Hence, we obtain that
\begin{align*}
\lim_{t\rightarrow 0^{+}}\bE\big(e^{izX_{t}/\sqrt{t}}\big)=\lim_{t\rightarrow 0^{+}}\big(\wh{\mu}(z/\sqrt{t})\big)^{t}=e^{-\sigma^{2}z^{2}/2},\quad z\in\bR,
\end{align*}
and, thus, the result of Lemma \ref{lem:NonZeroBMPosdLevySmallTimeAsymNormal}.
\end{remark}

We are now ready to prove Theorem \ref{thm:NonZeroBMPosdAmerCritPriceConvRate}. The proof is divided into two steps which are presented in the following two subsections, respectively.

\vspace{0.2cm}

\subsection{Step 1: The rate of convergence of $b_{e}(t)$}$\,$

\vspace{0.3cm}
\noindent
Using \eqref{eq:ExpLevyModel}, \eqref{eq:EuroPutPrice}, \eqref{eq:PropEuroCritPrice} as well as the martingale property of $(e^{X_{t}})_{t\in\bR_{+}}$, we first have
\begin{align*}
K-b_{e}(t)&=P_{e}\big(t,b_{e}(t)\big)=e^{-r\tau}\bE\Big(\big(K-b_{e}(t)e^{(r-\delta)\tau+X_{\tau}}\big)^{+}\Big)\\
&=Ke^{-r\tau}-b_{e}(t)e^{-\delta\tau}+e^{-r\tau}\bE\Big(\big(b_{e}(t)e^{(r-\delta)\tau+X_{\tau}}-K\big)^{+}\Big),
\end{align*}
where $\tau=T-t$, and so
\begin{align*}
\frac{K}{b_{e}(t)}\big(1-e^{-r\tau}\big)-\big(1-e^{-\delta\tau}\big)=e^{-r\tau}\bE\left(\bigg(e^{(r-\delta)\tau+X_{\tau}}-\frac{K}{b_{e}(t)}\bigg)^{+}\right).
\end{align*}
Denote by
\begin{align}\label{eq:Defzeta}
\zeta(\tau):=\frac{K}{b_{e}(T-\tau)}-1.
\end{align}
By Theorem \ref{thm:LimitAmerCritPrice}-(a), we deduce that, as $\tau\rightarrow 0^{+}$,
\begin{align}\label{eq:AsymEuroPutPrice}
(r-\delta)\tau=e^{-r\tau}\bE\Big(\big(e^{(r-\delta)\tau+X_{\tau}}-1-\zeta(\tau)\big)^{+}\Big)+o(\tau)=\bE\Big(\big(e^{(r-\delta)\tau+X_{\tau}}-1-\zeta(\tau)\big)^{+}\Big)+o(\tau),\quad\,\,\,
\end{align}
where the second equality above follows from the fact that
\begin{align}\label{eq:LimitExprdeltaXZeta}
\lim_{\tau\rightarrow 0^{+}}\bE\Big(\big(e^{(r-\delta)\tau+X_{\tau}}-1-\zeta(\tau)\big)^{+}\Big)=0.
\end{align}
The following proposition shows the rate of convergence of $b_{e}(t)$ to $K$, as $t\rightarrow T^{-}$, by characterizing the rate of convergence of $\zeta(\tau)$ to $0$ as $\tau\rightarrow 0^{+}$.

\begin{proposition}\label{prop:NonZeroBMPosdEuroCritPriceConvRate}
Assume that $\sigma>0$ and $d>0$. Then we have
\begin{align*}
\lim_{\tau\rightarrow 0^{+}}\frac{\zeta(\tau)}{\sqrt{-\tau\ln\tau}}=\sigma.
\end{align*}
\end{proposition}

The first step of the proof of Proposition \ref{prop:NonZeroBMPosdEuroCritPriceConvRate} is the following lemma, the proof of which is deferred to Appendix \ref{sec:AppendixA}.

\begin{lemma}\label{lem:NonZeroBMPosdEuroCritPrice}
Assume that $\sigma>0$ and $d>0$. Then we have
\begin{align*}
\lim_{\tau\rightarrow 0^{+}}\frac{\zeta(\tau)}{\sqrt{\tau}}=\infty.
\end{align*}
\end{lemma}

The next two lemmas are seeking for a small-time large deviations result for $(X_{\tau}^{W}+J_{\tau}^{-})/\sqrt{\tau}$, the proof of which are again deferred to Appendix \ref{sec:AppendixA}.

\begin{lemma}\label{lem:NonZeroBMPosdTailProbBMNegJump}
Assume that $\sigma>0$ and $d>0$. For any $\varepsilon\in(0,1)$ and $r\in(0,\infty)$, let
\begin{align}\label{eq:ConstLtau0}
L=L(\varepsilon;\sigma):=\sigma^{2}+2\int_{[-\varepsilon,0)}z^{2}\nu(dz),\quad\tau_{0}=\tau_{0}(r,\sigma,\varepsilon):=\frac{r^{2}}{\big(\nu((-\infty,-\varepsilon))-\sigma^{2}/2\big)^{2}}.
\end{align}
Then, for any $\tau\in(0,\tau_{0}]$, we have
\begin{align*}
\bP\Big(X_{\tau}^{W}+J_{\tau}^{-}\geq r\sqrt{\tau}\Big)\leq e^{-Lp^{2}/2},
\end{align*}
where
\begin{align}\label{eq:Constp}
p=p(\tau;r,\sigma,\varepsilon)=\frac{1}{L}\bigg[r-\sqrt{\tau}\bigg(\nu((-\infty,-\varepsilon))-\frac{\sigma^{2}}{2}\bigg)\bigg]\in\bR_{+}.
\end{align}
\end{lemma}

\begin{lemma}\label{lem:NonZeroBMPosdLargeDevBMNegJump}
Assume that $\sigma>0$ and $d>0$. Then for any $f:\bR_{+}\rightarrow\bR_{+}$ with $\lim_{\tau\rightarrow 0^{+}}f(\tau)=\infty$,
\begin{align*}
\lim_{\tau\rightarrow 0^{+}}\frac{1}{f^{2}(\tau)}\ln\bP\Big(X_{\tau}^{W}+J_{\tau}^{-}\geq\sqrt{\tau}f(\tau)\Big)= -\frac{1}{2\sigma^{2}}.
\end{align*}
\end{lemma}

\noindent
\textbf{Proof of Proposition \ref{prop:NonZeroBMPosdEuroCritPriceConvRate}.} In view of \eqref{eq:NonZeroBMPosdDecompLevy} and the independence among $X^{W}$, $J^{+}$, and $J^{-}$, for any $\tau\in\bR_{+}$, by conditioning on $X^{W}_{\tau}+J^{-}_{\tau}$ and noting that,
\begin{align*}
\bE\big(e^{J_{\tau}^{+}}\big)=\exp\bigg(\tau\int_{(0,\infty)}\big(e^{z}-1\big)\,\nu(dz)\bigg),
\end{align*}
we first have
\begin{align*}
\bE\Big(\big(e^{(r-\delta)\tau+X_{\tau}}\!-\!1\!-\!\zeta(\tau)\big)^{+}\Big)&\geq\bE\!\left(\!\left(e^{(r-\delta)\tau+X_{\tau}^{W}+J_{\tau}^{-}}\!\exp\!\bigg(\!\!-\!\tau\!\!\int_{\bR_{0}}\!\!\big(e^{z}\!-\!1\big)^{+}\nu(dz)\!\bigg)\bE\big(e^{J_{\tau}^{+}}\big)\!-\!1\!-\!\zeta(\tau)\!\right)^{+}\right)\\
&=\bE\Big(\big(e^{(r-\delta)\tau+X_{\tau}^{W}+J_{\tau}^{-}}-1-\zeta(\tau)\big)^{+}\Big)\geq\bE\Big(\big(X_{\tau}^{W}+J_{\tau}^{-}-\zeta(\tau)\big)^{+}\Big),
\end{align*}
where we have used Jensen's inequality in the first inequality  and $e^{z}\geq 1+z$ in the last inequality. Together with \eqref{eq:AsymEuroPutPrice}, we deduce that
\begin{align*}
\ln\bE\left(\bigg(\frac{X_{\tau}^{W}+J_{\tau}^{-}}{\sqrt{\tau}}-\frac{\zeta(\tau)}{\sqrt{\tau}}\bigg)^{+}\right)\leq\frac{1}{2}\ln\tau+\ln(r-\delta)+o(1),\quad\tau\rightarrow 0^{+},
\end{align*}
which, together with Lemma \ref{lem:NonZeroBMPosdEuroCritPrice}, implies that
\begin{align}\label{eq:LimitInfScallnExpBMNegJump}
\liminf_{\tau\rightarrow 0^{+}}\frac{\tau}{\zeta^{2}(\tau)}\ln\bE\left(\bigg(\frac{X_{\tau}^{W}+J_{\tau}^{-}}{\sqrt{\tau}}-\frac{\zeta(\tau)}{\sqrt{\tau}}\bigg)^{+}\right)\leq\liminf_{\tau\rightarrow 0^{+}}\frac{\tau\ln\tau}{2\zeta^{2}(\tau)}.
\end{align}
Moreover, by Markov's inequality, for any $\beta\in(1,\infty)$ and $\tau\in\bR_{+}$ we have
\begin{align*}
\bP\Big(X_{\tau}^{W}+J_{\tau}^{-}\geq\beta\zeta(\tau)\Big)\leq\frac{\sqrt{\tau}}{(\beta-1)\zeta(\tau)}\bE\left(\bigg(\frac{X_{\tau}^{W}+J_{\tau}^{-}}{\sqrt{\tau}}-\frac{\zeta(\tau)}{\sqrt{\tau}}\bigg)^{+}\right).
\end{align*}
This, together with Lemma \ref{lem:NonZeroBMPosdLargeDevBMNegJump} (with $f(\tau)=\beta\zeta(\tau)/\sqrt{\tau}$) and \eqref{eq:LimitInfScallnExpBMNegJump}, implies that
\begin{align*}
\liminf_{\tau\rightarrow 0^{+}}\frac{\tau\ln\tau}{2\zeta^{2}(\tau)}\geq\lim_{\tau\rightarrow 0^{+}}\frac{\tau}{\zeta^{2}(\tau)}\ln\bP\Big(X_{\tau}^{W}+J_{\tau}^{-}\geq\beta\zeta(\tau)\Big)+\lim_{\tau\rightarrow 0^{+}}\frac{\ln\big((\beta-1)\zeta(\tau)/\sqrt{\tau}\big)}{\zeta^{2}(\tau)/\tau}= -\frac{\beta^{2}}{2\sigma^{2}},
\end{align*}
or equivalently,
\begin{align*}
\liminf_{\tau\rightarrow 0^{+}}\frac{\zeta(\tau)}{\sqrt{-\tau\ln\tau}}\geq\frac{\sigma}{\beta}.
\end{align*}
Since $\beta\in(1,\infty)$ is arbitrary, by taking $\beta\rightarrow 1^{+}$ above we obtain that
\begin{align*}
\liminf_{\tau\rightarrow 0^{+}}\frac{\zeta(\tau)}{\sqrt{-\tau\ln\tau}}\geq\sigma.
\end{align*}

In order to derive an upper bound for $\limsup_{\tau\rightarrow 0^{+}}\zeta(\tau)/\sqrt{-\tau\ln\tau}$, we first deduce from \eqref{eq:AsymEuroPutPrice} a lower bound for $\bE((e^{X_{\tau}^{W}+J_{\tau}^{-}}-1-\zeta(\tau))^{+})$ as $\tau\rightarrow 0^{+}$. Indeed, for any $\tau\in\bR_{+}$, we have
\begin{align}
\bE\Big(\big(e^{(r-\delta)\tau+X_{\tau}}-1-\zeta(\tau)\big)^{+}\Big)&=\bE\Big(e^{(r-\delta)\tau+X_{\tau}}{\bf 1}_{\{X_{\tau}\geq\ln(1+\zeta(\tau))+(r-\delta)\tau\}}\Big)\\
\label{eq:DecompExpExprdeltaXZeta} &\quad -\big(1+\zeta(\tau)\big)\bP\big(X_{\tau}\geq\ln(1+\zeta(\tau))+(r-\delta)\tau\big).
\end{align}
Note that Lemmas \ref{lem:NonZeroBMPosdLevySmallTimeAsymNormal} and \ref{lem:NonZeroBMPosdEuroCritPrice} imply that
\begin{align*}
\lim_{\tau\rightarrow 0^{+}}\bP\big(X_{\tau}\geq\ln(1+\zeta(\tau))+(r-\delta)\tau\big)=\lim_{\tau\rightarrow 0^{+}}\bP\bigg(\frac{X_{\tau}}{\sqrt{\tau}}\geq\frac{\ln(1+\zeta(\tau))}{\sqrt{\tau}}+(r-\delta)\sqrt{\tau}\bigg)=0.
\end{align*}
Hence, we obtain from \eqref{eq:LimitExprdeltaXZeta} and \eqref{eq:DecompExpExprdeltaXZeta} that
\begin{align*}
\lim_{\tau\rightarrow 0^{+}}\bE\Big(e^{X_{\tau}}{\bf 1}_{\{X_{\tau}\geq\ln(1+\zeta(\tau))+(r-\delta)\tau\}}\Big)=\lim_{\tau\rightarrow 0^{+}}\bE\Big(e^{(r-\delta)\tau+X_{\tau}}{\bf 1}_{\{X_{\tau}\geq\ln(1+\zeta(\tau))+(r-\delta)\tau\}}\Big)=0,
\end{align*}
and so
\begin{align*}
\bE\Big(\big(e^{(r-\delta)\tau+X_{\tau}}-1-\zeta(\tau)\big)^{+}\Big)&=\bE\Big(e^{X_{\tau}}{\bf 1}_{\{X_{\tau}\geq\ln(1+\zeta(\tau))+(r-\delta)\tau\}}\Big)\\
&\quad -\big(1+\zeta(\tau)\big)\bP\big(X_{\tau}\geq\ln(1+\zeta(\tau))+(r-\delta)\tau\big)+o(\tau),\quad\tau\rightarrow 0^{+}.
\end{align*}
This, together with \eqref{eq:AsymEuroPutPrice}, implies that, as $\tau\rightarrow 0^{+}$,
\begin{align}\label{eq:AsymLowerBoundExpXZeta}
(r-\delta)\tau=\bE\Big(\big(e^{X_{\tau}}\!-\!1\!-\!\zeta(\tau)\big){\bf 1}_{\{X_{\tau}\geq\ln(1+\zeta(\tau))+(r-\delta)\tau\}}\Big)+o(\tau)\leq\bE\Big(\big(e^{X_{\tau}}\!-\!1\!-\!\zeta(\tau)\big)^{+}\Big)+o(\tau).\qquad
\end{align}
Using \eqref{eq:NonZeroBMPosdDecompLevy} and \eqref{eq:MGFJMinus} as well as the independence among $X^{W}$, $J^{+}$, and $J^{-}$, the expectation on the right-hand side of \eqref{eq:AsymLowerBoundExpXZeta} can be bounded by
\begin{align*}
\bE\Big(\!\big(e^{X_{\tau}}\!\!-\!1\!-\!\zeta(\tau)\big)^{+}\!\Big)\!&\leq\bE\Big(\!\big(e^{X_{\tau}^{W}\!+J_{\tau}^{+}\!+J_{\tau}^{-}}\!\!-\!1\!-\!\zeta(\tau)\big)^{+}\!\Big)\!\!\leq\!\bE\Big(\!\big(e^{J_{\tau}^{+}}\!\!-\!1\big)e^{X_{\tau}^{W}\!+J_{\tau}^{-}}\Big)\!\!+\!\bE\Big(\!\big(e^{X_{\tau}^{W}\!+J_{\tau}^{-}}\!\!-\!1\!-\!\zeta(\tau)\big)^{+}\!\Big)\\
&=\bE\big(e^{J_{\tau}^{+}}-1\big)\bE\big(e^{X_{\tau}^{W}}\big)\bE\big(e^{J_{\tau}^{-}}\big)+\bE\Big(\big(e^{X_{\tau}^{W}+J_{\tau}^{-}}-1-\zeta(\tau)\big)^{+}\Big)\\
&=\exp\bigg(\tau\int_{(0,\infty)}\big(e^{z}-1\big)\nu(dz)\bigg)-1+\bE\Big(\big(e^{X_{\tau}^{W}+J_{\tau}^{-}}-1-\zeta(\tau)\big)^{+}\Big)\\
&=\tau\int_{(0,\infty)}\big(e^{z}-1\big)\nu(dz)+\bE\Big(\big(e^{X_{\tau}^{W}+J_{\tau}^{-}}-1-\zeta(\tau)\big)^{+}\Big)+o(\tau),\quad\tau\rightarrow 0^{+}.
\end{align*}
Together with \eqref{eq:Defd} and \eqref{eq:AsymLowerBoundExpXZeta}, we deduce that
\begin{align}\label{eq:AsymLowerBoundExpXWZeta}
d\tau\leq\bE\Big(\big(e^{X_{\tau}^{W}+J_{\tau}^{-}}-1-\zeta(\tau)\big)^{+}\Big)+o(\tau),\quad\tau\rightarrow 0^{+}.
\end{align}

Next, we will derive an upper bound for $\bE((e^{X_{\tau}^{W}+J_{\tau}^{-}}-1-\zeta(\tau))^{+})$ as $\tau\rightarrow 0^{+}$. For any $\varepsilon\in(0,1)$, in view of Lemma \ref{lem:NonZeroBMPosdEuroCritPrice}, we take $\tau\in\bR_{+}$ small enough so that $\tau\leq\ln^{2}(1+\zeta(\tau))/[\tau(\nu((-\infty,-\varepsilon))-\sigma^{2}/2)^{2}]$. By Lemma \ref{lem:NonZeroBMPosdTailProbBMNegJump}, we have
\begin{align*}
\bE\Big(\!\big(e^{X_{\tau}^{W}\!+J_{\tau}^{-}}\!\!-\!1\!-\!\zeta(\tau)\big)^{+}\!\Big)\!&=\!\!\int_{\ln(1+\zeta(\tau))}^{\infty}\!\!e^{y}\bP\big(X_{\tau}^{W}\!\!+\!J_{\tau}^{-}\!\!\geq\!y\big)dy\!=\!\sqrt{\tau}\!\!\int_{\ln(1+\zeta(\tau))/\sqrt{\tau}}^{\infty}\!\!e^{\sqrt{\tau}z}\bP\big(X_{\tau}^{W}\!\!+\!J_{\tau}^{-}\!\!\geq\!\!\sqrt{\tau}z\big)dz\\
&\leq\sqrt{\tau}\int_{\ln(1+\zeta(\tau))/\sqrt{\tau}}^{\infty}e^{\sqrt{\tau}z}\exp\bigg(\!\!-\!\frac{1}{2L}\bigg[z-\sqrt{\tau}\bigg(\nu((-\infty,-\varepsilon))-\frac{\sigma^{2}}{2}\bigg)\bigg]^{2}\bigg)dz\\
&\leq\sqrt{L\tau}\big(1+O(\sqrt{\tau})\big)\int_{[\ln(1+\zeta(\tau))-\tau(L+\nu((-\infty,-\varepsilon))-\sigma^{2}/2)]/\sqrt{L\tau}}^{\infty}e^{-x^{2}/2}\,dx,
\end{align*}
where we used change of variable ${x}=\sqrt{L}\omega+\sqrt{\tau}(L+\nu((-\infty,-\varepsilon))-\sigma^{2}/2)$ in the last inequality, and $L=L(\varepsilon;\sigma)$ is given as in \eqref{eq:Constp}. In view of Lemma \ref{lem:NonZeroBMPosdEuroCritPrice}, the lower limit of the last integral above explodes as $\tau\rightarrow 0^{+}$. Hence, by \eqref{eq:BoundsStdNormalTailProb} we obtain that, as $\tau\rightarrow 0^{+}$,
\begin{align}\label{eq:AsymUpperBoundExpXWZeta}
\bE\Big(\!\big(e^{X_{\tau}^{W}\!+J_{\tau}^{-}}\!\!-\!1\!-\!\zeta(\tau)\big)^{+}\!\Big)\!\leq\!\sqrt{2\pi L\tau}\big(1\!+\!O(\sqrt{\tau})\big)\phi\bigg(\frac{\ln\!\big(1\!+\!\zeta(\tau)\big)\!-\!\tau\big(L\!+\!\nu((-\infty,-\varepsilon))\!-\!\sigma^{2}\!/2\big)}{\sqrt{L\tau}}\!\bigg).\qquad
\end{align}

Combining \eqref{eq:AsymLowerBoundExpXWZeta} and \eqref{eq:AsymUpperBoundExpXWZeta}, we deduce that
\begin{align*}
d\sqrt{\tau}+o\big(\sqrt{\tau}\big)\leq\sqrt{2\pi L}\big(1+O(\sqrt{\tau})\big)\phi\bigg(\frac{\ln\big(1+\zeta(\tau)\big)-\tau\big(L+\nu((-\infty,-\varepsilon))-\sigma^{2}/2\big)}{\sqrt{L\tau}}\bigg),\quad\tau\rightarrow 0^{+},
\end{align*}
and so
\begin{align*}
\limsup_{\tau\rightarrow 0^{+}}\frac{\ln\tau}{2\zeta^{2}(\tau)/\tau}\leq -\lim_{\tau\rightarrow 0^{+}}\frac{\big[\ln\big(1+\zeta(\tau)\big)-\tau\big(L+\nu((-\infty,-\varepsilon))-\sigma^{2}/2\big)\big]^{2}}{2L\zeta^{2}(\tau)}= -\frac{1}{2L},
\end{align*}
or equivalently,
\begin{align*}
\limsup_{\tau\rightarrow 0^{+}}\frac{\zeta(\tau)}{\sqrt{-\tau\ln\tau}}\leq\sqrt{L}.
\end{align*}
Finally, by taking $\varepsilon\rightarrow 0^{+}$ above and noting from \eqref{eq:ConstLtau0} that $L=L(\varepsilon;\sigma)\rightarrow\sigma^{2}$, we obtain that
\begin{align*}
\limsup_{\tau\rightarrow 0^{+}}\frac{\zeta(\tau)}{\sqrt{-\tau\ln\tau}}\leq\sigma,
\end{align*}
which completes the proof of the proposition.\hfill $\Box$

\vspace{0.2cm}

\subsection{Step 2: The difference between European and American critical prices}$\,$

\vspace{0.3cm}
\noindent
To study the asymptotic behavior of the difference between the European and American critical prices, we first recall some regularity results on the American put price $P$ defined as in \eqref{eq:AmerPutPrice}. It is more convenient to state those results after a logarithmic change of variable. Define
\begin{align}\label{eq:DefScalAmerPutPrice}
\wt{P}(t,x):=P\big(t,e^{x}\big),\quad (t,x)\in[0,T]\times\bR,
\end{align}
Since $x\mapsto e^{x}$ is increasing and convex on $\bR$, it follows from the analogous properties of $P(t,\cdot)$ that $\wt{P}(t,\cdot)$ is non-increasing and convex on $\bR$, for any $t\in[0,T]$. Let $\wt{\sA}$ be the infinitesimal generator of $\wt{X}:=(\wt{X}_{t})_{t\in\bR_{+}}$, where $\wt{X}_{t}:=(r-\delta)t+X_{t}$, namely,
\begin{align*}
\wt{\sA}f(x):=\bigg(r-\delta-\frac{\sigma^{2}}{2}\bigg)f'(x)+\frac{\sigma^{2}}{2}f''(x)+\int_{\bR_{0}}\big(f(x+z)-f(x)-f'(x)\big(e^{z}-1\big)\big)\nu(dz).
\end{align*}
The following result (cf. \cite[Theorem 3.3]{LambertonMikou:2008}) shows that the American put price satisfies a variational inequality in the sense of distributions.

\begin{theorem}\label{thm:VarIneqAmerPutPriceExpLevy}
The distribution $(\partial/\partial t+\wt{\sA}-r)\wt{P}$ is a nonpositive measure on $(0,T)\times\bR$. Moreover, $(\partial/\partial t+\wt{\sA}-r)\wt{P}=0$ on the continuation region $\wt{\cC}:=\{(t,x)\in(0,T)\times\bR:\wt{P}(t,x)>(K-e^{x})^{+}\}$.
\end{theorem}

The following result provides the estimation of the difference $b_{e}(t)-b(t)$ near maturity.

\begin{proposition}\label{prop:NonZeroBMPosdDiffEuroAmerCritPrices}
Assume $\sigma>0$ and $d>0$. Then we have
\begin{align*}
\limsup_{t\rightarrow T^{-}}\frac{b_{e}(t)-b(t)}{\sqrt{T-t}}<\infty.
\end{align*}
\end{proposition}

\noindent
\textbf{Proof.} Since $\wt{P}(\cdot,x)$ is non-increasing on $[0,T]$, for each $x\in\bR$, it follows from Theorem \ref{thm:VarIneqAmerPutPriceExpLevy} that
\begin{align}\label{eq:VarIneq}
\wt{\sA}\,\wt{P}=r\wt{P}-\frac{\partial\wt{P}}{\partial t}\geq r\wt{P}\geq 0\,\,\,\text{ on }\,\wt{\cC},
\end{align}
in the sense of distribution, or, equivalently,
\begin{align*}
\bigg(r-\delta-\frac{\sigma^{2}}{2}\bigg)\frac{\partial\wt{P}}{\partial x}+\frac{\sigma^{2}}{2}\frac{\partial^{2}\wt{P}}{\partial x^{2}}+\int_{\bR_{0}}\bigg(\wt{P}(\cdot\,,\cdot+z)-\wt{P}(\cdot\,,\cdot)-\frac{\partial\wt{P}}{\partial x}(\cdot\,,\cdot)\big(e^{z}-1\big)\bigg)\nu(dz)\geq 0\,\,\,\text{ on }\,\wt{\cC}.
\end{align*}
Since $\wt{P}(t,\cdot)$ is convex on $\bR$ for any $t\in[0,T]$, the right partial derivative of $\wt{P}(t,\cdot)$,
\begin{align}\label{eq:RightPartDerwtP}
\frac{\partial_{+}\wt{P}(t,x)}{\partial x}:=\lim_{h\rightarrow 0^{+}}\frac{\wt{P}(t,x+h)-\wt{P}(t,x)}{h},
\end{align}
is a well-defined function on $[0,T]\times\bR$, and we also have $\partial\wt{P}/\partial x=\partial_{+}\wt{P}/\partial x$ on $[0,T]\times\bR$ in the sense of distribution. With the notion of $d$ defined as {in \eqref{eq:Defd}}, we deduce that
\begin{align}\label{eq:IneqIntPosNegJumps}
\frac{\sigma^{2}}{2}\bigg(\frac{\partial^{2}\wt{P}}{\partial x^{2}}-\frac{\partial\wt{P}}{\partial x}\bigg)+\cI\geq -d\frac{\partial_{+}\wt{P}}{\partial x}-\int_{(0,\infty)}\sup_{y\in(\ln b(t),\ln K)}\big|\wt{P}(\cdot\,,y+z)-\wt{P}(\cdot\,,y)\big|\,\nu(dz)\,\,\,\text{ on }\,\wt{\cC},\qquad
\end{align}
in the sense of distribution, where
\begin{align*}
\cI(t,x):=\int_{(-\infty,0)}\bigg(\wt{P}(t,x+z)-\wt{P}(t,x)-\frac{\partial_{+}\wt{P}}{\partial x}(t,x)\big(e^{z}-1\big)\bigg)\nu(dz),\quad (t,x)\in\wt{\cC}.
\end{align*}
Recall that for each $t\in(0,T)$, $P(t,\cdot)$ is non-increasing and Lipschitz on $\bR_{+}$ (see Proposition \ref{prop:RegPropAmerPutPrice}), and thus, for any $y\in(\ln b(t),\ln K)$ and $z\in(0,\infty)$,
\begin{align*}
0\leq\wt{P}(t,y)-\wt{P}(t,y+z)=P\big(t,e^{y}\big)-P\big(t,e^{y+z}\big)\leq e^{y}\big(e^{z}-1\big)\leq K\big(e^{z}-1\big),
\end{align*}
and
\begin{align*}
0\leq\wt{P}(t,y)-\wt{P}(t,y+z)=P\big(t,e^{y}\big)-P\big(t,e^{y+z}\big)\leq P\big(t,b(t)\big)=\big(K-b(t)\big)\rightarrow 0,\quad\text{as }\,t\rightarrow T^{-}.
\end{align*}
Hence, the dominated convergence theorem ensures that
\begin{align}\label{eq:LimitIntSupDiffAmerPutPrice}
\lim_{t\rightarrow T^{-}}\int_{(0,\infty)}\sup_{y\in(\ln b(t),\ln K)}\big|\wt{P}(t,y+z)-\wt{P}(t,y)\big|\,\nu(dz)=0.
\end{align}

Now for any $t\in(0,T)$ and $x\in(\ln b(t),\ln b_{e}(t))$, using \eqref{eq:DefScalAmerPutPrice}, the convexity of $P(t,\cdot)$, and the fact that $(P-P_{e})(t,\cdot)$ is non-increasing on $\bR_{+}$ (cf. \cite[Corollary 3.1]{LambertonMikou:2013}), we have (with $\tau=T-t$)
\begin{align*}
e^{-x}\frac{\partial_{+}\wt{P}}{\partial x}(t,x)=\frac{\partial_{+}P}{\partial s}(t,e^{x})\leq\frac{\partial_{-}P}{\partial s}\big(t,b_{e}(t)\big)\leq\frac{\partial_{-}P_{e}}{\partial s}\big(t,b_{e}(t)\big)=-\bE\bigg(e^{-\delta\tau+X_{\tau}}{\bf 1}_{\{b_{e}(t)e^{(r-\delta)\tau+X_{\tau}}\leq K\}}\!\bigg),
\end{align*}
where the left-derivatives $\partial_{-}{P}/\partial x$ and $\partial_{-}{P}_{e}/\partial x$ are defined analogously to \eqref{eq:RightPartDerwtP}, but taking $h\rightarrow 0^{-}$. Using \eqref{eq:Defzeta}, Lemmas \ref{lem:NonZeroBMPosdLevySmallTimeAsymNormal} and \ref{lem:NonZeroBMPosdEuroCritPrice}, and the martingale property of $(e^{X_{t}})_{t\in\bR_{+}}$, we have
\begin{align}
-\!\!\lim_{t\rightarrow T^{-}}\!\!\frac{\partial_+ P}{\partial s}(t,e^{x})&\geq\!\lim_{\tau\rightarrow 0^{+}}\!\!\bE\bigg(e^{-\delta\tau+X_{\tau}}{\bf 1}_{\{b_{e}(t)e^{(r-\delta)\tau+X_{\tau}}\leq K\}}\!\bigg)=\!\lim_{\tau\rightarrow 0^{+}}\!\!\bE\bigg(e^{-\delta\tau+X_{\tau}}{\bf 1}_{\{(r-\delta)\tau+X_{\tau}\leq\ln(1+\zeta(\tau))\}}\!\bigg)\nonumber\\
\label{eq:LowerBoundLimitParDerSAmerPutPrice} &=\lim_{\tau\rightarrow 0^{+}}\bE\bigg(e^{-\delta\tau+X_{\tau}}{\bf 1}_{\{((r-\delta)\tau+X_{\tau})/\sqrt{\tau}\leq\ln(1+\zeta(\tau))/\sqrt{\tau}\}}\bigg)=\lim_{\tau\rightarrow 0^{+}}\bE\big(e^{X_{\tau}}\big)=1.
\end{align}
By combining \eqref{eq:LimitIntSupDiffAmerPutPrice} and \eqref{eq:LowerBoundLimitParDerSAmerPutPrice}, we have
\begin{align*}
&\liminf_{t\rightarrow T^{-}}\inf_{x\in(\ln b(t),\ln b_{e}(t))}\bigg(\!-d\,\frac{\partial_{+}\wt{P}}{\partial x}(t,x)-\int_{(0,\infty)}\sup_{y\in(\ln b(t),\ln K)}\big|\wt{P}(t,y+z)-\wt{P}(t,y)\big|\,\nu(dz)\bigg)\\
&\quad=\liminf_{t\rightarrow T^{-}}\inf_{s\in(b(t),b_{e}(t))}\bigg(-ds\frac{\partial_{+}P}{\partial s}(t,s)-\int_{(0,\infty)}\sup_{u\in(b(t),K)}\big|P(t,ue^{z})-P(t,u)\big|\,\nu(dz)\bigg)\geq dK.
\end{align*}
Hence, we can choose $\rho\in(0,\infty)$ such that for all $t\in(T-\rho,T)$ and $x\in(\ln b(t),\ln b_{e}(t))$,
\begin{align*}
-d\,\frac{\partial_{+}\wt{P}}{\partial x}(t,x)-\int_{(0,\infty)}\sup_{y\in(\ln b(t),\ln K)}\big|\wt{P}(t,y+z)-\wt{P}(t,y)\big|\,\nu(dz)\geq\frac{dK}{2}.
\end{align*}
Together with \eqref{eq:IneqIntPosNegJumps}, we obtain that, in the sense of distribution,
\begin{align}\label{eq:LowerBoundLimitInfcIts}
\frac{\sigma^{2}}{2}\bigg(\frac{\partial^{2}\wt{P}}{\partial x^{2}}-\frac{\partial\wt{P}}{\partial x}\bigg)+\cI\geq\frac{dK}{2}\,\,\,\text{ on }\,\wt{\cC}_{\rho},
\end{align}
where
\begin{align*}
\wt{\cC}_{\rho}:=\big\{(t,x)\in\bR_{+}\times\bR:t\in(T-\rho,T),x\in(\ln b(t),\ln b_{e}(t))\big\}.
\end{align*}

Next we will derive an upper bound for $\cI(t,x)$ on $\wt{\cC}$. To begin with, by \eqref{eq:DefScalAmerPutPrice} we first have
\begin{align*}
\cI(t,x)=\left(\int_{(-\infty,\ln b(t)-x]}+\int_{(\ln b(t)-x,0)}\right)\bigg(P(t,e^{x+z})-P(t,e^{x})-e^{x}\frac{\partial_{+}P}{\partial s}(t,e^{x})\big(e^{z}-1\big)\bigg)\nu(dz),
\end{align*}
for any $(t,x)\in\wt{\cC}$. Note that, for any $z\in(-\infty,\ln b(t)-x]$, we have
\begin{align*}
&P(t,e^{x+z})-P(t,e^{x})-e^{x}\frac{\partial_{+}P}{\partial s}(t,e^{x})\big(e^{z}-1\big)=\big(K-e^{x+z}\big)-P(t,e^{x})-e^{x}\frac{\partial_{+}P}{\partial s}(t,e^{x})\big(e^{z}-1\big)\\
&\quad\leq\big(K-e^{x+z}\big)-\big(K-e^{x}\big)-e^{x}\frac{\partial_{+}P}{\partial s}(t,e^{x})\big(e^{z}-1\big)=e^{x}\bigg(\frac{\partial_{+}P}{\partial s}(t,e^{x})+1\bigg)\big(1-e^{z}\big),
\end{align*}
while for any $z\in(\ln b(t)-x,0)$, we have from the convexity of $P(t,\cdot)$ that
\begin{align*}
P(t,e^{x+z})-P(t,e^{x})-e^{x}\frac{\partial_{+}P}{\partial s}(t,e^{x})\big(e^{z}-1\big)\leq e^{x}\big(e^{z}-1\big)\bigg(\frac{\partial_{+}P}{\partial s}\big(t,e^{x+z}\big)-\frac{\partial_{+}P}{\partial s}(t,e^{x})\bigg).
\end{align*}
Hence, we deduce that, for any $t\in(0,T)$ and $x\in(\ln b(t),\ln b_{e}(t))$,
\begin{align}
\cI(t,x)&\leq e^{x}\bigg(\frac{\partial_{+}P}{\partial s}(t,e^{x})+1\bigg)\int_{(-\infty,\ln b(t)-x]}\big(1-e^{z}\big)\,\nu(dz)\\
\label{eq:UpperBoundcItsP} &\quad +e^{x}\int_{(\ln b(t)-x,0)}\big(e^{z}-1\big)\bigg(\frac{\partial_{+}P}{\partial s}\big(t,e^{x+z}\big)-\frac{\partial_{+}P}{\partial s}(t,e^{x})\bigg)\nu(dz).
\end{align}
For any $t\in(0,T)$ and $\xi\in(0,\ln(b_{e}(t)/b(t)))$, we set $g_{t}(\xi):=P(t,b(t)e^{\xi})$ and its right-derivative is
\begin{align}\label{eq:Derivgt}
g_{t+}'(\xi):=\frac{d_{+}g_{t}(\xi)}{d\xi}=b(t)e^{\xi}\,\frac{\partial_{+}P}{\partial s}\big(t,b(t)e^{\xi}\big),
\end{align}
Due to the smooth-pasting property when $\sigma>0$ (cf. \cite[Proposition 4.1 \& Theorem 4.1]{LambertonMikou:2012}), $g_{t}'(0)$ exists and $g_{t}'(0)=g_{t+}'(0)= -b(t)$. By the non-increasing property and the convexity of $P(t,\cdot)$ on $\bR_{+}$ as well as the smooth-pasting property, we also have
\begin{align}\label{eq:BoundDerivgt}
\big|g_{t+}'(\xi)\big|\leq b_{e}(t)\bigg(\!-\!\frac{\partial_{+}P}{\partial s}\big(t,b(t)e^{\xi}\big)\bigg)\leq b_{e}(t)\bigg(\!-\!\frac{\partial_{+}P}{\partial s}\big(t,b(t)\big)\bigg)\leq b_{e}(t)\leq K.
\end{align}
Using these notations and with $x=\ln b(t)+\xi$, we can rewrite \eqref{eq:UpperBoundcItsP} as
\begin{align*}
\cI\big(t,x\big)&\leq\big(g_{t+}'(\xi)-g_{t}'(0)e^{\xi}\big)\!\int_{(-\infty,-\xi]}\!\big(1-e^{z}\big)\,\nu(dz)+\!\int_{(-\xi,0)}\!\big(g_{t+}'(\xi)-g_{t+}'(\xi+z)e^{-z}\big)\big(1-e^{z}\big)\,\nu(dz).
\end{align*}
Note that for any $\varepsilon\in(0,\infty)$, since $\xi\in(0,\ln(b_{e}(t)/b(t)))$,
\begin{align*}
\big(e^{\xi}-1\big)\int_{(-\infty,-\xi]}\!\big(1-e^{z}\big)\,\nu(dz)&\leq\big(e^{\xi}-1\big)\nu((-\infty,-\varepsilon])+\int_{(-\varepsilon,0)}\big(e^{-z}-1\big)\big(1-e^{z}\big)\,\nu(dz)\\
&\leq\bigg(\frac{b_{e}(t)}{b(t)}-1\bigg)\nu((-\infty,-\varepsilon])+\int_{(-\varepsilon,0)}\big(e^{-z}-1\big)\big(1-e^{z}\big)\,\nu(dz),
\end{align*}
and so, by \eqref{eq:RelEuroAmerCritPrice} and Theorem \ref{thm:LimitAmerCritPrice}-(a), we have
\begin{align*}
\lim_{t\rightarrow T^{-}}\sup_{\xi\in(0,\ln(b_{e}(t)/b(t)))}\big(e^{\xi}-1\big)\int_{(-\infty,-\xi]}\!\big(1-e^{z}\big)\,\nu(dz)=0.
\end{align*}
Therefore, we obtain that, for any $x\in(\ln b(t),\ln b_{e}(t))$,
\begin{align}\label{eq:UpperBoundcItsgxi}
\cI\big(t,x\big)\leq\cJ_{1}(t,x)+\cJ_{2}(t,x)+o(1),\quad t\rightarrow T^{-},
\end{align}
where, for any $\xi\in(0,\ln(b_{e}(t)/b(t)))$,
\begin{align*}
\cJ_{1}(t,x)&:=\wt{\cJ}_{1}(t,x-\ln b(t)),\quad {\wt{\cJ}_{1}(t,\xi)}:=\big(g_{t+}'(\xi)-g_{t}'(0)\big)\int_{(-\infty,-\xi]}\big(1-e^{z}\big)\,\nu(dz),\\
\cJ_{2}(t,x)&:=\wt{\cJ}_{2}(t,x-\ln b(t)),\quad {\wt{\cJ}_{2}(t,\xi)}:=\int_{(-\xi,0)}\big(g_{t+}'(\xi)-g_{t+}'(\xi+z)e^{-z}\big)\big(1-e^{z}\big)\,\nu(dz).\quad
\end{align*}
By combining \eqref{eq:LowerBoundLimitInfcIts} and \eqref{eq:UpperBoundcItsgxi}, there exists $\rho\in(0,T)$ such that, in the sense of distribution,
\begin{align}\label{eq:LowerBound2ndDerivwtPJ1J2}
\frac{dK}{4}\leq\cJ_{1}+\cJ_{2}+\frac{\sigma^{2}}{2}\bigg(\frac{\partial^{2}\wt{P}}{\partial x^{2}}-\frac{\partial\wt{P}}{\partial x}\bigg)\,\,\,\text{ on }\,\wt{\cC}_{\rho}.
\end{align}
Using the continuity of $P:[0,T]\times\bR_{+}\rightarrow\bR$, and the convexity of $P(t,\cdot)$, we can prove that (see Appendix \ref{sec:AppendixA}), for any $t\in(T-\rho,T)$ and $a\in(0,\ln(b_{e}(t)/b(t)))$,
\begin{align}
\frac{dKa}{4}&\leq\int_{\ln b(t)}^{a+\ln b(t)}\cJ_{1}(t,x)\,dx+\int_{\ln b(t)}^{a+\ln b(t)}\cJ_{2}(t,x)\,dx+\frac{\sigma^{2}b(t)e^{a}}{2}\bigg(\frac{\partial_{+}P(t,b(t)e^{a})}{\partial s}+1\bigg)\nonumber\\
\label{eq:IneqIntJ1J2} &=\int_{0}^{a}\wt{\cJ}_{1}(t,\xi)\,d\xi+\int_{0}^{a}\wt{\cJ}_{2}(t,\xi)\,d\xi+\frac{\sigma^{2}b(t)e^{a}}{2}\bigg(\frac{\partial_{+}P(t,b(t)e^{a})}{\partial s}+1\bigg).
\end{align}

We now estimate the first two integrals on the right-hand side of \eqref{eq:IneqIntJ1J2}. To begin with, in order to estimate the integral of $\wt{\cJ}_{1}$, we first have, for any $\xi\in(0,a)$,
\begin{align*}
g_{t+}'(\xi)=b(t)e^{\xi}\frac{\partial_{+}P}{\partial s}\big(t,b(t)e^{\xi}\big)\leq b(t)e^{\xi}\frac{\partial_{+}P}{\partial s}\big(t,b(t)e^{a}\big)=e^{\xi-a}g_{t+}'(a)\leq e^{-a}g_{t+}'(a),
\end{align*}
where the first inequality follows from the convexity of $P(t,\cdot)$ and the last inequality follows from the fact that $g_{t+}'(a)\leq 0$ (since $\partial_{+}P/\partial s\leq 0$). In addition, with the help of the convexity of $P(t,\cdot)$ and the smooth-pasting property, we also have, for any $a>0$,
\begin{align*}
e^{-a}g_{t+}'(a)-g_{t+}'(0)=b(t)\bigg(1+\frac{\partial_{+}P}{\partial s}\big(t,b(t)e^{a}\big)\bigg)\geq b(t)\bigg(1+\frac{\partial_{+}P}{\partial s}\big(t,b(t)\big)\bigg)=0,
\end{align*}
Hence, we deduce that
\begin{align*}
\int_{0}^{a}\wt{\cJ}_{1}(t,\xi)\,d\xi&\leq\big(e^{-a}g_{t+}'(a)-g_{t}'(0)\big)\int_{0}^{a}\bigg(\int_{(-\infty,-\xi]}\big(1-e^{z}\big)\,\nu(dz)\bigg)d\xi\\
&=\big(e^{-a}g_{t+}'(a)-g_{t}'(0)\big)\bigg(a\int_{(-\infty,-a)}\big(1-e^{z}\big)\,\nu(dz)+\int_{[-a,0)}z\big(e^{z}-1\big)\,\nu(dz)\bigg)\\
&\leq\big(e^{-a}g_{t+}'(a)-g_{t}'(0)\big)\bigg(\!-\!a\int_{(-\infty,-a)}z\,\nu(dz)+\int_{[-a,0)}z^{2}\,\nu(dz)\bigg).
\end{align*}
Since
\begin{align*}
0\leq\limsup_{a\rightarrow 0^{+}}\int_{(-\infty,-a)}(-az)\,\nu(dz)&=\lim_{\varepsilon\rightarrow 0^{+}}\limsup_{a\rightarrow 0^{+}}\bigg(\int_{(-\infty,-\varepsilon)}+\int_{[-\varepsilon,-a)}\bigg)(-az)\,\nu(dz)\\
&\leq\lim_{\varepsilon\rightarrow 0^{+}}\limsup_{a\rightarrow 0^{+}}\bigg(\int_{(-\infty,-\varepsilon)}(-az)\,\nu(dz)+\int_{[-\varepsilon,0)}z^{2}\,\nu(dz)\bigg)=0,
\end{align*}
we obtain that, as $a\rightarrow 0^{+}$,
\begin{align}\label{eq:EstIntcJ1}
\int_{0}^{a}\wt{\cJ}_{1}(t,\xi)\,d\xi\leq\big(e^{-a}g_{t+}'(a)-g_{t}'(0)\big)o(1)\leq g_{t+}'(a)-g_{t}'(0)+o(a).
\end{align}

As for the integral of $\wt{\cJ}_{2}$ in \eqref{eq:IneqIntJ1J2}, we first have
\begin{align*}
&\int_{0}^{a}\wt{\cJ}_{2}(t,\xi)\,d\xi\leq\int_{0}^{a}\bigg(\int_{(-\xi,0)}\big(g_{t+}'(\xi)-g_{t+}'(\xi+z)e^{-z}\big)(-z)\,\nu(dz)\bigg)d\xi\\
&\quad =\int_{(-a,0)}(-z)\bigg(\int_{-z}^{a}\big(g_{t+}'(\xi)-g_{t+}'(\xi+z)e^{-z}\big)\,d\xi\bigg)\nu(dz)\\
&\quad =\int_{(-a,0)}(-z)\big(g_{t}(a)-g_{t}(a+z)e^{-z}-g_{t}(-z)+g_{t}(0)e^{-z}\big)\,\nu(dz)\\
&\quad =\int_{(-a,0)}\!\!(-z)\big(g_{t}(a)\!-\!g_{t}(a\!+\!z)\!-\!g_{t}(-z)\!+\!g_{t}(0)\big)\nu(dz)+\!\int_{(-a,0)}\!\!(-z)\big(g_{t}(0)\!-\!g_{t}(a\!+\!z)\big)\big(e^{-z}\!-\!1\big)\nu(dz).
\end{align*}
By \eqref{eq:Derivgt} and the nonincreasing property and the convexity of $P(t,\cdot)$, we have, for any $z\in(-a,0)$,
\begin{align*}
g_{t}(a)-g_{t}(a+z)&=\int_{z}^{0}g_{t+}'(a+y)\,dy=\int_{z}^{0}b(t)e^{a+y}\frac{\partial_{+}P}{\partial s}\big(t,b(t)e^{a+y}\big)\,dy\\
&\leq b(t)\frac{\partial_{+}P}{\partial s}\big(t,b(t)e^{a}\big)\int_{z}^{0}e^{a+y}\,dy\leq b(t)\frac{\partial_{+}P}{\partial s}\big(t,b(t)e^{a}\big)(-z)=g_{t+}'(a)e^{-a}(-z),\\
g_{t}(-z)-g_{t}(0)&=\int_{0}^{-z}g_{t+}'(y)\,dy=\int_{0}^{-z}b(t)e^{y}\frac{\partial_{+}P}{\partial s}\big(t,b(t)e^{y}\big)\,dy\\
&\geq b(t)\frac{\partial_{+}P}{\partial s}\big(t,b(t)\big)\int_{0}^{-z}e^{y}\,dy\geq b(t)\frac{\partial_{+}P}{\partial s}\big(t,b(t)\big)(-z)e^{-z}=g_{t}'(0)(-z)e^{-z}.
\end{align*}
Together with \eqref{eq:BoundDerivgt}, we deduce that, as $a\rightarrow 0^{+}$,
\begin{align}
&\int_{0}^{a}\wt{\cJ}_{2}(t,\xi)\,d\xi\!\leq\!\!\int_{(-a,0)}\!\!z^{2}\big(g_{t+}'(a)e^{-a}\!-\!g_{t}'(0)e^{-z}\big)\nu(dz)\!+\!\!\sup_{z\in(0,a)}\!\big|g_{t+}'(z)\big|\!\int_{(-a,0)}\!\!(-z)(a\!+\!z)\big(e^{-z}\!-\!1\big)\nu(dz)\nonumber\\
&\quad\leq\big(e^{-a}g_{t+}'(a)\!-\!g_{t}'(0)\big)\!\int_{(-a,0)}\!z^{2}\nu(dz)-g_{t}'(0)\!\int_{(-a,0)}\!z^{2}\big(e^{-z}\!-\!1\big)\nu(dz)-Ka\!\int_{(-a,0)}\!z\big(e^{-z}\!-\!1\big)\nu(dz)\nonumber\\
\label{eq:EstIntcJ2} &\quad =\big(e^{-a}g_{t+}'(a)-g_{t}'(0)\big)o(1)+o(a)\leq g_{t+}'(a)-g_{t}'(0)+o(a).
\end{align}

By combining \eqref{eq:IneqIntJ1J2}, \eqref{eq:EstIntcJ1}, and \eqref{eq:EstIntcJ2}, we obtain that there exists $\rho\in(0,\infty)$ such that, when $t\in(T-\rho,T)$, for any $x\in(\ln b(t),\ln b_{e}(t))$,
\begin{align*}
\frac{d\big(x-\ln b(t)\big)}{6}\leq g_{t+}'\big(x-\ln b(t)\big)-g_{t}'(0)+\frac{\sigma^{2}e^{x}}{4}\bigg(\frac{\partial_{+}P(t,e^{x})}{\partial s}+1\bigg).
\end{align*}
thus, setting $a_{t}:=\ln(b_{e}(t)/b(t))$,
\begin{align*}
&\frac{d\big(\ln b_{e}(t)-\ln b(t)\big)^{2}}{12}\leq\int_{\ln b(t)}^{\ln b_{e}(t)}\Big(g_{t+}'\big(x-\ln b(t)\big)-g_{t}'(0)\Big)dx+\frac{\sigma^{2}}{4}\int_{\ln b(t)}^{\ln b_{e}(t)}e^{x}\bigg(\frac{\partial_{+}P(t,e^{x})}{\partial s}+1\bigg)dx\\
&\quad =g_{t}(a_{t})-g_{t}(0)-g_{t}'(0)a_{t}+\frac{\sigma^{2}}{4}\big(P(t,b_{e}(t))-P(t,b(t))+b_{e}(t)-b(t)\big)\\
&\quad =P\big(t,b_{e}(t)\big)-P\big(t,b(t)\big)+b(t)\ln\bigg(\frac{b_{e}(t)}{b(t)}\bigg)+\frac{\sigma^{2}}{4}\big(P(t,b_{e}(t))-P(t,b(t))+b_{e}(t)-b(t)\big)\\
&\quad\leq\bigg(1+\frac{\sigma^{2}}{4}\bigg)\Big( P\big(t,b_{e}(t)\big)-\big(K-b(t)\big)+b_{e}(t)-b(t)\Big)=\bigg(1+\frac{\sigma^{2}}{4}\bigg)\Big(P\big(t,b_{e}(t)\big)-P_{e}\big(t,b_{e}(t)\big)\Big),
\end{align*}
where we used $\ln(1+x)\leq x$ for $x>0$ in the second inequality. Therefore, by the early exercise premium formula (cf. \cite[Theorem 3.1 \& Remark 3.1]{LambertonMikou:2013}), we obtain that, for any $t\in(T-\rho,T)$,
\begin{align*}
\frac{d\big(\ln b_{e}(t)-b(t)\big)^{2}}{12}\leq rK(T-t)\bigg(1+\frac{\sigma^{2}}{4}\bigg),
\end{align*}
and thus
\begin{align*}
\limsup_{t\rightarrow T^{-}}\frac{b_{e}(t)-b(t)}{\sqrt{T-t}}=\limsup_{t\rightarrow T^{-}}\frac{\ln(b_{e}(t)/b(t))}{\sqrt{T-t}}<\infty,
\end{align*}
which completes the proof of the proposition.\hfill $\Box$

\medskip
\noindent
\textbf{Proof of Theorem 4.1.} In view of Proposition \ref{prop:NonZeroBMPosdDiffEuroAmerCritPrices}, clearly we have
\begin{align*}
K-b(t)=K-b_{e}(t)+b_{e}(t)-b(t)=K-b_{e}(t)+O\big(\sqrt{T-t}\big),\quad t\rightarrow T^{-}.
\end{align*}
Moreover, by \eqref{eq:Defzeta} and Proposition \ref{prop:NonZeroBMPosdEuroCritPriceConvRate} and noting that $\lim_{t\rightarrow T^{-}}b_{e}(t)=K$, we have
\begin{align*}
\lim_{t\rightarrow T^{-}}\frac{K-b_{e}(t)}{K\sqrt{-(T-t)\ln(T-t)}}=\lim_{t\rightarrow T^{-}}\frac{K-b_{e}(t)}{b_{e}(t)\sqrt{-(T-t)\ln(T-t)}}=\lim_{\tau\rightarrow 0^{+}}\frac{\zeta(\tau)}{\sqrt{-\tau\ln\tau}}=\sigma,
\end{align*}
which completes the proof of the theorem.\hfill $\Box$

\vspace{0.2cm}

\section{New Results on the Rate of Convergence of the Critical Price when \texorpdfstring{$d<0$}{}}\label{sec:NonZeroBMNegdAmerCritPriceConvRate}

\vspace{0.4cm}
\noindent
In this section, we consider the rate of convergence of critical boundary $b$ near maturity when $\sigma>0$ and $d<0$. We first assume that $X$ has a jump component of finite variation, i.e., \eqref{eq:FinVarLevy} holds true. In this case, the model \eqref{eq:ExpLevyModel} can be written as
\begin{align}\label{eq:DefwtX}
S_{t}=S_{0}\,e^{(r-\delta)t+X_{t}}=S_{0}\,e^{\wt{X}_{t}},\quad\wt{X}_{t}:=(r-\delta)t+X_{t}=\bigg(\gamma_{0}-\frac{\sigma^{2}}{2}\bigg)t+\sigma W_{t}+Z_{t},\quad t\in\bR_{+},
\end{align}
where
\begin{align}\label{eq:Defgamma0ProcZ}
\gamma_{0}:=r-\delta-\int_{\bR_{0}}\big(e^{x}-1\big)\nu(dx),\quad Z_{t}=\int_{0}^{t}\int_{\bR_{0}}z\,N(ds,dz).
\end{align}

\begin{assumption}\label{assump:LevyMeas}
Throughout this section, we make the following standard assumption:
\begin{align*}
\nu(dz)=s(z)\,dz,\quad\text{ for some }\,s\in C(\bR_{0}).
\end{align*}
\end{assumption}

We begin with the following lemma which provides an estimation of the expectation of $L_{t}^{K}$, the local time of the process $S$ at $K$ until time $t$, for small $t>0$. The proof is deferred to Appendix \ref{sec:AppendixB}.

\begin{lemma}\label{lem:SmallTimeEstExpLocalTime}
Let $S_{0}=b(T)e^{a\sqrt{\theta}}$ with $a\in(-\infty,0)$. If $b(T)<K$, then we have, for any $\bF$-stopping time $\tau$ taking values in $[0,\theta]$,
\begin{align}\label{eq:SmallTimeEstExpLocalTime}
\bE\big(L_{\tau}^{K}\big)=2K\,\bE\Big(\Big(\big(\!-\!a\sqrt{\theta}-\sigma W_{\tau}\big)^{+}-\big(\!-\!a\sqrt{\theta}-\sigma W_{\wh{T}_{1}}\big)^{+}\Big){\bf 1}_{\{\wh{T}_{1}<\tau\}}\Big)+o\big(\theta^{3/2}\big)\leq\omega_{0}\,\theta^{3/2},
\end{align}
as $\theta\rightarrow 0^{+}$, where $\wh{T}_{1}:=\inf\{s\in\bR_{+}:\Delta L_{s}=\ln(K/b(T))\}$ and $\omega_{0}\in(0,\infty)$ is independent of $a$.
\end{lemma}

The following theorem provides a second-order near-maturity expansion for the American put price $P$ around $b(T)$ along a certain parabolic branch, which serves as a key step to derive the convergence rate of the critical price. This is similar to Theorem 3.1 in \cite{BouselmiLamberton:2016}, where only jumps of finite-activity were considered.

\begin{theorem}\label{thm:SmallTimeExpanAmerPutPrice}
Let $d<0$. For any $a\in(-\infty,0)$, we have
\begin{align*}
P\big(T-\theta,b(T)e^{a\sqrt{\theta}}\big)=\big(K-b(T)e^{a\sqrt{\theta}}\big)^{+}+\sigma b(T)\bar{\delta}e^{\lambda}v_{\lambda,\beta}(a/\sigma)\theta^{3/2}+o\big(\theta^{3/2}\big),
\end{align*}
as $\theta\rightarrow 0^{+}$, where $\lambda=\nu(\{\ln(K/b(T))\})$, $\bar{\delta}=\delta+\int_{(\ln(K/b(T)),\infty)}e^{z}\nu(dz)$, and $v_{\lambda,\beta}$ is defined as in \eqref{eq:DefFuntvlambdabeta} with $\beta=K/(b(T)\bar{\delta})$.\end{theorem}

\noindent
\textbf{Proof.} By It\^{o}-Meyer's formula (see, e,g,, Theorem 70 in \cite[Chapter IV]{Protter:2005}) and the product formula for semimartingales, for any $\bF$-stopping time $\tau\in\sT_{0,\theta}$ and with $S_{0}=b(T)e^{a\sqrt{\theta}}$, we have
\begin{align}\label{eq:DecompAmerPutPrice}
\bE\big(e^{-r\tau}(K-S_{\tau})^{+}\big)-\big(K-b(T)e^{a\sqrt{\theta}}\big)^{+}=\cI^{a}(\tau)+\cJ^{a}(\tau),
\end{align}
where (recalling $\Phi(y,z):=(K-ye^{z})^{+}-(K-y)^{+}$)
\begin{align*}
\cI^{a}(\tau)\!:=\!\bE\!\left(\int_{0}^{\tau}\!e^{-rt}\!\bigg(\!{\bf 1}_{\{S_{t}\leq K\}}\!\big(r(S_{t}\!-\!K)\!-\!\gamma_{0}S_{t}\big)\!+\!\!\int_{\bR_{0}}\!\!\Phi(S_{t},z)\nu(dz)\!\bigg)dt\!\right)\!,\,\,\,\cJ^{a}(\tau)\!:=\!\frac{1}{2}\bE\bigg(\!\int_{0}^{\tau}\!e^{-rs}dL_{s}^{K}\!\bigg).
\end{align*}
By Lemma \ref{lem:SmallTimeEstExpLocalTime} we deduce that\footnote{Note that the second term in \eqref{eq:DecompcJa} is $o(\theta^{3/2})$. Indeed, since $s\to L_{s}^{K}$ is nondecreasing, $0\leq 1-e^{-rs}\leq rs$, and $\tau\leq\theta$, by Lemma \ref{lem:SmallTimeEstExpLocalTime} we have $\bE\big(\int_{0}^{\tau}\big(1-e^{-rs}\big)dL_{s}^{K}\big)\leq r \theta\,\bE\big(L_{\tau}^{K}\big)=O(\theta^{5/2})$.}
\begin{align}\label{eq:DecompcJa}
\cJ^{a}(\tau)&=\frac{1}{2}\,\bE\big(L_{\tau}^{K}\big)-\frac{1}{2}\,\bE\bigg(\int_{0}^{\tau}\big(1-e^{-rs}\big)dL_{s}^{K}\bigg)\\
\label{eq:EstcJatau} &=K\,\bE\Big({\bf 1}_{\{\wh{T}_{1}<\tau\}}\Big(\big(\!-a\sqrt{\theta}-\sigma W_{\tau}\big)^{+}-\big(\!-a\sqrt{\theta}-\sigma W_{\wh{T}_{1}}\big)^{+}\Big)\Big)+o\big(\theta^{3/2}\big),\quad\theta\rightarrow 0^{+},\quad
\end{align}
where we recall $\wh{T}_{1}=\inf\{t\in\bR_{+}:\Delta L_{t}=\ln(K/b(T))\}$.

Next, we analyze $\cI^{a}(\tau)$. To begin with, for any $\varepsilon\in(0,\infty)$ and recalling $T_{1}^{\varepsilon}=\inf\{t\in\bR_{+}:|\Delta X_{t}|>\varepsilon\}$, we first have
\begin{align*}
&\bE\bigg(\int_{0}^{\tau}e^{-rt}{\bf 1}_{\{S_{t}>K\}}\int_{\bR_{0}}\Phi(S_{t},z)\,\nu(dz)\,dt\bigg)=\bE\bigg(\int_{0}^{\tau}e^{-rt}{\bf 1}_{\{S_{t}>K\}}\int_{\bR_{0}}\big(K-S_{t}e^{z}\big)^{+}\nu(dz)dt\bigg)\\
&\leq K\int_{\bR_{0}}\!\big|1\!-\!e^{z}\big|\nu(dz)\int_{0}^{\theta}\!\bP\big(S_{t}\!>\!K\big)dt\leq K\int_{\bR_{0}}\!\big|1\!-\!e^{z}\big|\nu(dz)\int_{0}^{\theta}\!\Big(\bP\big(S_{t}\!>\!K,T_{1}^{\varepsilon}\!>\!\theta\big)+\bP\big(T_{1}^{\varepsilon}\!\leq\!\theta\big)\Big)dt\\
&\leq K\theta\int_{\bR_{0}}\big|1-e^{z}\big|\nu(dz)\bigg(\bP\Big(\sup_{t\in[0,\theta]}S_{t}^{\varepsilon}>K\Big)+\bP\big(T_{1}^{\varepsilon}\leq\theta\big)\bigg)\\
&=K\theta\int_{\bR_{0}}\big|1-e^{z}\big|\nu(dz)\Big(\bP\big(\tau_{K}^{\varepsilon}\leq\theta\big)+1-e^{-\theta\nu([-\varepsilon,\varepsilon]^{c})}\Big)=O(\theta^{2}),\quad\theta\rightarrow 0^{+},
\end{align*}
where the last equality follows from \eqref{eq:EstProbtauKeps}. Together with \eqref{eq:Defgamma0ProcZ} we deduce that
\begin{align*}
\cI^{a}(\tau)&=\bE\left(\int_{0}^{\tau}e^{-rt}{\bf 1}_{\{S_{t}\leq K\}}\bigg(\big(r(S_{t}-K)-\gamma_{0}S_{t}\big)+\int_{\bR_{0}}\Phi(S_{t},z)\,\nu(dz)\bigg)dt\right)+o\big(\theta^{3/2}\big)\\
&=\bE\left(\int_{0}^{\tau}e^{-rt}{\bf 1}_{\{S_{t}\leq K\}}\bigg(\!-rK+\delta S_{t}+\int_{\bR_{0}}\big(S_{t}e^{z}-K\big)^{+}\nu(dz)\bigg)dt\right)+o\big(\theta^{3/2}\big)\\
&=\bE\left(\int_{0}^{\tau}\big(e^{-rt}-1\big){\bf 1}_{\{S_{t}\leq K\}}\bigg(\!-rK+\delta S_{t}+\int_{\bR_{0}}\big(S_{t}e^{z}-K\big)^{+}\nu(dz)\bigg)dt\right)\\
&\quad +\bE\left(\int_{0}^{\tau}{\bf 1}_{\{S_{t}\leq K\}}\bigg(\!-rK+\delta S_{t}+\int_{\bR_{0}}\big(S_{t}e^{z}-K\big)^{+}\nu(dz)\bigg)dt\right)+o\big(\theta^{3/2}\big)\\
&=\bE\left(\int_{0}^{\tau}{\bf 1}_{\{S_{t}\leq K\}}\bigg(\!-rK+\delta S_{t}+\int_{\bR_{0}}\big(S_{t}e^{z}-K\big)^{+}\nu(dz)\bigg)dt\right)+o\big(\theta^{3/2}\big),\quad\theta\rightarrow 0^{+},
\end{align*}
where we note that
\begin{align*}
&\left|\bE\left(\int_{0}^{\tau}\big(e^{-rt}-1\big){\bf 1}_{\{S_{t}\leq K\}}\bigg(\!-rK+\delta S_{t}+\int_{\bR_{0}}\big(S_{t}e^{z}-K\big)^{+}\nu(dz)\bigg)dt\right)\right|\\
&\quad\leq\bigg((r+\delta)K+K\int_{\bR_{0}}\big|e^{z}-1\big|\nu(dz)\bigg)\int_{0}^{\theta}\big(1-e^{-rt}\big)dt=O(\theta^{2}).
\end{align*}
Denoting by
\begin{align*}
h(x):={\bf 1}_{\{x\leq\ln K\}}\bigg(\!-rK+\delta e^{x}+\int_{\bR_{0}}\big(e^{x+z}-K\big)^{+}\nu(dz)\bigg),
\end{align*}
and recalling \eqref{eq:DefwtX} as well as $S_{0}=b(T)e^{a\sqrt{\theta}}$, we thus have
\begin{align}\label{eq:EstcIatau1}
\cI^{a}(\tau)=\bE\left(\int_{0}^{\tau}h\big(\ln b(T)+a\sqrt{\theta}+\wt{X}_{t}\big)\,dt\right)+o\big(\theta^{3/2}\big),\quad\theta\rightarrow 0^{+}.
\end{align}

Now we will try to express the expectation in \eqref{eq:EstcIatau1} in a more appropriate form. Notice that, for any (fixed) $\varepsilon\in(0,\ln(K/b(T)))$, we have
\begin{align*}
&\big|h(x)-h(y)\big|\leq\big|h(x)-h(y)\big|{\bf 1}_{\{x\vee y\leq\ln K\}}+\big|h(x)\big|{\bf 1}_{\{x\leq\ln K<y\}}+\big|h(y)\big|{\bf 1}_{\{y\leq\ln K<x\}}\\
&\quad =\big|h(x)-h(y)\big|\big({\bf 1}_{\{x\vee y\leq\ln K-\varepsilon\}}+{\bf 1}_{\{x\leq\ln K-\varepsilon<y\leq\ln K\}}+{\bf 1}_{\{y\leq\ln K-\varepsilon<x\leq\ln K\}}+{\bf 1}_{\{\ln K-\varepsilon<x,y\leq\ln K\}}\big)\\
&\qquad +\big|h(x)\big|{\bf 1}_{\{x\leq\ln K<y\}}+\big|h(y)\big|{\bf 1}_{\{y\leq\ln K<x\}}\\
&\quad\leq A_{0}^{\varepsilon}\,\big|e^{x}-e^{y}\big|{\bf 1}_{\{x\vee y\leq\ln K-\varepsilon\}}+\Big(A_{0}^{\varepsilon}\,\big|e^{x}-e^{y}\big|+A_{1}^{\varepsilon}\Big)\big({\bf 1}_{\{x\leq\ln K-\varepsilon<y\leq\ln K\}}+{\bf 1}_{\{y\leq\ln K-\varepsilon<x\leq\ln K\}}\big)\\
&\qquad +\Big(A_{0}^{\varepsilon}\,\big|e^{x}-e^{y}\big|+2A_{1}^{\varepsilon}\Big){\bf 1}_{\{\ln K-\varepsilon<x,y\leq\ln K\}}+K(r\vee|d|)\big({\bf 1}_{\{x\leq\ln K<y\}}+{\bf 1}_{\{y\leq\ln K<x\}}\big)\\
&\quad\leq A_{0}^{\varepsilon}\big|e^{x}-e^{y}\big|+2A_{1}^{\varepsilon}{\bf 1}_{\{\ln K-\varepsilon<x\wedge y\}}+K(r\vee|d|)\big({\bf 1}_{\{\ln K<y\}}+{\bf 1}_{\{\ln K<x\}}\big),
\end{align*}
where
\begin{align*}
A_{0}^{\varepsilon}:=\delta+\int_{(\varepsilon,\infty)}e^{z}\nu(dz),\quad A_{1}^{\varepsilon}:=K\int_{(0,\varepsilon)}\big(e^{z}-1\big)\nu(dz).
\end{align*}
Hence, we deduce that, for any $t\in[0,\theta]$,
\begin{align}
&\bE\Big(\Big|h\big(\ln b(T)+a\sqrt{\theta}+\wt{X}_{t}\big)-h\big(\ln b(T)+a\sqrt{\theta}+\sigma W_{t}\big)\Big|\Big)\nonumber\\
&\quad\leq A_{0}^{\varepsilon}\,b(T)e^{a\sqrt{\theta}}\bE\Big(\big|e^{\wt{X}_{t}}-e^{\sigma W_{t}}\big|\Big)+2A_{1}^{\varepsilon}\bP\big(\wt{X}_{t}\wedge(\sigma W_{t})>\ln K-\ln b(T)-\varepsilon-a\sqrt{\theta}\big)\nonumber\\
&\qquad +K(r\vee|d|)\Big(\bP\big(\wt{X}_{t}>\ln K-\ln b(T)-a\sqrt{\theta}\big)+\bP\big(\sigma W_{t}>\ln K-\ln b(T)-a\sqrt{\theta}\big)\Big)\nonumber\\
&\quad\leq A_{0}^{\varepsilon}\,b(T)\bE\big(e^{\sigma W_{t}}\big)\bE\Big(\big|e^{(\gamma_{0}-\sigma^{2}/2)t+Z_{t}}-1\big|\Big)+\big(2A_{1}^{\varepsilon}+K(r\vee|d|)\big)\bP\bigg(W_{t}>\frac{\ln K-\ln b(T)-\varepsilon}{\sigma}\bigg)\nonumber\\
\label{eq:DecompDiffExphwtXhW} &\qquad +K(r\vee|d|)\bP\big(\wt{X}_{t}>\ln K-\ln b(T)\big).
\end{align}
As $\theta\rightarrow 0^{+}$, we have
\begin{align}\label{eq:EstDiffExphwtXhW1}
&\bE\big(e^{\sigma W_{t}}\big)\bE\Big(\big|e^{(\gamma_{0}-\sigma^{2}/2)t+Z_{t}}\!-\!1\big|\Big)\leq e^{\sigma^{2}t/2}\big|e^{(\gamma_{0}-\sigma^{2}/2)t}\!-\!1\big|\bE\big(e^{Z_{t}}\big)+e^{\sigma^{2}t/2}\bE\Big(\big|e^{Z_{t}}\!-\!1\big|\Big)=O(\theta),\qquad\\
\label{eq:EstDiffExphwtXhW2} &\bP\bigg(\sup_{t\in[0,\theta]}W_{t}>\frac{\ln K-\ln b(T)-\varepsilon}{\sigma}\bigg)\leq e^{-(\ln K-\ln b(T)-\varepsilon)^{2}/(2\sigma^{2}\theta)}=o(\theta^{n}),\quad\text{for any }n\in\bN,
\end{align}
where we used Doob's martingale inequality in the last inequality, and by \eqref{eq:DefwtX}, \eqref{eq:DefZepsPlusMinus}, and \eqref{eq:DefOverlineZepsSepsNeps},
\begin{align}
\bP\bigg(\sup_{t\in[0,\theta]}\!\wt{X}_{t}\!>\!\ln\!\bigg(\frac{K}{b(T)}\bigg)\!\bigg)&\leq\bP\bigg(\sup_{t\in[0,\theta]}\!W_{t}^{\varepsilon}\!>\!\frac{A_{2}(\theta)}{3\sigma}\bigg)+\bP\bigg(\sup_{t\in[0,\theta]}\!Z_{t}^{\varepsilon}\!>\!\frac{A_{2}(\theta)}{3}\bigg)+\bP\bigg(\sup_{t\in[0,\theta]}\!\overline{Z}_{t}^{\varepsilon}\!>\!\frac{A_{2}(\theta)}{3}\bigg)\nonumber\\
\label{eq:EstDiffExphwtXhW3} &\leq e^{-A_{2}^{2}(\theta)/(2\sigma^{2}\theta)}+o(\theta^{n})+\bP\big(T_{1}^{\varepsilon}\leq\theta\big)=O(\theta),
\end{align}
where we again used Doob's martingale inequality as well as \cite[Remark 3.1]{FigueroaLopezHoudre:2009} in the second inequality, and denoted by $A_{2}(\theta):=\ln K-\ln b(T)-|\gamma_{0}-\sigma^{2}/2|\theta$. By combining \eqref{eq:EstcIatau1}, \eqref{eq:DecompDiffExphwtXhW}, \eqref{eq:EstDiffExphwtXhW1}, \eqref{eq:EstDiffExphwtXhW2}, and \eqref{eq:EstDiffExphwtXhW3}, we deduce that
\begin{align}\label{eq:EstcIatau2}
\cI^{a}(\tau)=\bE\left(\int_{0}^{\tau}h\big(\ln b(T)+a\sqrt{\theta}+\sigma W_{t}\big)\,dt\right)+o\big(\theta^{3/2}\big),\quad\theta\rightarrow 0^{+}.
\end{align}

Note that the function $h$ is convex on $(-\infty,\ln K)$, and thus it is right- and left-differentiable. In particular, for $x\in(-\infty,\ln K)$, we have
\begin{align}\label{eq:LeftRightDerivh}
h'_{+}(x)=e^{x}\bigg(\delta+\int_{[\ln K-x,\infty)}e^{z}\nu(dz)\bigg),\quad h'_{-}(x)=e^{x}\bigg(\delta+\int_{(\ln K-x,\infty)}e^{z}\nu(dz)\bigg).
\end{align}
Hence, with $x_{0}:=\ln b(T)$, we can write
\begin{align*}
h'_{+}(x_{0})(x-x_{0})^{+}-h'_{-}(x_{0})(x-x_{0})^{-}\leq h(x)-h(x_{0})\leq h'_{-}(x)(x-x_{0})^{+}-h'_{+}(x)(x-x_{0})^{-},
\end{align*}
and thus
\begin{align*}
0&\leq h(x)-h(x_{0})-h'_{+}(x_{0})(x-x_{0})^{+}+h'_{-}(x_{0})(x-x_{0})^{-}\\
&\leq\big(h'_{-}(x)\!-\!h'_{+}(x_{0})\big)(x-x_{0})^{+}+\big(h'_{-}(x_{0})\!-\!h'_{+}(x)\big)(x-x_{0})^{-}=\big(h'_{-}(x\vee x_{0})\!-\!h'_{+}(x\wedge x_{0})\big)|x-x_{0}|.
\end{align*}
Noting that $h(x_{0})=h(\ln b(T))=0$ in light of \eqref{eq:LimitAmerCritPriceNegd}, we deduce that
\begin{align}\label{eq:ExpanFunth}
h(x+x_{0})=h'_{-}(x_{0})x+\big(h'_{+}(x_{0})-h'_{-}(x_{0})\big)x^{+}+|x|\wt{R}(x),\quad x\in(-\infty,\ln K-x_{0}),
\end{align}
where $\wt{R}(x)\geq 0$ and $\wt{R}(x)\rightarrow 0$ as $x\rightarrow 0$. For $x\in(-\infty,0]$, clearly
\begin{align}\label{eq:BoundwtRNeg}
\wt{R}(x)\leq h'_{-}(x_{0})-h'_{+}(x+x_{0})\leq h'_{-}(x_{0})=:C_{-}.
\end{align}
For $x\in(0,\ln K-x_{0})$, we have
\begin{align*}
\wt{R}(x)&=\frac{1}{x}h(x+x_{0})-h'_{+}(x_{0})=\frac{1}{x}\big(h(x+x_{0})-h(x_{0})\big)-h'_{+}(x_{0})\\
&=\frac{\delta}{x}e^{x_{0}}\big(e^{x}-1\big)+\frac{1}{x}\int_{\bR_{0}}\Big(\big(e^{x+x_{0}+z}-K\big)^{+}-\big(e^{x_{0}+z}-K\big)^{+}\Big)\nu(dz)-h'_{+}(x_{0})\\
&\leq\delta e^{x_{0}+x}+\frac{1}{x}\bigg(\big(e^{x+x_{0}}-e^{x_{0}}\big)\int_{(\ln K-x_{0},\infty)}e^{z}\,\nu(dz)+\int_{(\ln K-x_{0}-x,\ln K-x_{0})}\big(e^{x_{0}+x+z}-K\big)\nu(dz)\bigg)\\
&\leq\delta e^{x_{0}+x}+e^{x_{0}+x}\int_{(\ln K-x_{0},\infty)}e^{z}\,\nu(dz)+\frac{K}{x}\int_{(\ln K-x_{0}-x,\ln K-x_{0})}\big(e^{z}-1\big)\nu(dz)\\
&\leq K\delta+K\int_{(\ln K-x_{0},\infty)}e^{z}\,\nu(dz)+\frac{K}{x}\int_{(\ln K-x_{0}-x,\ln K-x_{0})}\big(e^{z}-1\big)\nu(dz).
\end{align*}
By Assumption \ref{assump:LevyMeas} and the Fundamental theorem of calculus,
\begin{align*}
\lim_{x\rightarrow 0+}\frac{1}{x}\int_{(\ln K-x_{0}-x,\ln K-x_{0})}\big(e^{z}-1\big)s(z)\,dz=\bigg(\frac{K}{b(T)}-1\bigg)s\big(\ln K-x_{0}\big).
\end{align*}
Hence, there exists $\eta_{0}\in(0,\ln K-x_{0})$, such that for any $x\in(0,\eta_{0})$,
\begin{align*}
\frac{1}{x}\int_{(\ln K-x_{0}-x,\ln K-x_{0})}\big(e^{z}-1\big)s(z)\,dz\leq 2\bigg(\frac{K}{b(T)}-1\bigg)s\big(\ln K-x_{0}\big).
\end{align*}
Consequently, for any $x\in(0,\ln K-x_{0})$, we have
\begin{align}\label{eq:BoundwtRPos}
\wt{R}(x)\!\leq\!K\bigg(\!\delta\!+\!\!\int_{\ln K-x_{0}}^{\infty}\!\!e^{z}s(z)\,dz\!+\!\frac{1}{\eta_{0}}\!\int_{0}^{\ln K-x_{0}}\!\!\big(e^{z}\!-\!1\big)s(z)\,dz\!\bigg)\!+\!2\bigg(\!\frac{K}{b(T)}\!-\!1\!\bigg)s(\ln K\!-\!x_{0})\!=:\!C_{+}.\qquad
\end{align}
By combining \eqref{eq:BoundwtRNeg} and \eqref{eq:BoundwtRPos}, we obtain that
\begin{align}\label{eq:BoundwtR}
\wt{R}(x)\leq C_{-}+C_{+}<\infty,\quad\text{for all }\,x\in(-\infty,\ln K-x_{0}).
\end{align}

Coming back to the estimation of $I^{a}(\tau)$, we deduce from \eqref{eq:ExpanFunth} that
\begin{align}
&\bE\bigg(\int_{0}^{\tau}h\big(x_{0}+a\sqrt{\theta}+\sigma W_{t}\big)\,dt\bigg)=\bE\bigg(\int_{0}^{\tau}h\big(x_{0}+a\sqrt{\theta}+\sigma W_{t}\big){\bf 1}_{\{x_{0}+a\sqrt{\theta}+\sigma W_{t}<\ln K\}}\,dt\bigg)\nonumber\\
&\quad =\bE\bigg(\int_{0}^{\tau}\Big(h'_{-}(x_{0})\big(a\sqrt{\theta}+\sigma W_{t}\big)+\big(h'_{+}(x_{0})-h'_{-}(x_{0})\big)\big(a\sqrt{\theta}+\sigma W_{t}\big)^{+}\Big){\bf 1}_{\{x_{0}+a\sqrt{\theta}+\sigma W_{t}<\ln K\}}\,dt\bigg)\nonumber\\
\label{eq:DecompExpInthW} &\qquad +\bE\bigg(\int_{0}^{\tau}\big|a\sqrt{\theta}+\sigma W_{t}\big|\wt{R}\big(a\sqrt{\theta}+\sigma W_{t}\big){\bf 1}_{\{x_{0}+a\sqrt{\theta}+\sigma W_{t}<\ln K\}}\,dt\bigg),
\end{align}
where the first equality follows from the fact that
\begin{align*}
0&\leq\bE\bigg(\int_{0}^{\tau}h\big(x_{0}+a\sqrt{\theta}+\sigma W_{t}\big){\bf 1}_{\{x_{0}+a\sqrt{\theta}+\sigma W_{t}=\ln K\}}dt\bigg)\\
&\leq -dK\,\bP\big(\,\text{Leb}\big\{t\in[0,\theta]:\,\sigma W_{t}=\ln K-x_{0}-a\sqrt{\theta}\big\}\big)=0,
\end{align*}
since $h(\ln K)= -dK\geq 0$ in light of \eqref{eq:Defd}. For the first term in \eqref{eq:DecompExpInthW}, we have
\begin{align}
&\left|\bE\bigg(\int_{0}^{\tau}\Big(\big(h'_{+}(x_{0})-h'_{-}(x_{0})\big)\big(a\sqrt{\theta}+\sigma W_{t}\big)^{+}+h'_{-}(x_{0})\big(a\sqrt{\theta}+\sigma W_{t}\big)\Big){\bf 1}_{\{x_{0}+a\sqrt{\theta}+\sigma W_{t}\geq\ln K\}}\,dt\bigg)\right|\nonumber\\
&\quad\leq h'_{+}(x_{0})\sqrt{\theta}\,\bE\left(\int_{0}^{\theta}\bigg(|a|+\sqrt{\frac{t}{\theta}}\big|\sigma W_{1}\big|\bigg){\bf 1}_{\{|\sigma W_{1}|\geq -a+(\ln K-x_{0})/\sqrt{\theta}\}}\,dt\right)\nonumber\\
\label{eq:EstExpInthW1} &\quad\leq h'_{+}(x_{0})\,\theta^{3/2}\sqrt{\bE\Big(\big(|a|+|W_{1}|\big)^{2}\Big)}\sqrt{\bP\bigg(\sigma|W_{1}|\geq\frac{\ln K-x_{0}}{\sqrt{\theta}}\bigg)}=o(\theta^{n}),\quad\theta\rightarrow 0^{+},
\end{align}
for any $n\in\bN$. As for the second term in \eqref{eq:DecompExpInthW}, by \eqref{eq:BoundwtR} and bounded convergence, we have
\begin{align}
&\bE\bigg(\int_{0}^{\tau}\big|a\sqrt{\theta}+\sigma W_{t}\big|\wt{R}\big(a\sqrt{\theta}+\sigma W_{t}\big){\bf 1}_{\{x_{0}+a\sqrt{\theta}+\sigma W_{t}<\ln K\}}\,dt\bigg)\nonumber\\
\label{eq:EstExpInthW2} &\leq\theta^{3/2}\,\bE\bigg(\int_{0}^{1}\big|a+\sigma W_{s}\big|\wt{R}\big(\sqrt{\theta}(a+\sigma W_{s})\big){\bf 1}_{\{x_{0}+\sqrt{\theta}(a+\sigma W_{s})<\ln K\}}\,ds\bigg)=o(\theta^{3/2}),\quad\theta\rightarrow 0^{+}.\qquad
\end{align}
By combining \eqref{eq:EstcIatau2}, \eqref{eq:LeftRightDerivh}, \eqref{eq:DecompExpInthW}, \eqref{eq:EstExpInthW1}, and \eqref{eq:EstExpInthW2}, we obtain that
\begin{align}
\cI^{{\Red a}}(\tau)&=\bE\bigg(\int_{0}^{\tau}\Big(h'_{-}(x_{0})\big(a\sqrt{\theta}+\sigma W_{t}\big)+\big(h'_{+}(x_{0})-h'_{-}(x_{0})\big)\big(a\sqrt{\theta}+\sigma W_{t}\big)^{+}\Big)\,dt\bigg)+o(\theta^{3/2})\nonumber\\
\label{eq:EstcIatau} &=b(T)\overline{\delta}\,\bE\bigg(\int_{0}^{\tau}\Big(\big(a\sqrt{\theta}+\sigma W_{t}\big)+\lambda\beta\big(a\sqrt{\theta}+\sigma W_{t}\big)^{+}\Big)dt\bigg)+o(\theta^{3/2}),\quad\theta\rightarrow 0^{+},
\end{align}
where $\bar{\delta}=\delta+\int_{(\ln(K/b(T)),\infty)}e^{z}\nu(dz)$, $\lambda=\nu(\{\ln(K/b(T))\})$, and $\beta=K/(b(T)\overline{\delta})$.

Coming back to \eqref{eq:DecompAmerPutPrice} and using \eqref{eq:EstcJatau} and \eqref{eq:EstcIatau}, we deduce that
\begin{align*}
\bE\big(e^{-r\tau}(K\!-\!S_{\tau})^{+}\big)&=\big(K-b(T)e^{a\sqrt{\theta}}\big)^{+}+b(T)\overline{\delta}\,\bE\bigg(\int_{0}^{\tau}\Big(\big(a\sqrt{\theta}+\sigma W_{t}\big)+\lambda\beta\big(a\sqrt{\theta}+\sigma W_{t}\big)^{+}\Big)dt\bigg)\\
&\quad +K\,\bE\Big({\bf 1}_{\{\wh{T}_{1}<\tau\}}\Big(\big(\!-\!a\sqrt{\theta}\!-\!\sigma W_{\tau}\big)^{+}-\big(\!-\!a\sqrt{\theta}\!-\!\sigma W_{\wh{T}_{1}}\big)^{+}\Big)\Big)+o\big(\theta^{3/2}\big)\\
&=\big(K-b(T)e^{a\sqrt{\theta}}\big)^{+}+b(T)\overline{\delta}\,\bE\bigg(\int_{0}^{\tau}\Big(\big(a\sqrt{\theta}+\sigma W_{t}\big)+\lambda\beta\big(a\sqrt{\theta}+\sigma W_{t}\big)^{+}\Big)dt\bigg)\\
&\quad +K\,\bE\Big({\bf 1}_{\{\wh{T}_{1}<\tau\}}\Big(\big(a\sqrt{\theta}+\sigma W_{\tau}\big)^{+}-\big(a\sqrt{\theta}+\sigma W_{\wh{T}_{1}}\big)^{+}\Big)\Big)+o\big(\theta^{3/2}\big),\quad\theta\rightarrow 0^{+},
\end{align*}
with the $o(\theta^{3/2})$ term independent of $\tau$, where the last equality follows from the Tanaka's formula. Therefore, we obtain that
\begin{align*}
P\big(T-\theta,b(T)e^{a\sqrt{\theta}}\big)=\big(K-b(T)e^{a\sqrt{\theta}}\big)^{+}+\sigma b(T)\overline{\delta}\,\bar{v}_{\lambda,\beta,\theta}(a/\sigma)+o(\theta^{3/2}),\quad\theta\rightarrow 0^{+},
\end{align*}
where
\begin{align*}
\bar{v}_{\lambda,\beta,\theta}(y):=\sup_{\tau\in\sT_{0,\theta}}\bE\bigg(\int_{0}^{\tau}f_{\lambda\beta}\big(y\sqrt{\theta}+W_{t}\big)\,dt+\beta\,{\bf 1}_{\{\wh{T}_{1}<\tau\}}\Big(\big(y\sqrt{\theta}+W_{\tau}\big)^{+}-\big(y\sqrt{\theta}+W_{\wh{T}_{1}}\big)^{+}\Big)\bigg),
\end{align*}
with $f_{c}(x)=x+cx^{+}$, because $\bar{v}_{\lambda,\beta,\theta}(y)=\bar{v}_{\lambda,\beta}(y)\theta^{3/2}+o(\theta^{3/2})$ as shown at the end of the proof of Theorem 3.1 in \cite{BouselmiLamberton:2016}.\hfill $\Box$

\medskip

Thanks to Theorem \ref{thm:SmallTimeExpanAmerPutPrice}, we are now ready to present our main result in this section. The proof is similar to that of \cite[Theorem 3.2]{BouselmiLamberton:2016}, and is presented below for completeness.

\begin{theorem}\label{thm:NonZeroBMNegdAmerCritPriceConvRate}
Suppose that \eqref{eq:FinVarLevy} and Assumption \ref{assump:LevyMeas} hold. Assume that $\sigma>0$ and $d<0$, and let $y_{\lambda,\beta}$, $\lambda,\beta\in\Bbb{R}_{+}$, be given as in Theorem \ref{thm:FinJumpActAmerCritPriceConvRate}-(c).
\begin{itemize}[leftmargin=2.0em]

\vspace{0.1cm}
\item [(a)] If $\nu(\{\ln(K/b(T))\})=0$, then we have
    \begin{align*}
    \lim_{t\rightarrow T^{-}}\frac{b(T)-b(t)}{\sigma b(T)\sqrt{T-t}}=y_{0,0}.
    \end{align*}
\item[(b)] If $\nu(\{\ln(K/b(T))\})>0$, then we have
    \begin{align*}
    \lim_{t\rightarrow T^{-}}\frac{b(T)-b(t)}{\sigma b(T)\sqrt{T-t}}=y_{\lambda,\beta},
    \end{align*}
    where $\lambda=\nu(\{\ln(K/b(T))\})$, $\beta=K/(b(T)\bar{\delta})$, and $\bar{\delta}=\delta+\int_{(\ln(K/b(T)),\infty)}e^{z}\nu(dz)$.
\end{itemize}
\end{theorem}

The proof of Theorem \ref{thm:NonZeroBMNegdAmerCritPriceConvRate} requires the following technical lemma, the proof of which is deferred to Appendix \ref{sec:AppendixB}. Note that this lemma was proved in \cite{BouselmiLamberton:2016} (see Lemmas 2.2 therein) when $\nu(\bR_{0})<\infty$. Here we extend it to the case when the jump component of $X$ is of finite variation (i.e., \eqref{eq:FinVarLevy} holds).

\begin{lemma}\label{lem:AsymLowerBound2ndDerivAmerPutPrice}
Under the model assumptions, for any $s\in(b(t),b(T))$, we have
\begin{align*}
\int_{b(t)}^{s}(u-b(t))\frac{\partial^{2}P}{\partial s^{2}}(t,du)\geq\frac{h(s,T-t)}{\sigma^{2}b^{2}(T)}(s-b(t))^{2},\quad t\rightarrow T^{-},
\end{align*}
where
\begin{align*}
h(s,\theta):=\big(\bar{\delta}-\varepsilon(\theta)\big)(b(T)-s)-b(T)\bar{\delta}\lambda\beta\,\bE\Big(\big(\sigma W_{\theta}-\ln(b(T)/s)\big)^{+}\Big)+o(\sqrt{\theta}),\quad\theta\rightarrow 0^{+},
\end{align*}
with $\lim_{\theta\rightarrow 0^{+}}\varepsilon(\theta)=0$, and $\lambda$, $\beta$, and $\bar{\delta}$ are given as in Theorem \ref{thm:NonZeroBMNegdAmerCritPriceConvRate}-(b).
\end{lemma}

\noindent
\textbf{Proof of Theorem \ref{thm:NonZeroBMNegdAmerCritPriceConvRate}.} We first note that by \eqref{eq:DefFuntvlambdabeta}, when $\lambda=\nu(\{\ln(K/b(T))\})=0$, the function $v_{0,\beta}$ is independent of the value of $\beta\in\bR_{+}$, i.e., $v_{0,0}=v_{0,\beta}$ for any $\beta\in\bR_{+}$. Therefore, we can proceed the proof for parts (a) and (b) together by considering any $\lambda,\beta\in\bR_{+}$.

By \cite[Lemma 3.1]{BouselmiLamberton:2016}, we have $y_{\lambda,\beta}>0$. Also, from the definition of $v_{\lambda,\beta}$, we have $v_{\lambda,\beta}(a/\sigma)>0$ for all $a> -\sigma y_{\lambda,\beta}$. Hence, by Theorem \ref{thm:SmallTimeExpanAmerPutPrice} we have that, for any $a\in(-\sigma y_{\lambda,\beta},0)$,
\begin{align*}
P\big(T-\theta,b(T)e^{a\sqrt{\theta}}\big)>\big(K-b(T)e^{a\sqrt{\theta}}\big)^{+},\quad\text{for }\,\theta>0\,\,\,\text{small enough}.
\end{align*}
Taking $t=T-\theta$, it follow from the definition of the critical price that
\begin{align*}
\ln b(T)+a\sqrt{T-t}>\ln b(t).
\end{align*}
Using the nondecreasing property of $b$ on $[0,T]$ and the inequality $\ln(1+z)\leq z$, $z>-1$, we have
\begin{align*}
\frac{b(T)-b(t)}{b(t)\sqrt{T-t}}\geq\frac{1}{\sigma}\frac{\big(\ln b(T)-\ln b(t)\big)}{\sqrt{T-t}}>-a.
\end{align*}
By taking first $t\rightarrow T^{-}$ and then $a\rightarrow (-\sigma y_{\lambda,\beta})^{+}$, we obtain that
\begin{align}\label{eq:NonZeroBMNegdAmerCritPriceLiminf}
\liminf_{t\rightarrow T^{-}}\frac{b(T)-b(t)}{b(T)\sqrt{T-t}}\geq\sigma y_{\lambda,\beta}.
\end{align}

On the other hand, for any $a\leq -\sigma y_{\lambda,\beta}<0$, $v_{\lambda,\beta}(a/\sigma)=0$ and consequently,
\begin{align}\label{eq:DiffAmerPutPriceSmallo}
g(\theta):=P\big(T-\theta,b(T)e^{a\sqrt{\theta}}\big)-\big(K-b(T)e^{a\sqrt{\theta}}\big)^{+}=o\big(\theta^{3/2}\big),\quad\theta\rightarrow 0^{+}.
\end{align}
In addition, for any $t\in[0,T)$ and $s\in(b(t),b(T))$, we have
\begin{align*}
P(t,s)-P(t,b(t))-(s-b(t))\frac{\partial P}{\partial s}(t,b(t))=\int_{b(t)}^{s}(u-b(t))\frac{\partial^{2}P}{\partial s^{2}}(t,du),
\end{align*}
in the distributional sense (note that $(\partial^{2}P/\partial s^{2})(t,du)$ is a positive measure on $(0,\infty)$). Due to the smooth-fit property (cf. \cite[Proposition 4.1 \& Theorem 4.1]{LambertonMikou:2012}) and Lemma \ref{lem:AsymLowerBound2ndDerivAmerPutPrice}, we deduce that
\begin{align*}
P(t,s)-(K-s)=\int_{b(t)}^{s}(u-b(t))\frac{\partial^{2}P}{\partial s^{2}}(t,du)\geq\frac{h(s,T-t)}{\sigma^{2}b^{2}(T)}(s-b(t))^{2},\quad\text{for small }\,\theta=T-t>0.
\end{align*}
For any $t\in [0,T]$ with $\theta=T-t$ small enough, by \eqref{eq:NonZeroBMNegdAmerCritPriceLiminf} we can pick $a=a(t)< -\sigma y_{\lambda,\beta}$ big enough so that $s=s(t):=b(T)e^{a\sqrt{\theta}}\in(b(t),b(T))$ and that $a(t)\uparrow -\sigma y_{\lambda,\beta}$ as $\theta\rightarrow 0^{+}$. It follows that
\begin{align*}
h(s,\theta)&=b(T)\big(\bar{\delta}-\varepsilon(\theta)\big)\big(1-e^{a\sqrt{\theta}}\big)-b(T)\bar{\delta}\lambda\beta\,\bE\Big(\big(\sigma W_{\theta}+a\sqrt{\theta}\big)^{+}\Big)+o(\sqrt{\theta})\\
&\geq b(T)\big(\bar{\delta}-\varepsilon(\theta)\big)\bigg(-a\sqrt{\theta}-\frac{a^{2}\theta}{2}\bigg)-b(T)\bar{\delta}\lambda\beta\sigma\sqrt{\theta}\,\bE\big((W_{1}+a\sigma^{-1})^{+}\big)+o(\sqrt{\theta})\\
&=b(T)\bar{\delta}\sigma\sqrt{\theta}\bigg(-\frac{a}{\sigma}-\lambda\beta\,\bE\big((W_{1}+a\sigma^{-1})^{+}\big)\bigg)+o(\sqrt{\theta}),\quad\theta\rightarrow 0^{+}.
\end{align*}
Letting $C(z):=z-\lambda\beta\,\bE[(W_{1}-z)^{+}]$, $z\in\bR$, we have
\begin{align*}
P\big(t,b(T)e^{a\sqrt{\theta}}\big)-\big(K-b(T)e^{a\sqrt{\theta}}\big)\geq\big(b(T)e^{a\sqrt{\theta}}-b(t)\big)^{2}\left(\frac{\bar{\delta}\sqrt{\theta}\,C(-a\sigma^{-1})}{\sigma b(T)}+o(\sqrt{\theta})\right),\quad\theta\rightarrow 0^{+}.
\end{align*}
By \cite[Lemma 3.2]{BouselmiLamberton:2016}, $C(z)>0$ for all $z>y_{\lambda,\beta}$. Together with \eqref{eq:DiffAmerPutPriceSmallo}, we deduce that,
\begin{align*}
\big(b(T)e^{a\sqrt{\theta}}-b(t)\big)^{2}\leq\frac{2b(T)\sigma g(\theta)}{\bar{\delta}\sqrt{\theta}C(-a\sigma^{-1})}=o(\theta),
\end{align*}
for $\theta>0$ small enough, which implies that
\begin{align*}
\frac{b(T)-b(t)}{b(T)\sqrt{\theta}}\leq -a+o(1).
\end{align*}
Finally, by taking $t\rightarrow T^{-}$, we obtain that
\begin{align*}
\limsup_{t\rightarrow T^{-}}\frac{b(T)-b(t)}{b(T)\sqrt{T-t}}\leq\sigma y_{\lambda,\beta},
\end{align*}
which completes the proof of the theorem.\hfill $\Box$

\vspace{0.2cm}

\section{Conclusions}

\vspace{0.4cm}
The present work study the convergence rate of the critical price of an American put option near-maturity under an exponential L\'{e}vy model with a nonzero Brownian component. We focus on two important scenarios which were not investigated in the literature. Namely, we consider the models with negative jumps of infinite variation when the critical price converges to the strike price and the models with infinite activity jumps when the critical price tends to a value strictly lower than the strike price. In both scenarios, the convergence rate is shown to be of order $O(\sqrt{T-t})$ with explicit constants proportionality. As a by product, we obtain a second-order near-maturity expansion of the American put price around the critical price.

\appendix

\section{Proofs of Lemmas in Section \ref{sec:NonZeroBMPosdAmerCritPriceConvRate}}\label{sec:AppendixA}

\vspace{0.4cm}

\subsection*{Proof of Lemma \ref{lem:NonZeroBMPosdLevySmallTimeAsymNormal}} Clearly, we have
\begin{align*}
\frac{1}{\sqrt{t}}\bigg(X_{t}^{W}-t\int_{\bR_{0}}\big(e^{z}-1\big)^{+}\nu(dz)-t\int_{-\infty}^{0-}\big(e^{z}-1-z\big)\,\nu(dz)\bigg)\cd\sigma W_{1},\quad\text{as }\,t\rightarrow 0^{+}.
\end{align*}
Since $J^{+}$ is a L\'{e}vy process of type B (cf. \cite[Definition 11.9]{Sato:1999}), by \cite[Theorem 43.20]{Sato:1999} we also have
\begin{align*}
\lim_{t\rightarrow 0^{+}}\frac{J_{t}^{+}}{\sqrt{t}}\rightarrow 0\quad\bP-\text{a.}\,\text{s.}
\end{align*}
In view of \eqref{eq:NonZeroBMPosdDecompLevy}, it remains to analyze the behavior of the stochastic integral in $J^{-}$. For any $\eta\in(0,\infty)$, we define
\begin{align*}
J_{t}^{-\eta}:=\int_{0}^{t}\int_{[-\eta,0)}z\,\wt{N}(ds,dz),\quad\overline{J}_{t}^{-\eta}:=\int_{0}^{t}\int_{(-\infty,-\eta)}z\,N(ds,dz),\quad t\in\bR_{+},
\end{align*}
so that
\begin{align*}
J_{t}^{-}=J_{t}^{-\eta}+\overline{J}_{t}^{-\eta}-t\int_{-\infty}^{0}\big(e^{z}-1-z{\bf 1}_{[-\eta,0)}(z)\big)\,\nu(dz).
\end{align*}
Clearly the last integral above is of order $o(\sqrt{t})$ as $t\rightarrow 0^{+}$. We now show that
\begin{align*}
J_{t}^{-\eta}+\overline{J}_{t}^{-\eta}=o_{\bP}(\sqrt{t}),\quad\text{as }\,t\rightarrow 0^{+}.
\end{align*}
For any $\varepsilon,\kappa\in(0,\infty)$, choose $\eta\in(0,\infty)$ small enough so that $\int_{[-\eta,0)}z^{2}\nu(dz)\leq\varepsilon^{2}\kappa^{2}$. For any $t\in(0,\varepsilon^{2}\kappa^{2})$, by Markov inequality and Cauchy-Schwarz inequality, we have
\begin{align*}
\bP\big(J_{t}^{-\eta}+\overline{J}_{t}^{-\eta}>\kappa\sqrt{t}\big)&\leq\frac{1}{\kappa\sqrt{t}}\Big(\bE\big(|J_{t}^{-\eta}|\big)+\bE\big(|\overline{J}_{t}^{-\eta}|\big)\Big)\leq\frac{\sqrt{\bE\big(|J_{t}^{-\eta}|^{2}\big)}}{\kappa\sqrt{t}}+\frac{\sqrt{t}}{\kappa}\int_{(-\infty,-\eta)}|z|\,\nu(dz)\\
&=\frac{1}{\kappa}\bigg(\int_{[-\eta,0)}\!z^{2}\nu(dz)\bigg)^{1/2}\!\!+\frac{\sqrt{t}}{\kappa}\!\int_{(-\infty,-\eta)}\!|z|\,\nu(dz)\leq\varepsilon\bigg(1+\!\int_{(-\infty,-\eta)}\!|z|\,\nu(dz)\bigg).
\end{align*}
Hence, we conclude that
\begin{align}\label{eq:ScalNegJumpConvProb}
\frac{J_{t}^{-}}{\sqrt{t}}\cp 0,\quad t\rightarrow 0^{+}.
\end{align}
The result of the lemma follows from \eqref{eq:NonZeroBMPosdDecompLevy} and Slutsky's theorem.\hfill $\Box$

\subsection*{Proof of Lemma \ref{lem:NonZeroBMPosdEuroCritPrice}} Note that
\begin{align*}
\bigg|\bE\Big(\big(e^{(r-\delta)\tau+X_{\tau}}\!-1-\zeta(\tau)\big)^{+}\Big)-\bE\Big(\big(e^{X_{\tau}}\!-1-\zeta(\tau)\big)^{+}\Big)\bigg|\leq\big(e^{(r-\delta)\tau}\!-1\big)\bE\big(e^{X_{\tau}}\big)=O(\tau),\quad\tau\rightarrow 0^{+}.
\end{align*}
In view of \eqref{eq:AsymEuroPutPrice}, and since $e^{z}\geq 1+z$, we have
\begin{align*}
\bE\big((X_{\tau}-\zeta(\tau))^{+}\big)\leq\bE\Big(\big(e^{X_{\tau}}-1-\zeta(\tau)\big)^{+}\Big)=O(\tau),\quad\tau\rightarrow 0^{+}.
\end{align*}
Hence, we deduce that
\begin{align*}
\lim_{\tau\rightarrow 0^{+}}\bE\bigg(\bigg(\frac{X_{\tau}}{\sqrt{\tau}}-\frac{\zeta(\tau)}{\sqrt{\tau}}\bigg)^{+}\bigg)=0.
\end{align*}
Now if we had $\lambda:=\liminf_{\tau\rightarrow 0^{+}}\zeta(\tau)/\sqrt{\tau}<\infty$, then by Lemma \ref{lem:NonZeroBMPosdLevySmallTimeAsymNormal} and Fatou's lemma
we would have $\bE((\sigma W_{1}-\lambda)^{+})=0$, and so $\bP(\sigma W_{1}\leq\lambda)=1$. However, the support of $W_{1}$ is the whole real line. The lemma is proved by contradiction.\hfill $\Box$

\subsection*{Proof of Lemma \ref{lem:NonZeroBMPosdTailProbBMNegJump}} For any $\varepsilon\in(0,1)$, $r\in(0,\infty)$, and $t\in(0,t_{0}]$, by Markov inequality, the independence between $X^{W}$ and $J^{-}$, and \eqref{eq:MGFJMinus}, we have
\begin{align*}
\bP\Big(X_{\tau}^{W}+J_{\tau}^{-}\geq r\sqrt{\tau}\Big)&\leq e^{-pr}\,\bE\Big(e^{pX_{\tau}^{W}\!/\sqrt{\tau}}\Big)\bE\Big(e^{pJ_{\tau}^{-}\!/\sqrt{\tau}}\Big)\\
&=\exp\!\left(-pr+\frac{1}{2}\big(\sigma^{2}p^{2}\!-\!p\sigma^{2}\!\sqrt{\tau}\big)+\tau\!\int_{(-\infty,0)}\!\!\bigg(e^{pz/\sqrt{\tau}}\!-\!1\!-\!\frac{p}{\sqrt{\tau}}\big(e^{z}-1\big)\bigg)\nu(dz)\right)\\
&\leq\exp\bigg(\!\!-\!pr+\frac{1}{2}\big(\sigma^{2}p^{2}-p\sigma^{2}\!\sqrt{\tau}\big)+p\sqrt{\tau}\,\nu((-\infty,-\varepsilon))+p^{2}\int_{[-\varepsilon,0)}z^{2}\,\nu(dz)\bigg)\\
&=\exp\bigg(\!\!-\!pr+\frac{Lp^{2}}{2}+p\sqrt{\tau}\bigg(\nu((-\infty,-\varepsilon))-\frac{\sigma^{2}}{2}\bigg)\bigg)=e^{-Lp^{2}/2},
\end{align*}
which completes the proof of the lemma.\hfill $\Box$

\subsection*{Proof of Lemma \ref{lem:NonZeroBMPosdLargeDevBMNegJump}} For any $\varepsilon\in(0,1)$, note that $\tau\leq f^{2}(\tau)/(\nu((-\infty,-\varepsilon))-\sigma^{2}/2)^{2}$ for $\tau>0$ small enough. Taking $r=f(\tau)$ in Lemma \ref{lem:NonZeroBMPosdTailProbBMNegJump}, and noting from \eqref{eq:ConstLtau0} and \eqref{eq:Constp} that
\begin{align*}
p=p(\tau;f(\tau),\sigma,\varepsilon)\sim\frac{f(\tau)}{L},\quad\tau\rightarrow 0^{+},\quad\text{and}\quad L=L(\varepsilon;\sigma)\rightarrow\sigma^{2},\quad\varepsilon\rightarrow 0^{+},
\end{align*}
we deduce that
\begin{align*}
\limsup_{\tau\rightarrow 0^{+}}\frac{1}{f^{2}(\tau)}\ln\bP\Big(X_{\tau}^{W}+J_{\tau}^{-}\geq\sqrt{\tau}f(\tau)\Big)\leq\lim_{\varepsilon\rightarrow 0^{+}}\lim_{\tau\rightarrow 0^{+}} -\frac{L(\varepsilon;\sigma)p^{2}(\tau;f(\tau),\sigma,\varepsilon)}{2f^{2}(\tau)}= -\frac{1}{2\sigma^{2}}.
\end{align*}

On the other hand, using the independence between $X^{W}$ and $J^{-}$, for any $\tau\in(0,\infty)$ we have
\begin{align*}
\bP\Big(X_{\tau}^{W}+J_{\tau}^{-}\geq\sqrt{\tau}f(\tau)\Big)&\geq\bP\bigg(\sigma W_{1}+\frac{J_{\tau}^{-}}{\sqrt{\tau}}-\frac{\sigma^{2}\sqrt{\tau}}{2}\geq f(\tau),\,\bigg|\frac{J_{\tau}^{-}}{\sqrt{\tau}}-\frac{\sigma^{2}\sqrt{\tau}}{2}\bigg|\leq\varepsilon\bigg)\\
&\geq\bP\big(\sigma W_{1}\geq f(\tau)+\varepsilon\big)\bP\bigg(\bigg|\frac{J_{\tau}^{-}}{\sqrt{\tau}}-\frac{\sigma^{2}\sqrt{\tau}}{2}\bigg|\leq\varepsilon\bigg)\\
&\geq\frac{\sigma\big(f(\tau)+\varepsilon\big)\phi\big((f(\tau)+\varepsilon)/\sigma\big)}{\sigma^{2}+\big(f(\tau)+\varepsilon\big)^{2}}\bP\bigg(\bigg|\frac{J_{\tau}^{-}}{\sqrt{\tau}}-\frac{\sigma^{2}\sqrt{\tau}}{2}\bigg|\leq\varepsilon\bigg),
\end{align*}
where $\phi$ denotes the standard normal density, and we used the following bounds on the tail probability of standard normal distribution in the last inequality
\begin{align}\label{eq:BoundsStdNormalTailProb}
\frac{z\phi(z)}{1+z^{2}}\leq\int_{z}^{\infty}\phi(z)\,dz\leq\frac{\phi(z)}{z},\quad z\in(0,\infty).
\end{align}
In view of \eqref{eq:ScalNegJumpConvProb}, we obtain that
\begin{align*}
\liminf_{\tau\rightarrow 0^{+}}\frac{\ln\!\bP\Big(\!X_{\tau}^{W}\!\!+\!J_{\tau}^{-}\!\geq\!\!\sqrt{\tau}f(\tau)\!\Big)}{f^{2}(\tau)}\!\geq\!-\!\!\lim_{\tau\rightarrow 0^{+}}\!\!\left(\!\frac{(f(\tau)\!+\!\varepsilon)^{2}}{2\sigma^{2}f^{2}(\tau)}\!+\!\frac{\ln\!\sqrt{2\pi}}{f^{2}(\tau)}\!-\!\frac{1}{f^{2}(\tau)}\!\ln\!\bigg(\frac{\sigma\big(f(\tau)+\varepsilon\big)}{\sigma^{2}\!+\!\big(f(\tau)\!+\!\varepsilon\big)^{2}}\!\bigg)\!\!\right)\!\!=\!-\frac{1}{2\sigma^{2}},
\end{align*}
which completes the proof of the lemma.\hfill $\Box$

\subsection*{Proof of (\ref{eq:IneqIntJ1J2})} In view of \eqref{eq:LowerBound2ndDerivwtPJ1J2}, for any test function $\wt{\phi}\in C_{c}^{\infty}(\wt{\cC}_{\rho})$, we have
\begin{align*}
\iint_{\wt{C}_{\rho}}\bigg(\frac{dK}{4}-\cJ_{1}(t,x)-\cJ_{2}(t,x)\bigg)\wt{\phi}(t,x)\,dx\,dt\leq\frac{\sigma^{2}}{2}\iint_{\wt{C}_{\rho}}\wt{P}(t,x)\bigg(\frac{\partial^{2}\wt{\phi}}{\partial x^{2}}(t,x)+\frac{\partial\wt{\phi}}{\partial x}(t,x)\bigg)dx\,dt.
\end{align*}
In view of Proposition \ref{prop:RegPropAmerPutPrice}, $P$ is jointly continuous on $[0,T]\times\bR_{+}$, and so is $\wt{P}$ on $[0,T]\times\bR$. Together with the fact that $\wt{\phi}\in C_{c}^{\infty}(\wt{\cC}_{\rho})$, we deduce that, for any $t\in(T-\rho,T)$,
\begin{align}\label{eq:WtPIneqTestFunt}
\int_{\ln b(t)}^{\ln b_{e}(t)}\!\!\bigg(\frac{dK}{4}\!-\!\cJ_{1}(t,x)\!-\!\cJ_{2}(t,x)\!\bigg)\wt{\phi}(t,x)\,dx\leq\frac{\sigma^{2}}{2}\!\int_{\ln b(t)}^{\ln b_{e}(t)}\!\wt{P}(t,x)\bigg(\frac{\partial^{2}\wt{\phi}}{\partial x^{2}}(t,x)\!+\!\frac{\partial\wt{\phi}}{\partial x}(t,x)\!\bigg)dx.\quad
\end{align}
For any fixed $(t_{0},x_{0})\in\wt{\cC}_{\rho}$, we choose $(\psi_{n})_{n\in\bN}\subset C_{c}^{\infty}(\ln b(t_{0}),\ln b_{e}(t_{0}))$ such that $\psi_{n}\geq 0$ and $\psi_{n}\rightarrow{\bf 1}_{(\ln b(t_{0}),x_{0}]}$ pointwisely on $(\ln b(t_{0}),\ln b_{e}(t_{0}))$, as $n\rightarrow\infty$. Then we extend each $\psi_{n}$ to $\wt{\phi}_{n}\in C_{c}^{\infty}(\wt{\cC}_{\rho})$ such that $\wt{\phi}_{n}(t_{0},\cdot)=\psi_{n}(\cdot)$, and clearly $\wt{\phi}_{n}$ satisfies \eqref{eq:WtPIneqTestFunt}. Denote $F_{t_{0}}(s):=\partial_{+}P(t_{0},s)/\partial s$ and $\phi_{n}(t,s):=\wt{\phi}_{n}(t,\ln s)$, $s\in(b(t_{0}),b_{e}(t_{0}))$. Due to the convexity of $P(t_{0},\cdot)$, $F_{t_{0}}$ is monotone nondecreasing on $(b(t_{0}),b_{e}(t_{0}))$. Hence, we deduce that
\begin{align}
&\int_{\ln b(t_{0})}^{\ln b_{e}(t_{0})}\!\wt{P}(t_{0},x)\bigg(\frac{\partial^{2}\wt{\phi}_{n}}{\partial x^{2}}(t_{0},x)+\frac{\partial\wt{\phi}_{n}}{\partial x}(t_{0},x)\bigg)dx\nonumber\\
&\quad =\int_{\ln b(t_{0})}^{\ln b_{e}(t_{0})}P(t_{0},e^{x})\bigg(e^{2x}\frac{\partial^{2}\phi_{n}}{\partial s^{2}}(t_{0},e^{x})+2e^{x}\frac{\partial\phi_{n}}{\partial s}(t_{0},e^{x})\bigg)dx\nonumber\\
&\quad =\int_{b(t_{0})}^{b_{e}(t_{0})}P(t_{0},s)\bigg(s\frac{\partial^{2}\phi_{n}}{\partial s^{2}}(t_{0},s)+2\frac{\partial\phi_{n}}{\partial s}(t_{0},s)\!\bigg)ds=\int_{b(t_{0})}^{b_{e}(t_{0})}P(t_{0},s)\frac{\partial^{2}(s\phi_{n})}{\partial s^{2}}(t_{0},s)\,ds\nonumber\\
\label{eq:EstRHSWtPIneqTestFuntn} &\quad =-\int_{b(t_{0})}^{b_{e}(t_{0})}F_{t_{0}}(s)\frac{\partial(s\phi_{n})}{\partial s}(t_{0},s)\,ds=\int_{b(t_{0})}^{b_{e}(t_{0})}s\,\phi_{n}(t_{0},s)\,dF_{t_{0}}(s).
\end{align}
Combining \eqref{eq:WtPIneqTestFunt} (for $\wt{\phi}_{n}$ with $t=t_{0}$) and \eqref{eq:EstRHSWtPIneqTestFuntn}, we obtain from the dominated convergence that
\begin{align*}
&\int_{\ln b(t_{0})}^{x_{0}}\!\bigg(\frac{dK}{4}\!-\!\cJ_{1}(t_{0},x)\!-\!\cJ_{2}(t_{0},x)\bigg)dx=\lim_{n\rightarrow\infty}\int_{\ln b(t_{0})}^{\ln b_{e}(t_{0})}\!\bigg(\frac{dK}{4}\!-\!\cJ_{1}(t_{0},x)\!-\!\cJ_{2}(t_{0},x)\bigg)\wt{\phi}_{n}(t,x)dx\\
&\quad\leq\frac{\sigma^{2}}{2}\lim_{n\rightarrow\infty}\int_{b(t_{0})}^{b_{e}(t_{0})}s\,\phi_{n}(t_{0},s)\,dF_{t_{0}}(s)=\frac{\sigma^{2}}{2}\int_{b(t_{0})}^{e^{x_{0}}}s\,dF_{t_{0}}(s)\\
&\quad\leq\frac{\sigma^{2}e^{x_{0}}}{2}\big(F_{t_{0}}(e^{x_{0}})\!-\!F_{t_{0}}(b(t_{0}))\big)=\frac{\sigma^{2}e^{x_{0}}}{2}\bigg(\frac{\partial_{+}P(t,e^{x_{0}})}{\partial s}\!+\!1\!\bigg).
\end{align*}
Finally, by taking $t_0=t$ and $x_0=a+\ln b(t)$, we conclude \eqref{eq:IneqIntJ1J2}, for any $t\in(T-\rho,T)$ and $a\in(0,\ln(b_{e}(t)/b(t)))$.\hfill $\Box$

\vspace{0.2cm}

\section{Proofs of Lemmas in Section \ref{sec:NonZeroBMNegdAmerCritPriceConvRate}}\label{sec:AppendixB}

\vspace{0.4cm}

\subsection*{Proof of Lemma \ref{lem:SmallTimeEstExpLocalTime}} For any $\varepsilon\in(0,\ln(K/b(T)))$, let $T_{1}^{\varepsilon}:=\inf\{t\in\bR_{+}:|\Delta X_{t}|>\varepsilon\}$. Clearly
\begin{align}\label{eq:DecompLocalTimetauK}
L_{\tau}^{K}=L_{\tau\wedge T_{1}^{\varepsilon}}^{K}+\big(L_{\tau}^{K}-L_{\tau\wedge T_{1}^{\varepsilon}}^{K}\big).
\end{align}
For any $t\in\bR_{+}$, set
\begin{align}\label{eq:DefZepsPlusMinus}
Z_{t}^{\varepsilon}:=\!\int_{0}^{t}\!\int_{[-\varepsilon,\varepsilon]\setminus\{0\}}\!z\,N(ds,dz),\,\,\,\,Z_{t}^{\varepsilon,+}:=\!\int_{0}^{t}\!\int_{(0,\varepsilon]}\!z\,N(ds,dz),\,\,\,\,Z_{t}^{\varepsilon,-}:=\!\int_{0}^{t}\!\int_{[-\varepsilon,0)}\!z\,N(ds,dz),\quad
\end{align}
and
\begin{align}\label{eq:DefOverlineZepsSepsNeps}
\overline{Z}_{t}^{\varepsilon}:=Z_{t}-Z_{t}^{\varepsilon},\quad S_{t}^{\varepsilon}:=S_{0}\exp\bigg(\bigg(\gamma_{0}-\frac{\sigma^{2}}{2}\bigg)t+\sigma W_{t}+Z_{t}^{\varepsilon}\bigg),\quad\overline{N}_{t}^{\varepsilon}:=N([0,t]\times[-\varepsilon,\varepsilon]^{c}).
\end{align}
The local time of the process $S^{\varepsilon}:=(S_{t}^{\varepsilon})_{t\in\bR_{+}}$ at $K$ until time $t$ will be denoted by $L_{t}^{K,\varepsilon}$.

\medskip
\noindent
\textbf{Step 1. Estimating $\bE(L_{\tau\wedge T_{1}^{\varepsilon}}^{K})$.}

\medskip
\noindent
Note that the two processes $S$ and $S^{\varepsilon}$ coincide up to $T_{1}^{\varepsilon}$, and so
\begin{align*}
L_{\tau\wedge T_{1}^{\varepsilon}}^{K}\leq L_{\theta\wedge T_{1}^{\varepsilon}}^{K}=L_{\theta\wedge T_{1}^{\varepsilon}}^{K,\varepsilon}\leq L_{\theta}^{K,\varepsilon}=L_{\theta}^{K,\varepsilon}{\bf 1}_{\{\tau_{K}^{\varepsilon}<\theta\}}\quad\bP-\text{a.}\,\text{s.},
\end{align*}
where the last equality is due to the fact that the sample paths of $L^{K,\varepsilon}:=(L_{t}^{K,\varepsilon})_{t\in\bR_{+}}$ are strictly increasing only on $\{t\in\bR_{+}:S_{t}^{\varepsilon}=K\}$ and $\tau_{K}^{\varepsilon}:=\inf\{t\in\bR_{+}:S_{t}^{\varepsilon}\geq K\}$. Hence, by H\"{o}lder's inequality we have
\begin{align}\label{eq:EstExpLocalTimeEpstauT1Ini}
\bE\Big(L_{\tau\wedge T_{1}^{\varepsilon}}^{K}\Big)\leq\bE\Big(L_{\theta}^{K,\varepsilon}{\bf 1}_{\{\tau_{K}^{\varepsilon}<\theta\}}\Big)\leq\sqrt{\bE\Big(\big(L_{\theta}^{K,\varepsilon}\big)^{2}\Big)\,\bP\big(\tau_{K}^{\varepsilon}<\theta\big)}
\end{align}
The first expectation on the right-hand side of \eqref{eq:EstExpLocalTimeEpstauT1Ini} can be estimated using the Meyer-It\^{o} formula:
\begin{align}
&\bE\Big(\big(L_{\theta}^{K,\varepsilon}\big)^{2}\Big)=4\bE\!\left(\!\bigg(\!\big(K\!-\!S_{\theta}^{\varepsilon}\big)^{+}\!-(K\!-\!S_{0})^{+}\!+\!\int_{0}^{\theta}\!{\bf 1}_{\{S_{t-}^{\varepsilon}\leq K\}}S_{t-}^{\varepsilon}\big(\gamma_{0}dt\!+\!\sigma dW_{t}\big)-\!\!\sum_{t\in[0,\theta]}\!\!\Delta\big(K\!-\!S_{t}^{\varepsilon}\big)^{+}\bigg)^{2}\right)\nonumber\\
&\quad\leq 20\,\bE\left(2K^{2}+\bigg(\int_{0}^{\theta}\gamma_{0}{\bf 1}_{\{S_{t}^{\varepsilon}\leq K\}}S_{t}^{\varepsilon}\,dt\bigg)^{2}+\bigg(\int_{0}^{\theta}\sigma{\bf 1}_{\{S_{t}^{\varepsilon}\leq K\}}S_{t}^{\varepsilon}\,dW_{t}\bigg)^{2}+\bigg(\sum_{t\in[0,\theta]}\Delta\big(K-S_{t}^{\varepsilon}\big)^{+}\bigg)^{2}\right)\nonumber\\
&\quad\leq 20\,\bE\left(2K^{2}+\int_{0}^{\theta}\gamma_{0}^{2}\,{\bf 1}_{\{S_{t}^{\varepsilon}\leq K\}}\big(S_{t}^{\varepsilon}\big)^{2}dt+\int_{0}^{\theta}\sigma^{2}\,{\bf 1}_{\{S_{t}^{\varepsilon}\leq K\}}\big(S_{t}^{\varepsilon}\big)^{2}dt+\bigg(\sum_{t\in[0,\theta]}S_{t-}^{\varepsilon}\Big|e^{\Delta Z_{t}^{\varepsilon}}-1\Big|\bigg)^{2}\right)\nonumber\\
\label{eq:EstExpLocalTimeEpstheta} &\quad\leq 40K^{2}+20K^{2}\big(\gamma_{0}^{2}+\sigma^{2}\big)\theta+4\bE\left(\bigg(\sum_{t\in[0,\theta]}S_{t-}^{\varepsilon}\Big|e^{\Delta Z_{t}^{\varepsilon}}-1\Big|\bigg)^{2}\right)=O(1),\quad\theta\rightarrow 0^{+},
\end{align}
where the bound is independent of $a$, since the last expectation above can be bounded by (recalling the definition of the process $S^{\varepsilon}$ in \eqref{eq:DefOverlineZepsSepsNeps} and that $S_{0}=b(T)e^{a\sqrt{\theta}}\leq b(T)$)
\begin{align*}
&\bE\left(\bigg(\sum_{t\in[0,\theta]}S_{t-}^{\varepsilon}\Big|e^{\Delta Z_{t}^{\varepsilon}}-1\Big|\bigg)^{2}\right)=\bE\left(\bigg(\int_{0}^{\theta}\int_{(-\varepsilon,\varepsilon)\setminus\{0\}}S_{t-}^{\varepsilon}\big|e^{z}-1\big|\,N(dt,dz)\bigg)^{2}\right)\\
&\quad =\int_{0}^{\theta}\bE\Big(\big(S_{t-}^{\varepsilon}\big)^{2}\Big)\,dt\int_{(-\varepsilon,\varepsilon)\setminus\{0\}}\big(e^{z}-1\big)^{2}\nu(dz)+\bigg(\int_{0}^{\theta}\bE\big(S_{t-}^{\varepsilon}\big)\,dt\int_{(-\varepsilon,\varepsilon)\setminus\{0\}}\big|e^{z}-1\big|\,\nu(dz)\bigg)^{2}\\
&\quad\leq\int_{0}^{\theta}\bE\Big(b^{2}(T)e^{(2\gamma_{0}-\sigma^{2})t+2\sigma W_{t}+2Z_{t-}^{\varepsilon}}\Big)\,dt\int_{(-\varepsilon,\varepsilon)\setminus\{0\}}\big(e^{z}-1\big)^{2}\nu(dz)\\
&\qquad +\bigg(\int_{0}^{\theta}\bE\Big(b(T)e^{(\gamma_{0}-\sigma^{2}/2)t+\sigma W_{t}+Z_{t-}^{\varepsilon}}\Big)\,dt\int_{(-\varepsilon,\varepsilon)\setminus\{0\}}\!\!\big|e^{z}-1\big|\,\nu(dz)\bigg)^{2}=O(\theta),\quad\theta\rightarrow 0^{+},
\end{align*}
where the $O(\theta)$ term is independent of $a$. For the last probability on the right-hand side of \eqref{eq:EstExpLocalTimeEpstauT1Ini}, with $c_{\theta}:=\ln(K/S_{0})-|\gamma_{0}-\sigma^{2}/2|\theta$ and $z\in(0,\ln(K/b(T)))$, by \eqref{eq:DefZepsPlusMinus} and \eqref{eq:DefOverlineZepsSepsNeps} we have,
\begin{align}
&\bP\big(\tau_{K}^{\varepsilon}<\theta\big)=\bP\Big(\sup_{t\in[0,\theta]}S_{t}^{\varepsilon}\geq K\Big)\leq\bP\Big(\sup_{t\in[0,\theta]}\sigma W_{t}+Z_{\theta}^{\varepsilon,+}\geq c_{\theta}\Big)\\
\label{eq:EstProbtauKeps} &\!\!\!\!\!\leq\bP\bigg(\sup_{t\in[0,\theta]}\!W_{t}\geq\frac{c_{\theta}\!-\!z}{\sigma}\bigg)+\bP\big(Z_{\theta}^{\varepsilon,+}\!\geq z\big)\leq e^{-(c_{\theta}-z)^{2}/(2\sigma^{2}\theta)}+\bP\big(Z_{\theta}^{\varepsilon,+}\!\geq z\big)=o\big(\theta^{n}\big),\quad\theta\rightarrow 0^{+},\qquad
\end{align}
for any $n\in\bN$, where the last inequality follows from the Doob's martingale inequality, and the last equality is due to \cite[Remark 3.1]{FigueroaLopezHoudre:2009}. By combining \eqref{eq:EstExpLocalTimeEpstauT1Ini}, \eqref{eq:EstExpLocalTimeEpstheta}, and \eqref{eq:EstProbtauKeps}, we obtain that
\begin{align}\label{eq:EstExpLocalTimeEpstauT1}
\bE\big(L_{\tau\wedge T_{1}^{\varepsilon}}^{K}\big)=o\big(\theta^{n}\big),\quad\theta\rightarrow 0^{+},
\end{align}
for any $n\in\bN$, where the $o(\theta^{n})$ term is independent of $a$.

\medskip
\noindent
\textbf{Step 2. Estimating $\bE(L_{\tau}^{K}-L_{\tau\wedge T_{1}^{\varepsilon}}^{K})$.}

\medskip
\noindent
Denoting by $\bE_{x}(\cdot):=\bE(\cdot\,|\,S_{0}=x)$. The strong Markov property and time-homogeneity of $S$ imply
\begin{align}
\bE\Big(L_{\tau}^{K}-L_{\tau\wedge T_{1}^{\varepsilon}}^{K}\Big)&=\bE\Big({\bf 1}_{\{T_{1}^{\varepsilon}<\tau\}}\big(L_{\tau}^{K}-L_{\tau\wedge T_{1}^{\varepsilon}}^{K}\big)\Big)=\bE\Big({\bf 1}_{\{T_{1}^{\varepsilon}<\tau\}}\,\bE\Big(\big(L_{\tau}^{K}-L_{\tau\wedge T_{1}^{\varepsilon}}^{K}\big)\,\big|\,\sF_{\tau\wedge T_{1}^{\varepsilon}}\Big)\Big)\nonumber\\
\label{eq:EstExpDiffLocalTime} &=\bE\Big({\bf 1}_{\{T_{1}^{\varepsilon}<\tau\}}\,\bE_{S_{T_{1}^{\varepsilon}}}\big(L_{\tau-\tau\wedge T_{1}^{\varepsilon}}^{K}\big)\Big)\leq\bE\Big({\bf 1}_{\{T_{1}^{\varepsilon}<\theta\}}\,\bE_{S_{T_{1}^{\varepsilon}}}\big(L_{\theta}^{K}\big)\Big).
\end{align}

\medskip
\noindent
\textbf{2.1. Estimating $\bE_{x}(L^K_{\theta})$.}

\medskip
\noindent
By the Meyer-It\^{o} formula, we have
\begin{align*}
\frac{1}{2}\bE_{x}\big(L_{\theta}^{K}\big)=\bE_{x}\big((K-S_{\theta})^{+}\big)-(K-x)^{+}+\bE\bigg(\int_{0}^{\theta}\gamma_{0}{\bf 1}_{\{S_{t-}\leq K\}}S_{t-}\,dt+\int_{0}^{\theta}\int_{\bR_{0}}\Phi(S_{t-},z)\,\nu(dz)\,dt\bigg),
\end{align*}
where $\Phi(y,z):=(K-ye^{z})^{+}-(K-y)^{+}$ satisfies $|\Phi(y,z)|\leq|y(e^{z}-1)|$, and so by \eqref{eq:DefwtX},
\begin{align}\label{eq:ExpxLocalTimeKtheta}
\frac{1}{2}\bE_{x}\big(L_{\theta}^{K}\big)&=\bE_{x}\big((K-S_{\theta})^{+}\big)-(K-x)^{+}+xO(\theta)\\
&=\bE_{x}\bigg(\Big(K-x\,e^{(\gamma_{0}-\sigma^{2}/2)\theta+\sigma W_{\theta}+Z_{\theta}}\Big)^{+}\bigg)-(K-x)^{+}+xO(\theta),\quad\theta\rightarrow 0^{+},
\end{align}
where the $O(\theta)$ term is independent of $x$. By the independence of $W$ and $Z:=(Z_{t})_{t\in\bR_{+}}$, we have
\begin{align*}
\bE_{x}\Big(\Big|e^{(\gamma_{0}-\sigma^{2}/2)\theta+\sigma W_{\theta}+Z_{\theta}}-e^{\sigma W_{\theta}}\Big|\Big)=e^{\sigma^{2}\theta/2}\,\bE_{x}\Big(\Big|e^{(\gamma_{0}-\sigma^{2}/2)\theta+Z_{\theta}}-1\Big|\Big)=O(\theta),\quad\theta\rightarrow 0^{+}.
\end{align*}
Hence, using \cite[Lemma 3.3]{BouselmiLamberton:2016}, we deduce that, as $\theta\rightarrow 0^{+}$,
\begin{align*}
\frac{1}{2}\bE_{x}\big(L_{\theta}^{K}\big)&=\bE_{x}\big(\big(K\!-\!x\,e^{\sigma W_{\theta}}\big)^{+}\big)-(K\!-\!x)^{+}\!+xO(\theta)=\bE_{x}\big((K\!-\!x(1\!+\!\sigma W_{\theta}))^{+}\big)-(K\!-\!x)^{+}\!+xO(\theta)\\
&=x\sigma\!\left(\bE_{x}\!\left(\!\bigg(\frac{K\!-\!x}{x\sigma}\!-\!W_{\theta}\bigg)^{+}\!\right)\!-\!\bigg(\frac{K\!-\!x}{x\sigma}\bigg)^{+}\right)\!+xO(\theta)\leq\frac{x\sigma\sqrt{\theta}}{\sqrt{2\pi}}\exp\bigg(\!\!-\!\frac{(K\!-\!x)^{2}}{2x^{2}\sigma^{2}\theta}\bigg)\!+\!xO(\theta),
\end{align*}
where the $O(\theta)$ term is independent of $x$.

Next, plugging back intp \eqref{eq:EstExpDiffLocalTime}, we obtain that
\begin{align*}
\frac{1}{2}\bE\Big(L_{\tau}^{K}-L_{\tau\wedge T_{1}^{\varepsilon}}^{K}\Big)&\leq\frac{\sigma\sqrt{\theta}}{\sqrt{2\pi}}\bE\left({\bf 1}_{\{T_{1}^{\varepsilon}<\theta\}}S_{T_{1}^{\varepsilon}}\exp\bigg(\!-\frac{\big(K-S_{T_{1}^{\varepsilon}}\big)^{2}}{2\sigma^{2}\theta S_{T_{1}^{\varepsilon}}}\bigg)\right)+\bE\big({\bf 1}_{\{T_{1}^{\varepsilon}<\theta\}}S_{T_{1}^{\varepsilon}}\big)O(\theta),
\end{align*}
as $\theta\rightarrow 0^{+}$, where the $O(\theta)$ term is independent of $a$. Note that, conditionally on $\{T_{1}^{\varepsilon}<\theta\}$, $T_{1}^{\varepsilon}$ is uniformly distributed on $[0,\theta]$. Since $T_{1}^{\varepsilon}$, $W$, $Z^{\varepsilon}:=(Z_{t}^{\varepsilon})_{t\in\bR_{+}}$, and $\overline{Z}^{\varepsilon}:=(\overline{Z}_{t}^{\varepsilon})_{t\in\bR_{+}}$ are independent,
\begin{align*}
S_{T_{1}^{\varepsilon}}\big|\,\{T_{1}^{\varepsilon}<\theta\}\,\ed\,b(T)\,e^{a\sqrt{\theta}+(\gamma_{0}-\sigma^{2}/2)\theta U+W_{\theta U}+Z_{\theta U}^{\varepsilon}+V_{\varepsilon}}=:\Gamma_{\theta},
\end{align*}
where $U\sim\cU[0,1]$, $V_{\varepsilon}\ed\overline{Z}_{T_{1}^{\varepsilon}}^{\varepsilon}$, and $U$, $V_{\varepsilon}$, $W$, and $Z^{\varepsilon}$ are independent. Therefore, we conclude that
\begin{align}\label{eq:EstExpDiffLocalTimeCond}
&\bE\Big(L_{\tau}^{K}-L_{\tau\wedge T_{1}^{\varepsilon}}^{K}\Big)\leq\left(\frac{\sigma\sqrt{2\theta}}{\sqrt{\pi}}\bE\left(\Gamma_{\theta}\exp\bigg(\!-\frac{\big(K-\Gamma_{\theta}\big)^{2}}{2\sigma^{2}\theta \Gamma_{\theta}}\bigg)\right)+\bE\big(\Gamma_{\theta}\big)O(\theta)\right)\bP\big(T_{1}^{\varepsilon}<\theta\big)\\
\label{eq:UpperBoundExpDiffLocalTimetauT1eps} &\quad\leq\bigg(\frac{\sqrt{2}\sigma}{\sqrt{\pi}}+O(\sqrt{\theta})\bigg)\bE\Big(b(T)\,e^{(\gamma_{0}-\sigma^{2}/2)\theta U+W_{\theta U}+Z_{\theta U}^{\varepsilon}+V_{\varepsilon}}\Big)\nu\big([-\varepsilon,\varepsilon]^{c}\big)\theta^{3/2},\quad\theta\rightarrow 0^{+},
\end{align}
where the $O(\sqrt{\theta})$ term is independent of $a$. Combining \eqref{eq:DecompLocalTimetauK}, \eqref{eq:EstExpLocalTimeEpstauT1}, and \eqref{eq:UpperBoundExpDiffLocalTimetauT1eps} leads to the inequality in \eqref{eq:SmallTimeEstExpLocalTime}.

\medskip
\noindent
\textbf{2.2. Estimating $\bE(L_{\tau}^{K}-L_{\tau\wedge T_{1}^{\varepsilon}}^{K})$ and proving the equality in \eqref{eq:SmallTimeEstExpLocalTime} when $\nu(\{\ln(K/b(T))\})=0$.}

\medskip
\noindent
As $\theta\rightarrow 0^{+}$, we have $\Gamma_{\theta}\rightarrow b(T)e^{V_{\varepsilon}}$. When $\nu(\{\ln(K/b(T))\})=0$, $b(T)e^{V_{\varepsilon}}\neq K$ $\bP$-a.$\,$s. Therefore, we deduce from the dominated convergence that
\begin{align*}
\lim_{\theta\rightarrow 0^{+}}\bE\left(\Gamma_{\theta}\exp\bigg(\!-\frac{\big(K-\Gamma_{\theta}\big)^{2}}{2\sigma^{2}\theta \Gamma_{\theta}}\bigg)\right)=0,
\end{align*}
By \eqref{eq:EstExpDiffLocalTimeCond}, this implies that
\begin{align}\label{eq:EstExpDiffLocalTimetauKT1epsZeronu}
\bE\Big(L_{\tau}^{K}-L_{\tau\wedge T_{1}^{\varepsilon}}^{K}\Big)=o\big(\theta^{3/2}\big),\quad\theta\rightarrow 0^{+}.
\end{align}
Together with \eqref{eq:DecompLocalTimetauK} and \eqref{eq:EstExpLocalTimeEpstauT1}, we obtain that
\begin{align*}
\bE\big(L_{\tau}^{K}\big)=o\big(\theta^{3/2}\big),\quad\theta\rightarrow 0^{+}.
\end{align*}

\smallskip
\noindent
\textbf{2.3. Estimating $\bE(L_{\tau}^{K}-L_{\tau\wedge T_{1}^{\varepsilon}}^{K})$ and proving the equality in \eqref{eq:SmallTimeEstExpLocalTime} when $\nu(\{\ln(K/b(T))\})\neq 0$.}

\medskip
\noindent
Now assume $\nu(\{\ln(K/b(T))\})\neq 0$. Introduce the processes $\wh{X}:=(\wh{X}_{t})_{t\in\bR_{+}}$ and $\wh{Z}:=(\wh{Z}_{t})_{t\in\bR_{+}}$ by
\begin{align}\label{eq:DefWhZX}
\wh{Z}_{t}:=\int_{0}^{t}\int_{\{\ln(K/b(T))\}}z\,N(ds,dz),\quad\wh{X}_{t}:=\wt{X}_{t}-\wh{Z}_{t},\quad t\in\bR_{+},
\end{align}
and let $\wh{T}_{1}:=\inf\{t\in\bR_{+}:\Delta\wh{Z}_{t}\neq 0\}$. Since $T_{1}^{\varepsilon}\leq\wh{T}_{1}$ (recalling $\varepsilon<\ln(K/b(T))$), we have
\begin{align}
&\bE\Big(L_{\tau}^{K}-L_{\tau\wedge T_{1}^{\varepsilon}}^{K}\Big)=\bE\Big(L_{\tau}^{K}-L_{\tau\wedge\wh{T}_{1}}^{K}\Big)+\bE\Big(L_{\tau\wedge\wh{T}_{1}}^{K}-L_{\tau\wedge T_{1}^{\varepsilon}}^{K}\Big)\\
\label{eq:DecompExpDiffLocalTimetauT1epsPostnu} &\quad =\bE\Big(L_{\tau}^{K}-L_{\tau\wedge\wh{T}_{1}}^{K}\Big)+\bE\Big({\bf 1}_{\{T_{1}^{\varepsilon}<\tau<\wh{T}_{1}\}}\big(L_{\tau}^{K}-L_{\tau\wedge T_{1}^{\varepsilon}}^{K}\big)\Big)+\bE\Big({\bf 1}_{\{T_{1}^{\varepsilon}<\wh{T}_{1}\leq\tau\}}\big(L_{\tau\wedge\wh{T}_{1}}^{K}-L_{\tau\wedge T_{1}^{\varepsilon}}^{K}\big)\Big).\qquad
\end{align}
Note that prior to $\wh{T}_{1}$, the process $\wt{X}$ matches with the process $\wh{X}$ whose L\'{e}vy measure does not charge the point $\ln(K/b(T))$. It follows from \eqref{eq:EstExpDiffLocalTimetauKT1epsZeronu} that
\begin{align}\label{eq:EstExpDiffLocalTimetauT1eps1}
\bE\Big({\bf 1}_{\{T_{1}^{\varepsilon}<\tau<\wh{T}_{1}\}}\big(L_{\tau}^{K}-L_{\tau\wedge T_{1}^{\varepsilon}}^{K}\big)\Big)=o\big(\theta^{3/2}\big),\quad\theta\rightarrow 0^{+}.
\end{align}
Also, on the event $\{T_{1}^{\varepsilon}<\wh{T}_{1}\leq\tau\}\subset\{T_{1}^{\varepsilon}<\wh{T}_{1}\leq\theta\}$ (since $\tau\leq\theta$), the process $\overline{Z}^{\varepsilon}$ jumps at least two times before $\theta$, and so
\begin{align*}
\bP\big(T_{1}^{\varepsilon}<\wh{T}_{1}\leq\theta\big)\leq\bP\bigg(\sum_{t\in[0,\theta]}{\bf 1}_{\{\Delta\overline{Z}^{\varepsilon}_{t}\neq 0\}}\geq 2\bigg)=O\big(\theta^{2}\big),\quad\theta\rightarrow 0^{+}.
\end{align*}
Hence, by the strong Markov property and time-homogeneity of $S$, we deduce that
\begin{align}
\bE\Big({\bf 1}_{\{T_{1}^{\varepsilon}<\wh{T}_{1}\leq\tau\}}\big(L_{\tau\wedge\wh{T}_{1}}^{K}-L_{\tau\wedge T_{1}^{\varepsilon}}^{K}\big)\Big)&\leq\bE\Big({\bf 1}_{\{T_{1}^{\varepsilon}<\wh{T}_{1}\leq\theta\}}\bE_{S_{\wh{T}_{1}}}\!\big(L_{\theta}^{K}\big)\Big)\\
\label{eq:EstExpDiffLocalTimetauT1eps2} &\leq\bE\Big({\bf 1}_{\{T_{1}^{\varepsilon}<\wh{T}_{1}\leq\theta\}}\big(2K+S_{\wh{T}_{1}}O(\theta)\big)\Big)=O\big(\theta^{2}\big),\quad\theta\rightarrow 0^{+},\quad
\end{align}
where we used \eqref{eq:ExpxLocalTimeKtheta} in the second inequality. Combining \eqref{eq:DecompExpDiffLocalTimetauT1epsPostnu}, \eqref{eq:EstExpDiffLocalTimetauT1eps1}, and \eqref{eq:EstExpDiffLocalTimetauT1eps2} leads to
\begin{align*}
\bE\Big(L_{\tau}^{K}-L_{\tau\wedge T_{1}^{\varepsilon}}^{K}\Big)=\bE\Big(L_{\tau}^{K}-L_{\tau\wedge\wh{T}_{1}}^{K}\Big)+o\big(\theta^{3/2}\big),\quad\theta\rightarrow 0^{+}.
\end{align*}

Next, by the strong Markov property and the time-homogeneity of $S$ as well as \eqref{eq:ExpxLocalTimeKtheta}, we have
\begin{align*}
&\frac{1}{2}\,\bE\Big(L_{\tau}^{K}-L_{\tau\wedge\wh{T}_{1}}^{K}\Big)=\frac{1}{2}\,\bE\Big({\bf 1}_{\{\wh{T}_{1}<\tau\}}\bE_{S_{\wh{T}_{1}}}\!\big(L_{\tau-\tau\wedge\wh{T}_{1}}^{K}\big)\Big)\\
&\quad =\bE\bigg({\bf 1}_{\{\wh{T}_{1}<\tau\}}\Big(\bE_{S_{\wh{T}_{1}}}\!\Big(\big(K-S_{\tau-\tau\wedge\wh{T}_{1}}\big)^{+}\Big)-\big(K-S_{\wh{T}_{1}}\big)^{+}+O\big(\theta^{2}\big)\Big)\bigg)\\
&\quad =\bE\Big({\bf 1}_{\{\wh{T}_{1}<\tau\}}\Big(\big(K-S_{\tau}\big)^{+}-\big(K-S_{\wh{T}_{1}}\big)^{+}\Big)\Big)+O\big(\theta^{2}\big)\\
&\quad =\bE\left({\bf 1}_{\{\wh{T}_{1}<\tau\}}\left(\bigg(K-\frac{KS_{0}}{b(T)}e^{\wt{X}_{\tau}+\wh{X}_{\wh{T}_{1}}-\wt{X}_{\wh{T}_{1}}}\bigg)^{+}-\bigg(K-\frac{KS_{0}\,e^{\wh{X}_{\wh{T}_{1}}}}{b(T)}\bigg)^{+}\right)\right)+O\big(\theta^{2}\big),\quad\theta\rightarrow 0^{+},
\end{align*}
where the last equality follows from the fact that $\wt{X}_{\wh{T}_{1}}-\wh{X}_{\wh{T}_{1}}=\ln(K/b(T))$. Since $\bP(\overline{N}_{\theta}^{\varepsilon}\geq 2)=O(\theta^{2})$, we deduce that, as $\theta\rightarrow 0^{+}$,
\begin{align}
\bE\Big(L_{\tau}^{K}\!-\!L_{\tau\wedge\wh{T}_{1}}^{K}\Big)&=2K\,\bE\!\left(\!{\bf 1}_{\{\overline{N}_{\theta}^{\varepsilon}=1\}}{\bf 1}_{\{\wh{T}_{1}<\tau\}}\!\left(\!\bigg(1-\frac{S_{0}\,e^{\wt{X}_{\tau}+\wh{X}_{\wh{T}_{1}}-\wt{X}_{\wh{T}_{1}}}}{b(T)}\bigg)^{+}\!\!-\bigg(1-\frac{S_{0}\,e^{\wh{X}_{\wh{T}_{1}}}}{b(T)}\bigg)^{+}\right)\!\right)\!+O\big(\theta^{2}\big)\nonumber\\
\label{eq:EstExpDiffLocalTimetauwhT1Ini} &=2K\,\bE\left({\bf 1}_{\{\overline{N}_{\theta}^{\varepsilon}=1\}}{\bf 1}_{\{\wh{T}_{1}<\tau\}}\left(\bigg(1-\frac{S_{0}\,e^{\wh{X}_{\tau}}}{b(T)}\bigg)^{+}\!-\bigg(1-\frac{S_{0}\,e^{\wh{X}_{\wh{T}_{1}}}}{b(T)}\bigg)^{+}\right)\right)+O\big(\theta^{2}\big).
\end{align}
To further estimate the first expectation in \eqref{eq:EstExpDiffLocalTimetauwhT1Ini}, we first have
\begin{align*}
&\bE\left({\bf 1}_{\{\overline{N}_{\theta}^{\varepsilon}=1\}}{\bf 1}_{\{\wh{T}_{1}<\tau\}}\bigg(1-\frac{S_{0}\,e^{\wh{X}_{\tau}}}{b(T)}\bigg)^{+}\right)=\bE\Big({\bf 1}_{\{\overline{N}_{\theta}^{\varepsilon}=1\}}{\bf 1}_{\{\wh{T}_{1}<\tau\}}{\bf 1}_{\{a\sqrt{\theta}+\wh{X}_{\tau}\leq 0\}}\big(1-e^{a\sqrt{\theta}+\wh{X}_{\tau}}\big)\Big)\\
&=\bE\Big({\bf 1}_{\{\overline{N}_{\theta}^{\varepsilon}=1\}}{\bf 1}_{\{\wh{T}_{1}<\tau\}}{\bf 1}_{\{a\sqrt{\theta}+\wh{X}_{\tau}\leq 0\}}\big(1\!-\!e^{a\sqrt{\theta}+\wh{X}_{\tau}}\!+\!a\sqrt{\theta}\!+\!\wh{X}_{\tau}\big)\Big)+\bE\Big({\bf 1}_{\{\overline{N}_{\theta}^{\varepsilon}=1\}}{\bf 1}_{\{\wh{T}_{1}<\tau\}}\big(\!-\!a\sqrt{\theta}\!-\!\wh{X}_{\tau}\big)^{+}\Big).
\end{align*}
Recalling that $\tau$ takes values in $[0,\theta]$ and $\varepsilon\in(0,\ln(K/b(T)))$, we see that $\overline{Z}_{\tau}^{\varepsilon}=\wh{Z}_{\tau}=\ln(K/b(T))$ and on the event $\{\overline{N}_{\theta}^{\varepsilon}=1,\wh{T}_{1}<\tau\}$. Using the independence between $\overline{N}^{\varepsilon}:=(\overline{N}^{\varepsilon}_{t})_{t\in\bR_{+}}$, $W$, and $Z^{\varepsilon}$ together with \eqref{eq:DefwtX}, \eqref{eq:DefOverlineZepsSepsNeps}, and \eqref{eq:DefWhZX}, we deduce that
\begin{align*}
&\bE\Big({\bf 1}_{\{\overline{N}_{\theta}^{\varepsilon}=1\}}{\bf 1}_{\{\wh{T}_{1}<\tau\}}{\bf 1}_{\{a\sqrt{\theta}+\wh{X}_{\tau}\leq 0\}}\big|1-e^{a\sqrt{\theta}+\wh{X}_{\tau}}+a\sqrt{\theta}+\wh{X}_{\tau}\big|\Big)\leq\bE\Big({\bf 1}_{\{\overline{N}_{\theta}^{\varepsilon}=1\}}{\bf 1}_{\{\wh{T}_{1}<\tau\}}\big(a\sqrt{\theta}+\wh{X}_{\tau}\big)^{2}\Big)\\
&=\bE\Big({\bf 1}_{\{\overline{N}_{\theta}^{\varepsilon}=1\}}{\bf 1}_{\{\wh{T}_{1}<\tau\}}\!\big(a\sqrt{\theta}+\!\wt{X}_{\tau}\!-\!\ln(K/b(T))\big)^{2}\Big)\!=\!\bE\Big({\bf 1}_{\{\overline{N}_{\theta}^{\varepsilon}=1\}}{\bf 1}_{\{\wh{T}_{1}<\tau\}}\!\big(a\sqrt{\theta}\!+\!(r\!-\!\delta)\tau\!+\!\sigma W_{\tau}\!+\!Z_{\tau}^{\varepsilon}\big)^{2}\Big)\\
&\leq 4(r-\delta)\theta^{2}+4\,\bP\big(\overline{N}_{\theta}^{\varepsilon}=1\big)\bigg(a^{2}\theta+\bE\Big(\sup_{t\in[0,\theta]}\sigma^{2}W_{t}^{2}+(Z_{\tau}^{\varepsilon})^{2}\Big)\bigg)=O\big(\theta^{2}\big),\quad\theta\rightarrow 0^{+},
\end{align*}
and that
\begin{align*}
&\bigg|\bE\Big({\bf 1}_{\{\overline{N}_{\theta}^{\varepsilon}=1\}}{\bf 1}_{\{\wh{T}_{1}<\tau\}}\big(\!-a\sqrt{\theta}-\wh{X}_{\tau}\big)^{+}\Big)-\bE\Big({\bf 1}_{\{\wh{T}_{1}<\tau\}}\big(\!-a\sqrt{\theta}-\sigma W_{\tau}\big)^{+}\Big)\bigg|\\
&\quad\leq\bE\Big({\bf 1}_{\{\overline{N}_{\theta}^{\varepsilon}=1\}}{\bf 1}_{\{\wh{T}_{1}<\tau\}}\big|\big(\!-a\sqrt{\theta}-\wh{X}_{\tau}\big)^{+}-\big(\!-a\sqrt{\theta}-\sigma W_{\tau}\big)^{+}\big|\Big)+\bE\Big({\bf 1}_{\{\overline{N}_{\theta}^{\varepsilon}\geq 2\}}\big(\!-a\sqrt{\theta}-\sigma W_{\tau}\big)^{+}\Big)\\
&\quad\leq\bE\left({\bf 1}_{\{\overline{N}_{\theta}^{\varepsilon}=1\}}\bigg|Z_{\tau}^{\varepsilon}+\bigg(\gamma_{0}-\frac{\sigma^{2}}{2}\bigg)\tau\bigg|\right)+\bE\Big({\bf 1}_{\{\overline{N}_{\theta}^{\varepsilon}\geq 2\}}\big(\!-a\sqrt{\theta}-\sigma W_{\tau}\big)^{+}\Big)\\
&\quad\leq\bP\big(\overline{N}_{\theta}^{\varepsilon}=1\big)\left(\bE\big(Z_{\tau}^{\varepsilon}\big)+\bigg|\gamma_{0}-\frac{\sigma^{2}}{2}\bigg|\theta\right)+\bP\big(\overline{N}_{\theta}^{\varepsilon}\geq 2\big)\bE\big(|a\sqrt{\theta}+\sigma W_{\tau}|\big)=O\big(\theta^{2}\big),\quad\theta\rightarrow 0^{+}.
\end{align*}
Hence, we obtain that
\begin{align}\label{eq:EstExpDiffLocalTimetauwhT11}
\bE\left(\!{\bf 1}_{\{\overline{N}_{\theta}^{\varepsilon}=1\}}{\bf 1}_{\{\wh{T}_{1}<\tau\}}\bigg(1\!-\!\frac{S_{0}\,e^{\wh{X}_{\tau}}}{b(T)}\bigg)^{+}\right)=\bE\Big({\bf 1}_{\{\wh{T}_{1}<\tau\}}\!\big(\!-\!a\sqrt{\theta}\!-\!\sigma W_{\tau}\big)^{+}\Big)+O\big(\theta^{2}\big),\quad\theta\rightarrow 0^{+}.\quad
\end{align}
Similar arguments lead to
\begin{align}\label{eq:EstExpDiffLocalTimetauwhT12}
\bE\left(\!{\bf 1}_{\{\overline{N}_{\theta}^{\varepsilon}=1\}}{\bf 1}_{\{\wh{T}_{1}<\tau\}}\bigg(1\!-\!\frac{S_{0}\,e^{\wh{X}_{\wh{T}_{1}}}}{b(T)}\bigg)^{+}\right)=\bE\Big({\bf 1}_{\{\wh{T}_{1}<\tau\}}\!\big(\!-\!a\sqrt{\theta}\!-\!\sigma W_{\wh{T}_{1}}\big)^{+}\Big)+O\big(\theta^{2}\big),\quad\theta\rightarrow 0^{+}.\qquad
\end{align}
By Combining \eqref{eq:EstExpDiffLocalTimetauwhT1Ini}, \eqref{eq:EstExpDiffLocalTimetauwhT11}, and \eqref{eq:EstExpDiffLocalTimetauwhT12}, we obtain that
\begin{align}\label{eq:EstExpDiffLocalTimetauwhT1}
\!\!\!\!\bE\Big(\!L_{\tau}^{K}\!-\!L_{\tau\wedge\wh{T}_{1}}^{K}\!\Big)\!=\!2K\bE\Big(\!{\bf 1}_{\{\wh{T}_{1}<\tau\}}\!\Big(\big(\!-\!a\sqrt{\theta}\!-\!\sigma W_{\tau}\big)^{+}\!\!-\!\big(\!-\!a\sqrt{\theta}\!-\!\sigma W_{\wh{T}_{1}}\big)^{+}\Big)\!\Big)\!+\!O\big(\theta^{2}\big),\quad\theta\rightarrow 0^{+}.\quad
\end{align}
Finally, by combining \eqref{eq:DecompLocalTimetauK}, \eqref{eq:EstExpLocalTimeEpstauT1}, and \eqref{eq:EstExpDiffLocalTimetauwhT1}, we deduce the inequality in \eqref{eq:SmallTimeEstExpLocalTime}.\hfill $\Box$

\subsection*{Proof of Lemma \ref{lem:AsymLowerBound2ndDerivAmerPutPrice}} The variational inequality \eqref{eq:VarIneq} gives
\begin{align*}
\frac{\sigma^{2}}{2}\bigg(\frac{\partial^{2}\wt{P}}{\partial x^{2}}-\frac{\partial\wt{P}}{\partial x}\bigg)&\geq r\wt{P}-(r-\delta)\frac{\partial\wt{P}}{\partial x}-\int_{\bR_{0}}\!\bigg(\wt{P}(\cdot\,,\cdot+z)-\wt{P}(\cdot\,,\cdot)-\frac{\partial\wt{P}}{\partial x}(\cdot\,,\cdot)\big(e^{z}-1\big)\bigg)\nu(dz)\\
&=r\wt{P}-(r-\delta)\frac{\partial_{+}\wt{P}}{\partial x}-\int_{\bR_{0}}\!\bigg(\wt{P}(\cdot\,,\cdot+z)-\wt{P}(\cdot\,,\cdot)-\frac{\partial_{+}\wt{P}}{\partial x}(\cdot\,,\cdot)\big(e^{z}-1\big)\bigg)\nu(dz)=:\tilde{g}
\end{align*}
on $\wt{\cC}=\{(t,x)\in(0,T)\times\bR:\wt{P}(t,x)>(K-e^{x})^{+}\}=\{(t,x)\in(0,T)\times\bR:b(t)<e^{x}\}$ in the sense of distribution. Taking any $\wt{\varphi}\in C_{c}^{\infty}(\wt{\cC})$ and $\psi\in C_{c}^{\infty}((0,T))$ and since $\wt{\varphi}\psi\in C_{c}^{\infty}(\wt{\cC})$, we have
\begin{align*}
\frac{\sigma^{2}}{2}\int_{\wt{\cC}}\wt{P}(t,x)\bigg(\frac{\partial^{2}}{\partial x^{2}}+\frac{\partial}{\partial x}\bigg)\wt{\varphi}(t,x)\psi(t)\,dx\,dt\geq\int_{\wt{\cC}}\tilde{g}(t,x)\wt{\varphi}(t,x)\psi(t)\,dx\,dt.
\end{align*}
Since $\psi$ is arbitrary, for any $t\in(0,T)$ with $\wt{\cC}_{t}:=\{x\in\bR:(t,x)\in\wt{C}\}$, we have
\begin{align*}
\frac{\sigma^{2}}{2}\int_{\wt{\cC}_{t}}\wt{P}(t,x)\bigg(\frac{\partial^{2}}{\partial x^{2}}+\frac{\partial}{\partial x}\bigg)\wt{\varphi}(t,x)\,dx\geq\int_{\wt{\cC}_{t}}\tilde{g}(t,x)\wt{\varphi}(t,x)\,dx.
\end{align*}
Define $g$ and $\varphi$ on $\cC=\{(t,s)\in(0,T)\times\bR_{+}:b(t)<s\}$ respectively via $g(t,s)=\tilde{g}(t,\ln s)$ and $\varphi(t,s)=\wt{\varphi}(t,\ln s)$. Then by using change of variable $s=e^{x}$, we deduce that
\begin{align*}
&\frac{\sigma^{2}}{2}\int_{\wt{\cC}_{t}}\wt{P}(t,x)\bigg(\frac{\partial^{2}}{\partial x^{2}}+\frac{\partial}{\partial x}\bigg)\wt{\varphi}(t,x)\,dx=\frac{\sigma^{2}}{2}\int_{\wt{\cC}_{t}}P(t,e^{x})\bigg(\frac{\partial^{2}}{\partial x^{2}}+\frac{\partial}{\partial x}\bigg)\varphi(t,e^{x})\,dx\\
&\quad =\frac{\sigma^{2}}{2}\!\int_{\wt{\cC}_{t}}P(t,e^{x})\bigg(e^{2x}\frac{\partial^{2}\varphi}{\partial s^{2}}(t,e^{x})+2e^{x}\frac{\partial\varphi}{\partial s}(t,e^{x})\bigg)\,dx=\frac{\sigma^{2}}{2}\!\int_{\cC_{t}}P(t,s)\bigg(s\frac{\partial^{2}\varphi}{\partial s^{2}}(t,s)+2\frac{\partial\varphi}{\partial s}(t,s)\bigg)\,ds\\
&\quad =\frac{\sigma^{2}}{2}\int_{\cC_{t}}P(t,s)\frac{\partial^{2}}{\partial s^{2}}\big(s\varphi(t,s)\big)\,ds=\frac{\sigma^{2}}{2}\int_{\cC_{t}}s\varphi(t,s)\frac{\partial^{2}P}{\partial s^{2}}(t,ds),
\end{align*}
where $\cC_{t}:=\{s\in\bR_{+}:(t,s)\in\cC\}$, and thus
\begin{align*}
\frac{\sigma^{2}}{2}\int_{\cC_{t}}s\varphi(t,s)\frac{\partial^{2}P}{\partial s^{2}}(t,ds)\geq\int_{\cC_{t}}\frac{1}{s}g(t,s)\varphi(t,s)\,ds.
\end{align*}

Now for any fixed $t\in(0,T)$ and $s\in(b(t),b(T))$, we can choose a nonnegative sequence $(\varphi_{n})_{n\geq 1}\subset C_{c}^{\infty}(\cC)$ such that $\varphi_{n}(t,u)\uparrow {\bf 1}_{[b(t),s]}(u)(u-b(t))$ for all $u\in\cC_{t}$. It follows that
\begin{align*}
\frac{\sigma^{2}}{2}\int_{b(t)}^{s}u(u-b(t))\frac{\partial^{2}P}{\partial s^{2}}(t,du)\geq\int_{b(t)}^{s}\frac{u-b(t)}{u}g(t,u)\,du,
\end{align*}
which implies that
\begin{align}\label{eq:LowerBoundInt2ndDeriP}
\int_{b(t)}^{s}(u-b(t))\frac{\partial^{2}P}{\partial s^{2}}(t,du)\geq\frac{2}{\sigma^{2}b^{2}(T)}\int_{b(t)}^{s}(u-b(t))g(t,u)\,du.
\end{align}

To estimate the function $g$ from below, we need to establish the following technical lemma. Denote the early exercise premium by $e(T-t,s):=P(t,s)-P_{e}(t,s)$, and set $\theta=T-t$ as usual.

\begin{lemma}\label{lem:AsymDerivAmerPutPrice}
Under the model assumptions, for any $s\in (0,b(T))$, we have
\begin{align*}
\text{(a)}\,\,\,\left|\frac{\partial_{+}e}{\partial s}(\theta,s)\right|=o(\sqrt{\theta}),\qquad\text{(b)}\,\,\,\frac{\partial_{+}P}{\partial s}(t,s)+1=o(\sqrt{\theta}),
\end{align*}
as $\theta=T-t\rightarrow 0^{+}$, with $o(\sqrt{\theta})$ uniform with respect to $s$.
\end{lemma}

\noindent
\textbf{Proof.} Clearly, for any $s\in(0,b(t))$ we have $(\partial_{+}P/\partial s)(t,s)+1=0$, so it suffices to consider $s\in(b(t),b(T))$. In view of \cite[Corollary 3.1]{LambertonMikou:2013}, the function $s\mapsto P(t,s)-P_{e}(t,s)$ is nonincreasing on $\bR_{+}$. Moreover, the convexity of $P(t,\cdot)$ ensures that the function $s\mapsto P(t,s)-(K-s)$ is nondecreasing. It follows that
\begin{align}\label{eq:1stDerivOrder}
0\leq\frac{\partial_{+}P}{\partial s}(t,s)+1\leq\frac{\partial_{+}P_{e}}{\partial s}(t,s)+1.
\end{align}
Let $\wt{Z}_{t}:=\wt{X}_{t}-\sigma W_{t}$, $t\in\bR_{+}$. For any $s\in(b(t),b(T))$, noting that $b(T)<K$ when $d<0$, we have
\begin{align*}
1+\frac{\partial_{+}P_{e}}{\partial s}(t,s)&=1-e^{-r\theta}\,\bE\Big(e^{\wt{X}_{\theta}}{\bf 1}_{\{\wt{X}_{\theta}<\ln(K/s)\}}\Big)=1-\bE\Big(e^{\sigma W_{\theta}}{\bf 1}_{\{\wt{X}_{\theta}<\ln(K/s)\}}\Big)+o(\sqrt{\theta})\\
&\leq\bE\Big(\big(1-e^{\sigma W_{\theta}}\big){\bf 1}_{\{\wt{X}_{\theta}<\ln(K/s)\}}\Big)+\bP\big(\wt{X}_{\theta}\geq\ln(K/b(T))\big)+o(\sqrt{\theta})\\
&=\bE\big(1-e^{\sigma W_{\theta}}\big)-\bE\Big(\big(1-e^{\sigma W_{\theta}}\big){\bf 1}_{\{\wt{X}_{\theta}\geq\ln(K/s)\}}\Big)+o(\sqrt{\theta})\\
&\leq\bE\Big(\big|1-e^{\sigma W_{\theta}}\big|{\bf 1}_{\{\sigma W_{\theta}\geq\ln(K/b(T))/2\}}{\bf 1}_{\{\wt{Z}_{\theta}\geq\ln(K/b(T))/2\}}\Big)+o(\sqrt{\theta}),\quad\theta\rightarrow 0^{+}.
\end{align*}
Using arguments similar to those leading to \eqref{eq:EstDiffExphwtXhW3} together with the independence between $(W_{t})_{t\in\bR_{+}}$ and $(\wt{Z}_{t})_{t\in\bR_{+}}$, we deduce that, as $\theta\rightarrow 0^{+}$,
\begin{align*}
1+\frac{\partial_{+}P_{e}}{\partial s}(t,s)\leq\bE\Big(\big|1-e^{\sigma W_{\theta}}\big|{\bf 1}_{\{\sigma W_{\theta}\geq\ln(K/b(T))/2\}}\Big)\bP\bigg(\wt{Z}_{\theta}\geq\frac{\ln(K/b(T))}{2}\bigg)+o(\sqrt{\theta})=o(\sqrt{\theta}),
\end{align*}
which, together with \eqref{eq:1stDerivOrder} and the definition of early exercise premium, completes the proof.\hfill $\Box$

\medskip
Coming back to the proof of Lemma \ref{lem:AsymLowerBound2ndDerivAmerPutPrice}, for any fixed $t\in(0,T)$ and $s\in(b(t),b_{e}(t)\wedge b(T))$, we will estimate $g(t,u)$ from below for $u\in[b(t),s]$. To begin with, we first have
\begin{align*}
g(t,u)&=rP(t,u)-(r-\delta)u\frac{\partial_{+}P}{\partial s}(t,u)-\int_{\bR_{0}}\bigg(P(t,ue^{z})-P(t,u)-u\big(e^{z}-1\big)\frac{\partial_{+}P}{\partial s}(t,u)\bigg)\nu(dz)\\
&\geq r(K-u)+(r-\delta)u-\int_{\bR_{0}}\left(P(t,ue^{z})-P(t,u)+u\big(e^{z}-1\big)\right)\nu(dz)\\
&\quad -u\bigg(\frac{\partial_{+}P}{\partial s}(t,u)+1\bigg)\bigg((r-\delta)-\int_{\bR_{0}}\big(e^{z}-1\big)\nu(dz)\bigg),
\end{align*}
where both integrals on the right-hand side of the second inequality are finite due to \eqref{eq:FinVarLevy}. Thanks to \eqref{eq:FinVarLevy} and Lemma \ref{lem:AsymDerivAmerPutPrice}-(a), we see that, as $\theta=T-t\rightarrow 0^{+}$,
\begin{align}\label{eq:IntDiffPPe}
\int_{\bR_{0}}\!\big(P(t,ue^{z})\!-\!P(t,u)\!-\!P_{e}(t,ue^{z})\!+\!P(t,u)\big)\nu(dz)=\!\int_{\bR_{0}}\!\big(e(\theta,ue^{z})\!-\!e(\theta,u)\big)\nu(dz)=o(\sqrt{\theta}).\qquad
\end{align}
Together with Lemma \ref{lem:AsymDerivAmerPutPrice}-(b), we obtain that
\begin{align}\label{eq:LowerBound1gtu}
g(t,u)\geq rK-\delta s-\int_{\bR_{0}}\left(P_{e}(t,ue^{z})-P_{e}(t,u)+u\big(e^{z}-1\big)\right)\nu(dz)+o(\sqrt{\theta}),\quad\theta\rightarrow 0^{+}.
\end{align}
In view of \eqref{eq:EuroPutPrice} and the martingale property of $(e^{X_{t}})_{t\in\bR_{+}}=(e^{-(r-\delta)t+\wt{X}_{t}})_{t\in\bR_{+}}$,
\begin{align}
&\int_{\bR_{0}}\left(P_{e}(t,ue^{z})-P_{e}(t,u)+u\big(e^{z}-1\big)\right)\nu(dz)\\
&\quad =e^{-r\theta}\int_{\bR_{0}}\bE\Big(\big(K-ue^{z}e^{\wt{X}_{\theta}}\big)^{+}-\big(K-ue^{\wt{X}_{\theta}}\big)^{+}+ue^{\delta\theta}\big(e^{z}-1\big)e^{\wt{X}_{\theta}}\Big)\nu(dz)\\
\label{eq:IntPe} &\quad =\int_{\bR_{0}}\bE\Big(\big(K-ue^{z}e^{\wt{X}_{\theta}}\big)^{+}-\big(K-ue^{\wt{X}_{\theta}}\big)^{+}+u\big(e^{z}-1\big)e^{\wt{X}_{\theta}}\Big)\nu(dz)+O(\theta),\quad\theta\rightarrow 0^{+},
\end{align}
we thus deduce that, for $u\in [b(t),s]$ where $s\in(b(t),b_{e}(t)\wedge b(T))$ and $t\in (0,T)$,
\begin{align}\label{eq:LowerBound2gtu}
g(t,u)\!\geq\!rK\!-\!\delta s\!-\!\!\int_{\bR_{0}}\!\!\bE\Big(\!\big(K\!-\!ue^{z}e^{\wt{X}_{\theta}}\big)^{+}\!\!\!-\!\big(K\!-\!ue^{\wt{X}_{\theta}}\big)^{+}\!\!\!+\!u\big(e^{z}\!-\!1\big)e^{\wt{X}_{\theta}}\Big)\nu(dz)\!+\!o(\sqrt{\theta}),\,\,\,\,\theta\rightarrow 0^{+}.\qquad
\end{align}

We will estimate the integral term above over various subsets of $\bR_{0}$. For this purpose, we set
\begin{align*}
I(A):=\int_{A}\bE\Big(\big(K-ue^{z}e^{\wt{X}_{\theta}}\big)^{+}-\big(K-ue^{\wt{X}_{\theta}}\big)^{+}+u\big(e^{z}-1\big)e^{\wt{X}_{\theta}}\Big)\nu(dz),\quad A\in\cB(\bR_{0}).
\end{align*}
First, for $A_{1}=(-\infty,0)$, we have
\begin{align}\label{eq:AsymIA1}
I(A_{1})=\int_{A_{1}}\bE\Big(\big(K-ue^{\wt{X}_{\theta}}\big){\bf 1}_{\{z+\wt{X}_{\theta}<\ln(K/u)\leq\wt{X}_{\theta}\}}+u\big(e^{z}-1\big)e^{\wt{X}_{\theta}}{\bf 1}_{\{\wt{X}_{\theta}\geq\ln(K/u)-z\}}\Big)\nu(dz)\leq 0.\qquad
\end{align}
Next, for $A_{2}=(0,\ln(K/b(T))/2)$, we have
\begin{align*}
I(A_{2})&=\int_{A_{2}}\bE\Big(\big(ue^{z}e^{\wt{X}_{\theta}}-K\big){\bf 1}_{\{\wt{X}_{\theta}<\ln(K/u)\leq\wt{X}_{\theta}+z\}}+u\big(e^{z}-1\big)e^{\wt{X}_{\theta}}{\bf 1}_{\{\wt{X}_{\theta}\geq\ln(K/u)\}}\Big)\nu(dz)\\
&\leq\int_{A_{2}}\bE\Big(\big(ue^{z}e^{\wt{X}_{\theta}}-ue^{\wt{X}_{\theta}}\big){\bf 1}_{\{\wt{X}_{\theta}<\ln(K/u)\leq\wt{X}_{\theta}+z\}}+u\big(e^{z}-1\big)e^{\wt{X}_{\theta}}{\bf 1}_{\{\wt{X}_{\theta}\geq\ln(K/u)\}}\Big)\nu(dz)\\
&\leq K\int_{A_{2}}(e^{z}-1)\nu(dz)\cdot 2\bE\Big(e^{\wt{X}_{\theta}}{\bf 1}_{\{\wt{X}_{\theta}\geq\ln(K/b(T))/2\}}\Big).
\end{align*}
Using an argument similar to those leading to \eqref{eq:EstDiffExphwtXhW3} together with H\"{o}lder's inequality, we deduce that, for some $p\in(1,2)$ and $q>2$ with $p^{-1}+q^{-1}=1$,
\begin{align}\label{eq:AsymIA2}
I(A_{2})\leq 2K\!\int_{A_{2}}\!(e^{z}-1)\nu(dz)\!\cdot\!\left(\bE\big(e^{q\wt{X}_{\theta}}\big)\right)^{1/q}\!\left(\bP\big(\wt{X}_{\theta}\geq\ln(K/b(T))/2\big)\right)^{1/p}\!\!=o(\sqrt{\theta}),\,\,\,\,\theta\rightarrow 0^{+}.\qquad\,
\end{align}
Finally, for $A_{3}=[\ln(K/b(T))/2,\ln(K/b(T)))$, by a similar argument leading to \eqref{eq:AsymIA2} we first have
\begin{align*}
I(A_{3})&=\int_{A_{3}}\bE\Big(\big(ue^{z}e^{\wt{X}_{\theta}}-K\big){\bf 1}_{\{\wt{X}_{\theta}<\ln(K/u)\leq\wt{X}_{\theta}+z\}}+u\big(e^{z}-1\big)e^{\wt{X}_{\theta}}{\bf 1}_{\{\wt{X}_{\theta}\geq\ln(K/u)\}}\Big)\nu(dz)\\
&\leq\int_{A_{3}}\bE\Big(\big(ue^{z}e^{\wt{X}_{\theta}}-ue^{z}\big){\bf 1}_{\{\wt{X}_{\theta}<\ln(K/u)\leq\wt{X}_{\theta}+z\}}\Big)\,\nu(dz)+o(\sqrt{\theta})\\
&\leq K\int_{A_{3}}e^{z}\,\bE\Big(\big(e^{\wt{X}_{\theta}}-1\big){\bf 1}_{\{\wt{X}_{\theta}<\ln(K/u)\leq\wt{X}_{\theta}+z\}}\Big)\,\nu(dz)+o(\sqrt{\theta}),\quad\theta\rightarrow 0^{+}.
\end{align*}
Using the independence between $(W_{t})_{t\in\bR_{+}}$ and $(\wt{Z}_{t})_{t\in\bR_{+}}$, we see that
\begin{align*}
\left|\bE\Big(\big(e^{\wt{X}_{\theta}}-e^{\sigma W_{\theta}}\big){\bf 1}_{\{\wt{X}_{\theta}<\ln(K/u)\leq\wt{X}_{\theta}+z\}}\Big)\right|\leq\bE\big(e^{\sigma W_{\theta}}\big)\bE\left(\big|e^{\wt{Z}_{\theta}}-1\big|\right)=O(\theta),\quad\theta\rightarrow 0^{+}.
\end{align*}
Hence, we deduce that
\begin{align}
I(A_{3})&\leq K\int_{A_{3}}e^{z}\,\bE\Big(\big|e^{\sigma W_{\theta}}-1\big|{\bf 1}_{\{\wt{X}_{\theta}<\ln(K/u)\leq\wt{X}_{\theta}+z\}}\Big)\,\nu(dz)+o(\sqrt{\theta})\\
&\leq\sigma K\sqrt{\theta}\int_{A_{3}}e^{z}\,\bE\Big(|W_{1}|{\bf 1}_{\{W_{1}\geq (\ln(K/b(T))-z-\wt{Z}_{\theta})/\sqrt{\theta}\}}\Big)\,\nu(dz)+o(\sqrt{\theta})\\
\label{eq:AsymIA3} &\leq\sigma K\sqrt{\theta}\!\int_{A_{3}}\!\!e^{z}\!\Big(\bP\big(W_{1}\!\geq\!(\ln(K/b(T))\!-\!z\!-\!\wt{Z}_{\theta})/\sqrt{\theta}\big)\Big)^{1/2}\!\!\nu(dz)\!+\!o(\sqrt{\theta})\!=\!o(\sqrt{\theta}),\,\,\,\theta\rightarrow 0^{+}.\qquad\,
\end{align}
Combining \eqref{eq:LowerBound2gtu}, \eqref{eq:AsymIA1}, \eqref{eq:AsymIA2}, and \eqref{eq:AsymIA3}, we obtain that, as $\theta\rightarrow 0^{+}$,
\begin{align*}
g(t,u)\geq rK-\delta s-\int_{[\ln(K/b(T)),\infty)}\bE\Big(\!\big(K\!-\!ue^{z}e^{\wt{X}_{\theta}}\big)^{+}\!\!\!-\!\big(K\!-\!ue^{\wt{X}_{\theta}}\big)^{+}\!\!\!+\!u\big(e^{z}\!-\!1\big)e^{\wt{X}_{\theta}}\Big)\nu(dz)+o(\sqrt{\theta}).
\end{align*}
Moreover, by \eqref{eq:IntDiffPPe}, \eqref{eq:LowerBound1gtu}, and \eqref{eq:IntPe} (with $\bR_{0}$ replaced by $[\ln(K/b(T)),\infty)$), and noting that $P(t,u)>(K-u)$ for $u>b(t)$, we have, for any $u\in[b(t),s]$, $s\in(b(t),b_{e}(t)\wedge b(T))$, and $t\in(0,T)$,
\begin{align}
g(t,u)&\geq rK-\delta s-\int_{[\ln(K/b(T)),\infty)}\left(P(t,ue^{z})-P(t,u)+u\big(e^{z}-1\big)\right)\nu(dz)+o(\sqrt{\theta})\\
\label{eq:LowerBound3gtu} &\geq rK-\delta s-\int_{[\ln(K/b(T)),\infty)}\left(P(t,ue^{z})-\big(K-ue^{z}\big)\right)\nu(dz)+o(\sqrt{\theta}),\quad\theta\rightarrow 0^{+}.
\end{align}
We are left to estimate the integral term above.

In view of the early exercise premium formula (cf. \cite[Theorem 3.2]{LambertonMikou:2013}), we have that
\begin{align*}
e(\theta,s)=\bE\bigg(\int_{0}^{\theta}e^{-ru}\,\Psi\big(t+u,se^{\wt{X}_{u}}\big)du\bigg),
\end{align*}
where
\begin{align*}
\Psi(t,s)=\bigg(rK-\delta s-\int_{(0,\infty)}\big(P(t,se^{z})-(K-se^{z})\big)\nu(dz)\bigg){\bf 1}_{\{s<b(t)\}},\quad (t,s)\in[0,T)\times\bR_{+}.
\end{align*}
By Theorem \ref{thm:VarIneqAmerPutPriceExpLevy}, the distribution $(\partial/\partial t+\wt{\sA}-r)\wt{P}$ is a nonpositive measure on $(0,T)\times\bR$. On the other hand, by \cite[Proposition 3.1]{LambertonMikou:2013},
\begin{align*}
\bigg(\frac{\partial}{\partial t}+\wt{\sA}-r\bigg)\wt{P}(t,x)= -\Psi(t,e^{x})\quad dt\,dx\text{-a.e.}\,\,\text{on }(0,T)\times\bR.
\end{align*}
Since $\Psi$ is continuous on $[0,T)\times\bR_{+}$, we deduce that $\Psi$ is nonnegative on $[0,T)\times\bR_{+}$. Noting that the integral term in $\Psi$ is nonnegative, we obtain that $0\leq\Psi(t,s)\leq rK$ for any $(t,s)\in[0,T)\times\bR_{+}$. It follows that $e(\theta,s)=O(\theta)$ as $\theta\rightarrow 0^{+}$, and hence we have
\begin{align}
&\int_{[\ln(K/b(T)),\infty)}P(t,ue^{z})\,\nu(dz)=\int_{[\ln(K/b(T)),\infty)}P_{e}(t,ue^{z})\,\nu(dz)+o(\sqrt{\theta})\\
&\quad =\int_{[\ln(K/b(T)),\infty)}\bE\left(\big(K-s\,e^{z+\wt{X}_{\theta}}\big)^{+}\right)\nu(dz)+o(\sqrt{\theta})\\
&\quad =\int_{[\ln(K/b(T)),\infty)}\bE\left(\big(K-s\,e^{z+\sigma W_{\theta}}\big)^{+}\right)\nu(dz)+o(\sqrt{\theta})\\
\label{eq:IntP} &\quad =\int_{[\ln(K/b(T)),\infty)}\bE\left(\big(K-se^{z}(1+\sigma W_{\theta})\big)^{+}\right)\nu(dz)+o(\sqrt{\theta}),\quad\theta\rightarrow 0^{+}.
\end{align}
On the set $(\ln(K/b(T)),\infty)$, we have $b(T)e^{z}>K$, and so
\begin{align*}
&\int_{(\ln(K/b(T)),\infty)}\bE\left(\big(K-se^{z}(1+\sigma W_{\theta})\big)^{+}\right)\nu(dz)\\
&\quad\leq\int_{(\ln(K/b(T)),\infty)}\bE\left(\big(b(T)e^{z}-se^{z}-se^{z}\sigma\sqrt{\theta}W_{1}\big){\bf 1}_{\{se^{z}\sigma\sqrt{\theta}W_{1}<(K-se^{z})\}}\big)\right)\nu(dz)\\
&\quad =(b(T)-s)\int_{(\ln(K/b(T)),\infty)}e^{z}\,\bP\bigg(W_{1}<\frac{Ke^{-z}s^{-1}-1}{\sigma\sqrt{\theta}}\bigg)\nu(dz)\\
&\qquad -\sqrt{\theta}\sigma s\int_{(\ln(K/b(T)),\infty)}e^{z}\,\bE\Big(W_{1}{\bf 1}_{\{W_{1}<(Ke^{-z}s^{-1}-1)/(\sigma\sqrt{\theta})\}}\Big)\nu(dz).
\end{align*}
Since $s>b(t)$, for all $z>\ln(K/b(T))$ we have $Ke^{-z}/s-1<Ke^{-z}/b(t)-1\leq 0$, which implies that
\begin{align*}
\lim_{\theta\rightarrow 0^{+}}\bP\bigg(W_{1}<\frac{Ke^{-z}s^{-1}-1}{\sigma\sqrt{\theta}}\bigg)=0,\quad\lim_{\theta\rightarrow 0^{+}}\bE\Big(|W_{1}|{\bf 1}_{\{W_{1}<(Ke^{-z}s^{-1}-1)/(\sigma\sqrt{\theta})\}}\Big)=0.
\end{align*}
Therefore, by dominated convergence we obtain that, as $\theta\rightarrow 0^{+}$,
\begin{align*}
\varepsilon(\theta):=\int_{(\ln(K/b(T)),\infty)}e^{z}\,\bP\bigg(W_{1}<\frac{Ke^{-z}s^{-1}-1}{\sigma\sqrt{\theta}}\bigg)\nu(dz)&\rightarrow 0,\\
\sqrt{\theta}\sigma s\int_{(\ln(K/b(T)),\infty)}e^{z}\,\bE\Big(W_{1}{\bf 1}_{\{W_{1}<(Ke^{-z}s^{-1}-1)/(\sigma\sqrt{\theta})\}}\Big)\nu(dz)&=o(\sqrt{\theta}).
\end{align*}
Consequently, we deduce that, with $\varepsilon(\theta)\rightarrow 0$ as $\theta\rightarrow 0^{+}$,
\begin{align}\label{eq:UpperBoundIntP1}
\int_{(\ln(K/b(T)),\infty)}\bE\left(\big(K-se^{z}(1+\sigma W_{\theta})\big)^{+}\right)\nu(dz)\leq (b(T)-s)\varepsilon(\theta)+o(\sqrt{\theta}),\quad\theta\rightarrow 0^{+}.
\end{align}
Moreover, on the singleton $\{\ln(K/b(T))\}$, since $s<b(T)$, we have
\begin{align}
&\int_{\{\ln(K/b(T))\}}\bE\left(\big(K-se^{z}(1+\sigma W_{\theta})\big)^{+}\right)\nu(dz)\\
&\quad =\int_{\{\ln(K/b(T))\}}\big(K-se^{z}\big)\nu(dz)+\int_{\{\ln(K/b(T))\}}\bE\left(\big(se^{z}\sigma W_{\theta}-(K-se^{z})\big)^{+}\right)\nu(dz)\\
&\quad =\int_{\{\ln(K/b(T))\}}\big(K-se^{z}\big)\nu(dz)+\frac{sK}{b(T)}\nu(\{\ln(K/b(T))\})\,\bE\Big(\big(\sigma W_{\theta}-(s^{-1}b(T)-1)\big)^{+}\Big)\\
\label{eq:UpperBoundIntP2} &\quad\leq\int_{\{\ln(K/b(T))\}}\big(K-se^{z}\big)\nu(dz)+K\nu(\{\ln(K/b(T))\})\,\bE\Big(\big(\sigma W_{\theta}-\ln(b(T)/s)\big)^{+}\Big).
\end{align}
Combining \eqref{eq:IntP}, \eqref{eq:UpperBoundIntP1}, and \eqref{eq:UpperBoundIntP2}, we obtain that, as $\theta\rightarrow 0^{+}$,
\begin{align}
\int_{[\ln(K/b(T)),\infty)}P(t,ue^{z})\,\nu(dz)&\leq (b(T)-s)\varepsilon(\theta)+\int_{\{\ln(K/b(T))\}}\big(K-se^{z}\big)\nu(dz)\\
\label{eq:UpperBoundIntP} &\quad +K\nu(\{\ln(K/b(T))\})\bE\Big(\big(\sigma W_{\theta}-\ln(b(T)/s)\big)^{+}\Big)+o(\sqrt{\theta}).\quad
\end{align}

Finally, combining \eqref{eq:LowerBound3gtu} and \eqref{eq:UpperBoundIntP}, and noting that $b(T)$ satisfies \eqref{eq:LimitAmerCritPriceNegd}, we obtain that
\begin{align*}
g(t,u)&\geq rK\!-\!\delta s\!+\!\!\int_{(\ln(K/b(T)),\infty)}\!\!\!\!\big(K\!-\!se^{z}\big)\nu(dz)\!-\!(b(T)\!-\!s)\varepsilon(\theta)\!-\!K\lambda\bE\Big(\!\big(\sigma W_{\theta}\!-\!\ln(b(T)/s)\big)^{+}\!\Big)\!+\!o(\sqrt{\theta})\\
&=(b(T)-s)\bigg(\delta+\int_{(\ln(K/b(T)),\infty)}e^{z}\nu(dz)-\varepsilon(\theta)\bigg)-K\lambda\bE\Big(\big(\sigma W_{\theta}-\ln(b(T)/s)\big)^{+}\Big)+o(\sqrt{\theta})\\
&=\big(\bar{\delta}-\varepsilon(\theta)\big)(b(T)-s)-b(T)\bar{\delta}\lambda\beta\,\bE\Big(\big(\sigma W_{\theta}-\ln(b(T)/s)\big)^{+}\Big)+o(\sqrt{\theta}),\quad\theta\rightarrow 0^{+},
\end{align*}
which, together with \eqref{eq:LowerBoundInt2ndDeriP}, completes the proof of the lemma.\hfill $\Box$

\bibliographystyle{plain}

\end{document}